\documentclass{cas-sc}
\usepackage{graphicx}
\usepackage{float}
\usepackage{amsmath,amsthm, amssymb, bm, subcaption}
\usepackage{placeins}
\usepackage{longtable,tabularx,booktabs}
\usepackage{soul}
\usepackage{miller}
\usepackage[most]{tcolorbox}
\usepackage{tikz}
\usepackage{enumitem}
\usepackage{atbegshi}
\usetikzlibrary{calc}
\usepackage[export]{adjustbox}  
\definecolor{C0}{HTML}{1F77B4}
\definecolor{C1}{HTML}{FF7F0E}
\definecolor{C2}{HTML}{2ca02c}
\definecolor{C3}{HTML}{d62728}
\definecolor{C4}{HTML}{9467bd}
\definecolor{C5}{HTML}{8c564b}

\ifdefined\usetodonotes
\usepackage{todonotes}
\newcommand{\tododone}[1]{\todo[disable]{#1}\addcontentsline{tdo}{todo}{\st{#1}}}

\else
\newcommand{\tododone}[1]{}
\fi

\ifdefined\usechanges
  
\usetikzlibrary{calc}
\usepackage{fontawesome}
\usepackage[commentmarkup=todo,authormarkup=none]{changes} 
\definechangesauthor[color=C0]{R1} 
\definechangesauthor[color=C1]{R2} 
\definechangesauthor[color=C3]{RA} 
\makeatletter
\@namedef{Changes@AuthorColor}{C3}
\colorlet{Changes@Color}{C3}
\renewcommand{\Changes@Markup@comment}[3]{%
  \IfStrEq{\Changes@optioncommentmarkup}{todo}%
		{\colorlet{Changes@todocolor}{authorcolor}\todo[color=Changes@todocolor!10, bordercolor=Changes@todocolor, linecolor=Changes@todocolor!70, nolist]{\faBookmarkO\ \textbf{#1}}}{}}
\makeatother
\newtcolorbox{addedbox}[3][]
{
  colframe=#2,
  colback=#2!10,
  boxrule=0pt,
  title={\ \ \ \ \faBookmarkO\ \textsf{\textbf{#3}}},
  boxsep=1mm,
  left=-1mm,
  right=-1mm,
  enhanced,
  width=\linewidth,
  #1,
}

\AtBeginShipout{%
    \AtBeginShipoutUpperLeft{%
        \begin{tikzpicture}[overlay,remember picture]
            \fill [gray!25] (current page.north east) rectangle
                ($ (current page.south east) - (3cm,0cm) $);
        \end{tikzpicture}
    }
}

\else
  
\NewDocumentEnvironment{addedbox}{ m m }{}{}
\usepackage[final,nocomment]{changes}
\fi

\usepackage[capitalize]{cleveref}

\makeatletter
\define@key{Gin}{Crop3D}[]{\setkeys{Gin}{trim=21mm 15mm 80mm 22mm,clip}}
\makeatother

\makeatletter                                                                   
\newlength{\bibsep}{\@listi \global\bibsep\itemsep \global\advance\bibsep by\parsep} 
\makeatother 

\begin{document}

\ifdefined\usetodonotes
\setcounter{page}{0}
\listoftodos
\fi

\shorttitle{}    
\shortauthors{Runnels}  
\title [mode = title]{Mechanics of incompatible asymmetric grain boundary migration}
\author
{Brandon Runnels}[orcid=0000-0003-3043-5227]
\ead{brunnels@iastate.edu}
\affiliation
{organization={Department of Aerospace Engineering, Iowa State University},
            city={Ames},
            state={IA},
            country={USA}}

\begin{abstract}
  Grain boundary (GB) migration governs microstructure evolution and can mediate plastic deformation through sliding or shear coupling.
  Numerous experimental and numerical studies have reported a wide range of behaviors associated with boundary migration, such as defect emission or mode switching.
  Notably, recent studies have reported directionally asymmetric migration rates under symmetric loading, attributing this behavior to intrinsically asymmetric mobility; however, a mechanistic mesoscale explanation for this behavior remains lacking.
  In this work, we introduce a constitutive flow rule for grain-boundary eigendeformation within a multiphase-field framework, in which interfacial shear evolves in response to its mechanically conjugate driving force through the phase field Allen-Cahn equations.
  The formulation systematically employs regularized grain boundary shear kinematics informed by crystallography, and enables elastic compatibility to modulate boundary motion.
  Migration thresholds, residual back-stress, and apparent directional asymmetry appear naturally as emergent mechanical behavior.
  \replaced[id=R1,comment={1.3}]{Simulations of symmetric and asymmetric tilt grain boundaries under mechanical, synthetic, and curvature-driven loading reveal persistent defect-like residuals following incompatible migration, transitions from planar motion to lamination at large inclinations, and driving-mode-dependent ``ratcheting'' behavior.}{Simulations of symmetric and asymmetric tilt grain boundaries under mechanical, synthetic, and curvature-driven loading reveal persistent defect-like residuals following incompatible migration, transitions from planar motion to lamination at large inclinations, and even ``ratcheting'' behavior.}
  These results provide a mechanically transparent explanation for behaviors such as effective mobility asymmetry and establish elastic compatibility as a constitutive mechanism in mesoscale models of boundary-mediated plasticity.
\end{abstract}


\maketitle

\section{Introduction}

Microstructure evolution in metals can lead to a wide range of material properties ranging from electric to mechanical.
For high symmetry crystalline materials, such as nickel, copper, aluminum, etc., the desirable behavior of ductility is largely carried by dislocation motion.
Microstructure evolution also  plays a key role in effecting plastic behavior \cite{bugas2024grain}.
In particular, materials such as those with a hexagonal close-packed (hcp) structure, are more likely to accommodate deformation through the nucleation and growth of twins.
Deformation-induced twinning is an example of a broader class of ``boundary-mediated plasticity'' which is the accommodation of permanent deformation through the evolution of phase or grain boundaries in microstructure.
Like crystal plasticity, boundary-mediated plasticity originates from lattice-scale defect migrations and admits mesoscale homogenization.
But despite its importance, a comprehensive coupling between general microstructure evolution and ductility remains elusive, and is an area of active investigation \cite{dai2025unveiling}.

Modeling approaches to study boundary-mediated plasticity include such atomistic methods as density functional theory \cite{suhane2022solute}, transition state theory \cite{song2023driving} and nudged elastic band \cite{rajabzadeh2013elementary,combe2016disconnections,joshi2025energetics}, molecular dynamics \cite{olmsted2009survey,rohrer2023grain}, phase field crystal \cite{xiao2025controlling,skogvoll2022hydrodynamic}, and coarse-grained atomistics such as Gaussian Phase Packet \cite{spinola2024finitetemperature}.
Atomistic simulations are reliable and repeatable (despite sometimes exhibiting significant potential dependence \cite{waters2023automated}), and are often the tool of choice for studying grain boundary phenomena.
Methods at this resolution are nevertheless limited to varying degree to atomistic time and length scales.
Moreover, aside from time and lengthscale limitations, atomistic methods cannot consider decoupled mechanical behavior, making it difficult to directly connect complex atomistic simulations to reduced-order mechanical models (such as plasticity or curvature-driven flow).

Mesoscale phase field methods offer an alternative perspective: mechanisms are introduced explicitly, allowing migration to be decomposed into distinct thermodynamic and mechanical contributions.
While this limits the accessible physics, it provides interpretability and makes phase field particularly suited to isolating the mechanical origins of complex behaviors such as mobility asymmetry.
\added[id=R1,comment={1.1}]{
The key link between the atomistic boundary migration mechanism and macroscale plasticity is the shear coupling factor $\beta$, originally defined by \cite{read1950dislocation} as the ratio between disconnection Burgers vector and step height, and later popularized and calculated for a wide range of boundaries \cite{cahn2006coupling}.
Originally $\beta$ was known to be enumerable for any rational symmetric boundary, and was more recently shown to be enumerable for any rational boundary as well \cite{admal2022interface}.
The transformation mechanisms corresponding to the different enumerations of $\beta$ for a given boundary are known as the shear coupling modes.
Of the countably infinite number of modes available to any rational boundary, usually only one or two are realized \cite{chen2020grain}, though the exact mechanism by which a mode is selected remains an open question.
}

\replaced[id=R1,comment={1.1}]{
A variety of complementary methods have been proposed in the modeling of grain boundary mediated plasticity.
A disconnection-based method for grain boundary migration was proposed by \cite{thomas2017reconciling,zhang2017equation,han2018grain} and since applied to polycrystals \cite{chen2024revealing}, the phase field method \cite{salvalaglio2022disconnection}, grain rotation \cite{qiu2024disconnection}, triple points \cite{thomas2019disconnection}, and multi-mode migration \cite{wei2019continuum}.
By considering a boundary as containing a statistical ensemble of geometrically necessary disconnections, this method uses disconnection theory to rigorously derive boundary migration.
Elastic coupling is achieved though the superposition of the linear elastic stress field associated with a single disconnection.
While this is effective in determining the resultant self-stress fields from disconnection theory, it does not represent swept shear as a finite-deformation, history-dependent eigendeformation field in a diffuse-interface continuum model.
Alternatively, other approaches to modeling grain boundary plasticity have been proposed that treat boundaries using a modification of disclination \cite{acharya2012coupled,acharya2015continuum} or dislocation \cite{joshi2022finite} defect theories.
By treating boundaries as arrays of geometrically necessary dislocations, it is possible to simulate shear coupling and traditional plasticity in a unified way \cite{he2022polycrystal}.
However, because most boundaries are not reducible to dislocation or disclination defects, it is not possible to capture the full range of boundary migration and shear coupling in this way.
Yet another approach was presented by the author which developed grain boundary shear coupling as a dissipative plastic mechanism in its own right, in which the flow rule is dictated by the crystallography-informed shear coupling factor and the accumulated slip is determined by the interface position.
This has been applied to bicrystals \cite{chesser2020continuum}, polycrystals \cite{bugas2024grain}, and phase field \cite{gokuli2021multiphase}, the latter of which was combined with grain boundary anisotropy to simulate ``phase field disconnections''.
This family of approaches, which form the basis of the present work, capture both the crystallography and the global nonlinear mechanics boundary migration, but require a global reference frame and grain-wise eigenstrains to capture the effect of swept shear.
}
{
However, recent developments in the mechanics of boundary mediated plasticity have yet to be fully accounted for in phase field models, particularly regarding the inclusion of grain boundary shear coupling matrices, the resulting eigenstrain, and the corresponding driving force.
}
Additional development is needed to properly model boundary-mediated plasticity at the mesoscale.

The central contribution of this work is a phase field formulation for boundary-mediated plasticity in which the eigendeformation jump across phase boundaries evolves according to a kinematic flow rule, providing a mechanical driving force conjugate to the boundary shear.
Unlike existing phase field approaches that prescribe shear coupling through a static grain-wise eigendeformation, this work presents a model that derives these effects naturally by regularizing the grain boundary eigendeformation and effecting history-dependent eigenstrain evolution.
As a result, such a framework can capture important behaviors including the emergence of migration thresholds \cite{yu2021survey}, back-stress \cite{stangebye2025grain}, defect emission \cite{race2015mechanisms}, lamination arising from driven incompatible boundaries \cite{dmitrieva2009lamination}, and apparent mobility asymmetry that manifests under symmetric kinetic laws \cite{qiu2024grain}.
By allowing these features to emerge without explicit prescription, this formulation provides a mechanically transparent mesoscale description of boundary-mediated plasticity.

This paper is structured in the following way.
\Cref{sec:model} presents the new grain boundary eigenstrain model in a familiar phase field model for microstructure evolution.
\Cref{sec:implementation} provides an overview of the salient implementation details.
\Cref{sec:examples} provides a set of three classes of results, each oriented towards addressing a particular gap in current phase field capability: (\ref{sec:stabilization}) stabilization of nucleation, (\ref{sec:atgb}) simulation of asymmetric tilt grain boundaries, and (\ref{sec:ratcheting}) a mechanistic explanation for the phenomenon of apparent orientation-dependence (``ratcheting'') in grain boundary migration.
Finally, the limitations of the model and implications for future work are summarized in \cref{sec:conclusion}.

\section{Model}\label{sec:model}

This section presents the phase field formulation used in this work.
The model is based on a standard multiphase field description of microstructure coupled to elasticity, and adds an interfacial eigendeformation whose evolution is governed by a constitutive flow rule.
After introducing the general setting and strain energy decomposition, the grain boundary flow rule and its associated mechanical driving force are derived.

\subsection{General setting: multiphase field model of microstructure}

As the general setting for this work, a multiphase field scheme is used for describing microstructure and its evolution under a mechanical load.
For $N$ grains, let $\eta$ be an $N-$component vector of order parameters with the support of $\eta_n$ corresponding to grain $n$.
Diffuseness of the boundaries requires that $\eta_n$ be at least Lipschitz; however most phase field models include a gradient term, placing the stronger requirement of a continuous second spatial derivative on $\eta_n$.
Models in which the boundary energy is anisotropic \cite{ribot2019new}, or that are evolved using Cahn-Hilliard kinetics, have even stronger continuity requirements; these are not considered here.

We restrict attention to curvature and mechanical driving, neglecting anisotropic boundary energy and higher-order regularization to isolate incompatibility effects.
As stated previously, the torque resulting from anisotropic boundary energy will be neglected, as well as any higher-order regularization, to isolate the effects of incompatibility in our presentation.
The free energy of the microstructure is
\begin{align}
  \mathrm{W}[\eta,\bm{\varphi}] = \int_\Omega \Big[ \frac{1}{2}\ell_{gb}\,\sigma\,|\nabla\eta|^2 + \mathrm{U}[\eta,\mathbf{F}]\Big]\,\mathrm{d}\bm{x},
\end{align}
where $\ell_{gb}$ describes the diffuse boundary width, $\sigma$ is the isotropic boundary energy, $\bm{\varphi}$ is the deformation mapping, $\mathbf{F}=\nabla\bm{\varphi}$ is the deformation gradient, and $\mathrm{U}$ is the elastic strain energy density.

The multiphase field description of microstructure requires that $\eta$ partition unity, and that exactly one component of $\eta$ be equal to one in the majority of the domain except at the grain boundaries.
In most systems this is realized through the addition of a multi-well chemical potential function of $\eta$ that is minimized when exactly one component of $\eta$ is unity and the others are zero.
Alternatively, it may be enforced as an explicit constraint on the system.
Here, the former is chosen.

Following \cite{chesser2020continuum,gokuli2021multiphase}, the kinetics of the system are derived from the principle of minimum dissipation potential for grain boundary migration,
\begin{gather}
  \inf_{\dot{\eta}}\Big\{\frac{\partial}{\partial t}\Big[\inf_{\bm{\varphi}\,\text{ adm.}}W[\eta,\mathbf{F}]\Big] + \phi^*(\dot\eta)\Big\} \notag\\
  \text{subject to }\, \int_\Omega\Gamma(\eta)\,d\bm{x} < \epsilon(\ell), \label{eq:min_diss_pot}
\end{gather}
where $\Gamma$ is a multi-well function with minima of $0$ when exactly one $\eta_n=1$,  and $\epsilon>0$ is a tolerance corresponding to the diffuse boundary width.
Note that setting $\epsilon=0$ would eliminate the existence of any non-trivial solution.
The function $\phi^*$ is a dual dissipation potential, having the form
\begin{align}
  \phi^*(\dot{\eta}) = \sum_n\Big(\phi_n^0|\dot\eta_n| + \frac{1}{2}\phi^1_n\dot\eta_n^2\Big)
\end{align}
where $\phi^0_n$ and $\phi^1_n$ are rate-independent and rate-dependent coefficients, respectively.
\added[id=R1,comment={1.2}]{The rate-independent coefficient $\phi^0_n$ is interpreted as the critical driving force necessary to drive the boundary; the rate-dependent coefficient $\phi^1_n$ is the ratio of the driving force to the boundary velocity (i.e. the reciprocal of the mobility).}
In principle, the value of these coefficients is dependent on boundary character; however, we consider them here to be isotropic along with the grain boundary energy.
The solution of \cref{eq:min_diss_pot} leads to the following kinetic relation for $\eta$:
\begin{align}
  \frac{\partial\eta_n}{\partial t} &=
  \lambda\,\frac{\partial\Gamma(\eta)}{\partial\eta_n}
  -\frac{1}{\phi^1_n}
  \begin{cases}
    \delta W^*/\delta\eta_n - \phi^0_n & \delta W^*/\delta\eta_n > +\phi^0_n\\
    \delta W^*/\delta\eta_n + \phi^0_n & \delta W^*/\delta\eta_n < -\phi^0_n\\
    0 & |\delta W^*/\delta\eta_n| \le \phi^0_n
  \end{cases},
\end{align}
\replaced[id={R1},comment={1.2}]{where $\lambda$ can be considered either a Lagrange multiplier enforcing the boundary localization penalty, or can be calculated based on the grain boundary energy $\sigma$.
Here, it is determined based on the latter.}{where $\lambda$ is a Lagrange multiplier whose value depends on $\epsilon$ (and consequently $\ell_{gb}$) and is chosen by calibration.}
$\delta/\delta\eta_n$ indicates the variational derivative, and the asterisk on $W$ indicates minimization with respect to $\bm{\phi}$ by quasi-static solution of mechanical equilibrium.
In the absence of rate-independent terms ($\phi^0_n=0$) and elastic terms, the above reduces to basic curvature-driven flow.
Rate-independent coefficients act as a threshold\footnote{Not to be confused with the ``threshold dynamics'' family of methods for solving similar problems}, rendering the kinetic relation nonlinear and producing a variety of interesting behaviors in phase field methods \cite{hu2025atomisticinformed,guin2023phasefield,liang2012nonlinear}.
In this work, the specific forms of $\sigma$, $\ell_{gb}$, $\Gamma$, etc., follow \cite{moelans2008quantitative,moelans2008quantitativea} and subsequent works, though the flow rule framework itself can be generalized to any phase field method.
\added[id=R1,comment={1.2}]{For completeness, a set of parameters and their interpretation are provided in \cref{tab:parameter_summary}, and input files for all simulations are supplied in the supplementary information.}

\subsection{Strain energy decomposition}\label{sec:grain_boundary_flow_rule}

This work is particularly concerned with the influence of mechanical loading on microstructure evolution.
Mechanical loading appears in the model through the strain energy functional $U$.
In a multiphase field context, the following decomposition for $U$ is often considered:
\begin{align}
  U(\eta,\mathbf{F}) = \sum_{j}^N g_j(\eta)\,U_j(\mathbf{F}) \label{eq:basic_u_mix}
\end{align}
where
\added[id=R1,comment={1.2}]{$U_j$ is the strain energy corresponding to grain $j$,}
$g_i:[0,1]^N\to[0,1]$ is an interpolation function such that $\{g_i(\eta)\}_{i=1}^N$ is guaranteed to partition unity (even if $\eta$ does not), and that satisfies $g_i=\eta_i$ when $\eta_i=\delta_{ij}$ for any $1\le j\le N$.
Importantly, it is also required that
\begin{align}\label{eq:dg_deta_zero}
  \frac{\partial g_i}{\partial \eta_i}\Big|_{\eta_i=0,1} = 0.
\end{align}
The derivative requirement eliminates linear forms for $g$, such as $g_i=\eta_i$ or even $g_i=\eta_i/\sum_{j}\eta_j$.
The simplest example, which will be used here, is $g_i(\eta) = \eta_i^2/\sum_j\eta_j^2$, though numerous other options are possible.
The mechanical driving force results from taking the variational derivative of \cref{eq:basic_u_mix}, yielding
\begin{align}
  \frac{\delta U}{\delta\eta_i} = \sum_{j} \frac{\partial g_j}{\partial\eta_i}\,U_j(\mathbf{F}),
\end{align}
showing that the mechanical driving force to transform from grain $i$ to grain $j$ is proportional to the difference in mechanical strain energies in the respective grains.
The need for \cref{eq:dg_deta_zero} is now apparent to ensure that the mechanical driving force is zero in grain interiors.

The strain energy difference is dependent upon the form of the mechanical models that are selected.
In a polycrystal, a driving force can result from anisotropy in the elastic moduli, or from different eigenstrains.
A common approach is to attach an eigenstrain to each phase, $U(\mathbf{F})\mapsto U(\mathbf{F}\mathbf{F}_0^{-1})$, such that each phase is considered to be altered from the ``reference state'' by its own unique deformation $\mathbf{F}_0$.
This approach is attractive due to its stability and its natural incorporation of eigenstrain through mechanical strain energy differences.
However it is limited in its ability to communicate information about the microstructure to the mechanical models $U_i$: each grain has a single immutable deformation regardless of history.
As such, there is no delineation between material that has been swept by a boundary and material that has not, making it impossible to fully capture the effect of shear coupling in the presence of incompatible deformation modes.

\subsection{Flow rule}

Mechanical compatibility acts as a mechanical selector that determines whether boundary-mediated shear can occur.
This subsection develops a regularization of the grain boundary eigendeformation jump that is amenable to incorporation in phase field.
To do so, the eigendeformation field is decoupled from individual phases, allowing it to evolve as a field variable.
In a finite deformation framework, a multiplicative decomposition is assumed for the deformation gradient in the style of the elastic-plastic decomposition:
\begin{align}
  \mathbf{F} = \mathbf{F}^e\mathbf{F}^{gb},
\end{align}
where $\mathbf{F}^e$ is the elastic deformation and $\mathbf{F}^{gb}$ is the eigendeformation resulting from boundary motion.
Crystal plasticity could be incorporated by extending the decomposition $\mathbf{F}=\mathbf{F}^e\mathbf{F}^{p}\mathbf{F}^{gb}$, but this lies outside the present scope.
The elastic strain energy is then expressed as 
\begin{align}
  U(\eta,\mathbf{F}) = \sum_{j=1}^N g_j(\eta)\,U_j(\mathbf{F}\mathbf{F}^{gb-1}_j).\label{eq:new_strain_energy}
\end{align}
Here, $\mathbf{F}^{gb}_j$ is tracked per grain, even though it is no longer a constant grain-wise parameter (hence the subscript on $\mathbf{F}^{gb}_j$).

To express the kinematics of shear coupling between phases, the rank-1 GB shear tensor $\Delta\mathbf{F}^{gb}_{ij}$ is used (which was extensively defined and used in \cite{bugas2024grain}), and which expresses the signed deformation induced by the positive motion of a boundary between \replaced[id=R1,comment={1.4}]{grains $i$ and $j$}{grain i and grain j}.
A boundary with no shear coupling has $\Delta\mathbf{F}^{gb}=\mathbf{0}$, and the signedness of the shear tensor implies that $\Delta\mathbf{F}^{gb}_{ij} = -\Delta\mathbf{F}^{gb}_{ji}$. 

The precise form of $\Delta\mathbf{F}^{gb}$ for any particular boundary can be inferred from crystallography.
From an atomistic perspective, shear coupling is understood to be mediated through the nucleation and motion of disconnections, which are grain boundary defects characterized by \replaced[id=R1,comment={1.4}]{Burgers vector}{burgers vector} as well as step height.
At the mesoscale, the shear mediated by the aggregate migration of disconnections can then be expressed as
\begin{align}
  \Delta\mathbf{F}^{gb}_{ij} = \mathbf{b}\otimes\mathbf{h},
\end{align}
where $\bm{b}$ is the Burgers vector and $\bm{h}$ a covector with magnitude equal to the inverse of the step height.
The deformations on either side of the boundary with normal $\mathbf{n}$ may then be expressed as
\begin{align}
  \mathbf{F}_1 &= \mathbf{F}_1^e\big(\mathbf{F}_0 + \frac{1}{2}\Delta\mathbf{F}_{12}\big)
  &
  \mathbf{F}_2 &= \mathbf{F}_2^e\big(\mathbf{F}_0 - \frac{1}{2}\Delta\mathbf{F}_{12}\big).\label{eq:F1_and_F2}
\end{align}
where $\mathbf{F}_0$ is a pre-existing eigendeformation.
The Hadamard compatibility condition requires that the difference $[[\mathbf{F}]]=\mathbf{F}_1-\mathbf{F}_2$ be rank-one connected, and that $[[\mathbf{F}]]\,\mathbf{t}=0$ for all $\mathbf{t}\cdot\mathbf{n}=0$. 
Substitution of \cref{eq:F1_and_F2} into the compatibility condition yields
\begin{align}
  (\mathbf{F}_1^e+\mathbf{F}_2^e)^{-1}(\mathbf{F}_1^e-\mathbf{F}_2^e)\,\mathbf{F}_0\,\mathbf{t}  = - \frac{1}{2}\Delta\mathbf{F}_{12}^{gb}\,\mathbf{t}.
\end{align}
If the elastic deformation is small, then $\mathbf{F}^e_i+\mathbf{F}^e_j\approx 2\mathbf{I} >> \mathbf{F}^e_i-\mathbf{F}^e_j$, and the expression simplifies
\begin{align}
  [[\mathbf{F}^e]]\,\mathbf{F}_0\,\mathbf{t}  \approx - \Delta\mathbf{F}_{12}^{gb}\,\mathbf{t}.\label{eq:backstressjump}
\end{align}
Thus, a jump in the elastic deformation (and the associated back-stress) is required when the shear coupling deformation is not rank-one compatible with the interface.
On the other hand, if the boundary is aligned with the step height vector, then it can be expressed as 
\begin{align}
  \Delta\mathbf{F}_{ij} = \beta\,\mathbf{t}\otimes\mathbf{n}, 
\end{align}
in which $\beta = \frac{|\mathbf{b}|}{|\mathbf{h}|}$ is the shear coupling factor.
Such a configuration naturally produces no back-stress, and corresponds to the classic case of pure shear coupling of symmetric tilt grain boundaries.

Now, consider the evolution of the grain boundary eigenstrain in phase field.
For simplicity, it is assumed that the boundary region $|\nabla\eta|$ has compact support in the $\epsilon$ neighborhood of the sharp-interface limit.
This allows the regularization of the jump condition in weak form,
\begin{align}
  \int_{\text{grain i}}^\text{grain j} d\mathbf{F}^{gb}
  = \Delta \mathbf{F}^{gb}_{ij}\int_0^1d g_i
  = \frac{1}{2}\Delta \mathbf{F}^{gb}_{ij}\int_0^1dg_i  + \frac{1}{2}\Delta \mathbf{F}^{gb}_{ji}\int_1^0dg_j.
\end{align}
where $g_i$ is an indicator function varying smoothly from 0 outside grain $i$ to 1 inside grain $i$.
Dropping the integral and expressing in differential form, employing the summation convention, gives
\begin{align}
  d\mathbf{F}^{gb}_i = \frac{1}{2}\Delta\mathbf{F}_{ij}^{gb} dg_j.
\end{align}
If the indicator variables $\mathbf{g}$ are interpolation functions on a set of order parameters $\mathbf{\eta}$ (as described above) then expansion of the differential yields:
\begin{align}
  d\mathbf{F}^{gb}_i = \frac{1}{2}\Delta\mathbf{F}_{ij}^{gb} \,\frac{\partial g_j}{\partial\eta_k}\,d\eta_k.
\end{align}
All rational boundaries possess a countably infinite number of shear coupling modes, and some boundaries are known to exhibit shear coupling with respect to different modes.
Then, the shear coupling matrix may be expressed as
\begin{align}
  \Delta\mathbf{F}^{gb}_{ij} = \alpha_n\Delta\mathbf{F}^{gb}_{ijn}\label{eq:fgb_flowrule}
\end{align}
where $\alpha_n$ controls mode selection over the available shear coupling modes indexed by $n$.
This leads to the final expression for evolution of grain boundary eigendeformation in grain $i$: 
\begin{align}
  d\mathbf{F}^{gb}_i 
  = \alpha_n\Delta\mathbf{F}^{gb}_{ijn}\frac{\partial g_j}{\partial\eta_k}\,d\eta_k.
\end{align}

\subsection{Mechanical driving force}
\replaced[id=R1,comment={1.4}]{Because of the spatial variation of the eigendeformation with $\eta$, the mechanical driving force must be taken into account explicitly.}{Because of the spatial variation of the eigendeformation with $\eta$, the mechanical driving force must take this into account explicitly.}
Taking the variational derivative of the driving force \cref{eq:new_strain_energy}, substituting in the flow rule (\cref{eq:fgb_flowrule}) in the differentiation of the argument, yields
\begin{align}
  \frac{\partial U}{\partial\eta_i}
  = \sum_{j} \Bigg(&\frac{\partial g_j}{\partial\eta_i}\,U_j(\mathbf{F}\mathbf{F}^{gb-1}_j)
                            + \frac{1}{2}g_j(\eta)\,\mathbf{P}_j:\Big(\mathbf{F}\frac{\partial\mathbf{F}^{gb-1}_j}{\partial\mathbf{F}_j^{gb}}\alpha_n\Delta\bm{F}^{gb}_{jin}\frac{\partial g_j}{\partial\eta_i}\Big)\Bigg), \label{eq:exact_df}
\end{align}
where $:$ indicates tensor double contraction.
Evidently, in \cref{eq:exact_df} there are two players in the mechanical driving force: the first term drives the boundary to relieve elastic modulus mismatch, while the second drives the boundary to relieve stress by accumulating shear.
In a typical phase field implementation, the driving force on $\eta$ is calculated as above, and then \cref{eq:fgb_flowrule} used to update the eigenstrain based on the evolving order parameter.
The calculation of the driving force is closed-form and straightforward.
It can be further simplified through the application of two simplifications that impose minimal error on the solution while streamlining the formulation.
\begin{enumerate}
\item
  {\bf Common normal disconnections.}
  Most grain boundary migrations calculations involve the activation of a single mode at a time.
  In the simulation of such simple systems, (especially simple bicrystals, as in all of the examples presented here), it may be safely assumed that the eigenstrain at any point is simple shear: 
  \begin{align}
    \mathbf{F}^{gb}_{i} = \bm{I} + \bm{a}\otimes\bm{n}
  \end{align}
  where $\bm{a}\perp\bm{n}$, which can accommodate the activation of multiple possible disconnection modes as long as they have a common normal.
  With this simplification, it is straightforward to show that $\mathbf{F}^{gb-1}_i=2\mathbf{I}-\mathbf{F}^{gb}$, and consequently that $d\mathbf{F}_i^{gb-1} = - d\mathbf{F}_i^{gb}$.
  Applying this approximation simplifies the inverse derivative second term in \cref{eq:exact_df}, simplifying the second part of the  driving force to
  \begin{align}
    - g_j(\eta)\,\mathbf{P}_j:\Big(\mathbf{F}\alpha_n\Delta\mathbf{F}^{gb}_{jin}\frac{\partial g_j}{\partial\eta_i}\Big).
  \end{align}
\item
  {\bf Linearization of elastic field about eigendeformation.}
  If there is not substantial rotation taking place, and as long as the elastic part of the deformation is small, it is admissible to use the following linearization of the elastic model,
  \begin{align}
    U_i(\mathbf{F}\mathbf{F}^{gb-1}_i) &\approx \tilde{U}(\mathbf{F}-\mathbf{F}^{gb}_i) = \frac{1}{2}(\mathbf{F} - \mathbf{F}^{gb}_i):\mathbb{C}_i(\mathbf{F} - \mathbf{F}^{gb}_i).
  \end{align}
  This replaces the multiplicative decomposition with the additive decomposition, which substantially simplifies the variational derivative, leaving
  \begin{align}
    \frac{\partial U}{\partial\eta_i}
    = \sum_{j}^N \Bigg(\frac{\partial g_j}{\partial\eta_i}\,\hat{U}_j(\mathbf{F}-\mathbf{F}^{gb}_j)
    - g_j(\eta)\,\mathbf{P}_j:\alpha_n\Delta\mathbf{F}^{gb}_{jin}\frac{\partial g_j}{\partial\eta_i}\Bigg). \label{eq:simp2_df}
  \end{align}
  From this formulation, it is clear that $\mathbf{P}:\Delta\mathbf{F}^{gb}$ drives shear-coupled motion, as expected \cite{chesser2020continuum}. 
\end{enumerate}

Since the eigendeformation evolves only when mechanically admissible, compatibility causes the generation of threshold behavior and back-stress without explicitly invoking yield criteria or asymmetric mobility.
Importantly, the flow rule and driving force themselves are symmetric, so any observed directional effects arise from elastic incompatibility and resulting back-stress.

\section{Implementation}\label{sec:implementation}

The computational phase field and quasistatic mechanical equilibrium equations are solved using Alamo, an in-house open-source multi-physics solver \cite{runnels2025alamo} which is based on the Block-Structured Adaptive Mesh Refinement (BSAMR) code AMReX \cite{zhang2019amrex}.
BSAMR is an AMR strategy that divides the domain into different levels of refinement, facilitating communication through averaging / interpolating between levels.
Each level is refined independently, and temporal subcyling is used to evolve each level at a different timestep to increase performance and stability.

Selective refinement ensures adequate resolution across regions of interest, without incurring excessive computational cost.
During regridding, the solution at each node is evaluated to determine if resolution is sufficient.
Criteria for refinement is based on a gradient condition for the solution variables (the order parameter $\eta$, the deformation gradient field $\mathbf{F}$).
The following criteria are used to determine regridding on level $n$:
\begin{align}
  \nabla\eta_i | \Delta \mathbf{x}_n| &> 0.1 & |\nabla\mathbf{F}||\Delta \mathbf{x}_n| > 0.01
\end{align}
where $|\Delta\mathbf{x}_n|$ is the length of the diagonal of the unit cell and $|\nabla\mathbf{F}|$ is the Hermitian norm of the third order Hessian of the deformation.
This ensures fidelity of the mechanical equilibrium solution as well as the phase field solution.
Regridding is generally set to occur more frequently than necessary, and gridding parameters are set to ensure a wide buffer region around regions of interest to ensure smoothness of the solution.

Grain migration is typically much slower than the speed of sound, so the phase field mechanical driving force is informed by the quasi-static solution to the mechanical equilibrium equations.
This requires an implicit, finite-deformation solution to be run each timestep, although testing has indicated that up to about 10 phase field integration timesteps can be taken between each elastic solve as long as the timestep is sufficiently small.
Alamo employs a unique strong-form method to solve the finite deformation equilibrium equations using both BSAMR and geometric multigrid (methods presented in \cite{runnels2021massively,agrawal2023robust}, extended to hyperelasticity in \cite{meier2024finite}).
Notably, the mechanical solver was extended in this work to account for periodic boundary conditions, which was not previously possible.

\section{Examples} \label{sec:examples}

\begin{table}
  \begin{addedbox}{C0}{1.2}
    \caption{Parameter summary for the simulations reported in \cref{sec:examples}. Numerical values are listed separately from material and crystallographic inputs because only the latter are intended as directly calibratable material quantities.}
    \label{tab:parameter_summary}
    \small
    \begin{tabularx}{\linewidth}{@{}p{0.3\linewidth}p{0.12\linewidth}p{0.25\linewidth}p{0.25\linewidth}@{}}
      \toprule
      \textbf{Quantity} & \textbf{Notation} & \textbf{Nondimensional stabilization} & \textbf{Dimensional Al bicrystals}\\
      \midrule

      \multicolumn{4}{@{}l}{\textbf{Multiphase-field material and kinetic inputs}}\\
      Chemical potential barrier height & $\gamma$ & $10.0$ \cite{moelans2008quantitative,moelans2008quantitativea} & $10.0$ \cite{moelans2008quantitative,moelans2008quantitativea}\\
      Diffuse boundary width & $\ell_{gb}$ & $0.05$ & $1\,\mathrm{nm}$\\
      Grain-boundary energy & $\sigma_0$ & $0.05$ & $0.443\,\mathrm{J/m^2}$ \cite{homer2022examination}\\
      Mobility & $M=1/\phi_1$ & $1.0$ & $1\times10^{-11}\,\mathrm{m/(s\,Pa)}$ \cite{qiu2024grain}\\
      Explicit threshold / dissipation & $\phi_0$ & $0$ & $0$ (unless specified)\\
      Synthetic driving force & $\psi_n$ & 0 & $\psi_{1,2}=-\psi_{2,1}=\pm80\times10^7\,\mathrm{J/m^3}$ (SDF only) \cite{qiu2024grain}\\

      \midrule
      \multicolumn{4}{@{}l}{\textbf{Elasticity and grain-boundary eigendeformation}}\\
      Elastic constants & $C_{11},C_{12},C_{44}$ & $1.71$, $1.22$, $0.75$ & $108.2$, $61.3$, $28.3\,\mathrm{GPa}$ \cite{qiu2024grain}\\
      Crystallography & $\beta,\phi$ & $\beta=0.2$, $\phi=0^\circ,45^\circ$ & See \cref{tab:asymmetry_crystallography} \\
      \bottomrule
    \end{tabularx}
  \end{addedbox}
\end{table}

The following examples are organized to isolate the mechanical consequences of the proposed model.
\Cref{sec:stabilization} demonstrates the stabilizing effect on inclusions such as grain or twin nuclei.
\Cref{sec:atgb} illustrates the effects of boundary incompatibility, showing the spontaneous emergence of defects and lamination in response to asymmetric driven boundary migration.
\Cref{sec:ratcheting} examines the emergence of apparent mobility asymmetry.
\added[id=R1,comment={1.2}]{
The parameters used in these examples are summarized in \cref{tab:parameter_summary}.
The intent is to separate model parameters with direct physical interpretation, such as boundary energy, mobility, elastic constants, and crystallographic shear coupling, from numerical or deliberately nondimensional choices used to isolate mechanisms.
This organization emphasizes that the dimensional aluminum examples use material and crystallographic inputs drawn from measurable quantities or atomistic reference data, while the nondimensional stabilization example is used only as a controlled mechanical demonstration.
}

\subsection{Stabilization}\label{sec:stabilization}

\newcommand{\imgwithlabel}[3][]{%
\begin{tikzpicture}
  \node[inner sep=0] (img) {\includegraphics[#1]{#2}};
  \node[black] at (img.south west) [anchor=south west,xshift=2mm,yshift=2mm] {#3};
\end{tikzpicture}%
}
\begin{figure}
  \begin{subfigure}{0.3\linewidth}
    \includegraphics[width=\linewidth]
{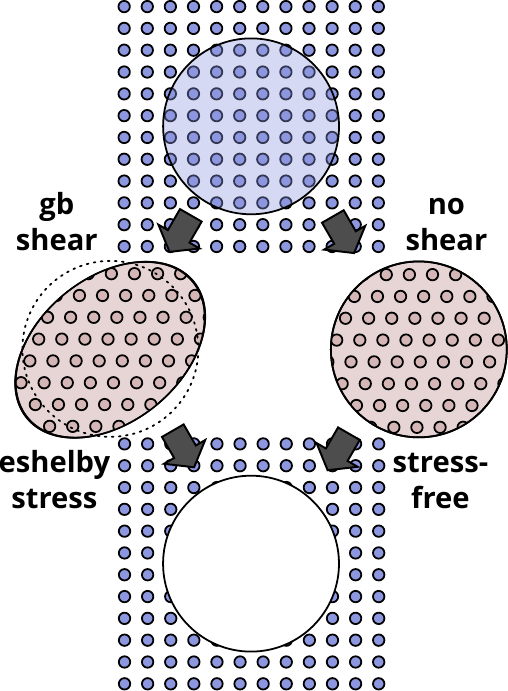}
    \caption{Schematic illustrating construction of eigenstrain-free inclusion}
    \label{fig:results/Inclusion/drawing}
  \end{subfigure}\hfill
  \begin{subfigure}{0.65\linewidth}
      \imgwithlabel[width=0.33\linewidth]
{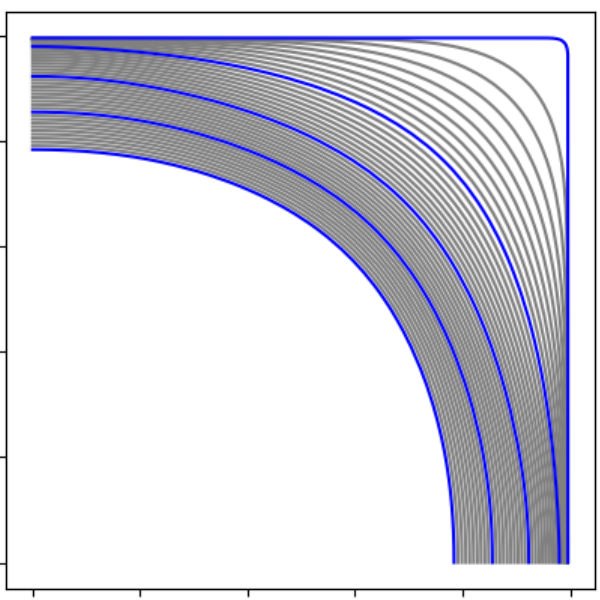}{$\theta=0^\circ$, fac=1}%
      \imgwithlabel[width=0.33\linewidth]
{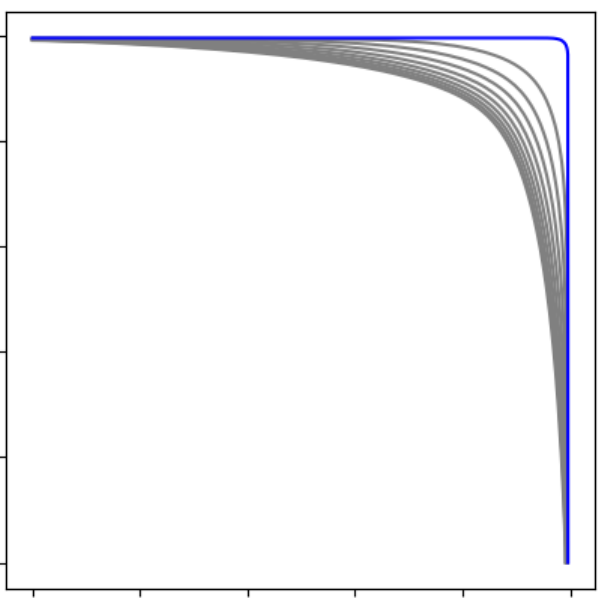}{$\theta=0^\circ$, fac=5}%
      \imgwithlabel[width=0.33\linewidth]
{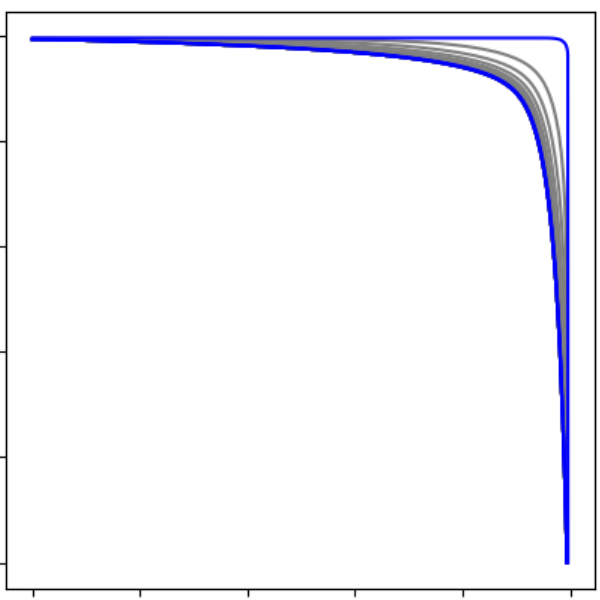}{$\theta=0^\circ$, fac=10}
      \imgwithlabel[width=0.33\linewidth]
{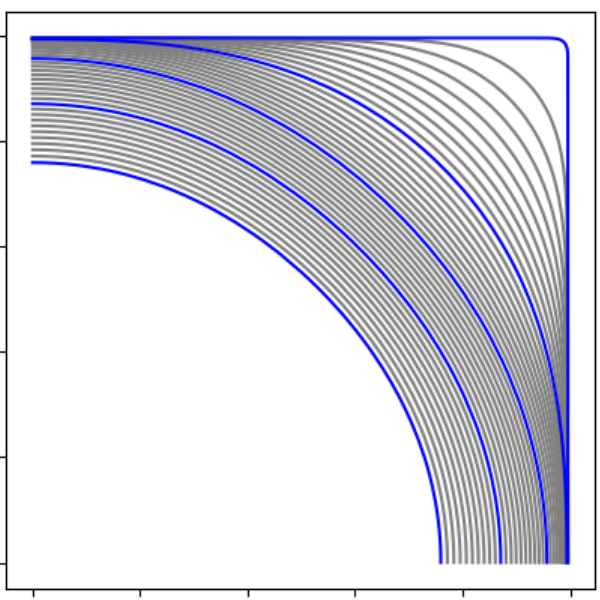}{$\theta=45^\circ$, fac=1}%
      \imgwithlabel[width=0.33\linewidth]
{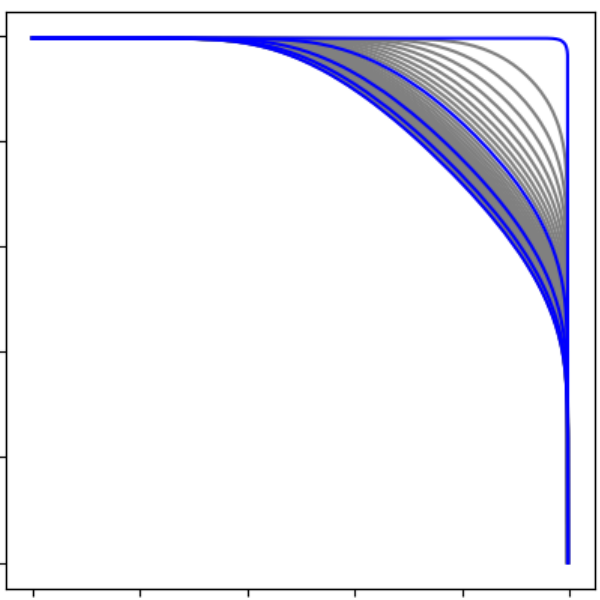}{$\theta=45^\circ$, fac=5}%
      \imgwithlabel[width=0.33\linewidth]
{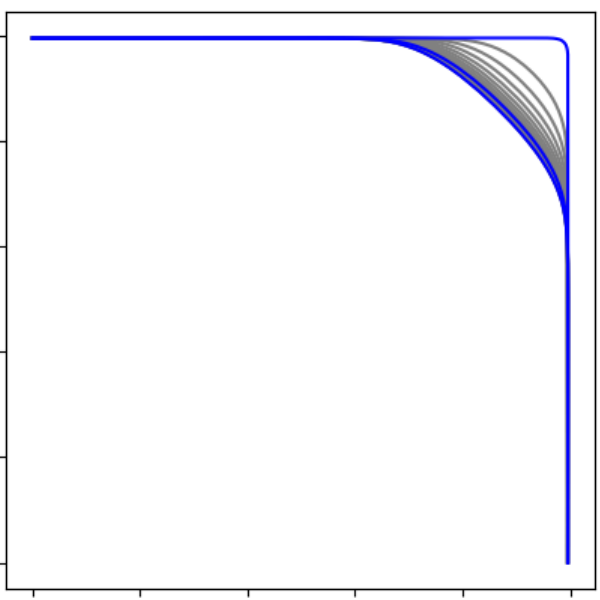}{$\theta=45^\circ$, fac=10}

      \caption{
      Stabilization of a square inclusion by formation of elastic back-stress.
      Contours correspond to equal time intervals, with each 10th in blue.
      }
      \label{fig:results/Stabilization}
  \end{subfigure}
  \caption{
  Stabilization of a square inclusion by compatibility-induced back-stress.
  No initial eigendeformation is imposed, so the inclusion is stress-free at initialization.
  Parameter $\theta$ denotes the twin-boundary orientation relative to the inclusion, and ``fac'' scales the elastic contribution to the driving force.
  Increasing ``fac'' arrests motion preferentially at incompatible corners.
  }
  \label{fig:stabilization}
\end{figure}

Modeling nucleation and mechanically-driven growth of phases or grains with shear coupling or transformation strains is challenging in phase field methods, because assigning a grain-wise eigenstrain turns the nucleus into an Eshelby inclusion (\cref{fig:results/Inclusion/drawing}).
The resulting stress can dominate the driving force and cause artificial shrinkage, inflating the apparent critical radius \cite{hu2025atomisticinformed}.

In the present approach, an inclusion is initialized by setting the order parameter without imposing an initial eigendeformation.
The initial state is therefore stress-free, and eigendeformation develops only as the boundary moves.
This makes the nucleus mechanically stable at initialization, and sets up a competition between curvature and mechanically induced back-stress.
This is tested by simulating the behavior of a square inclusion whose shear coupling relationship with the matrix
\begin{align}
  \Delta\mathbf{F}^{gb} = \begin{bmatrix} 0 & 0.2 \\ 0 & 0 \end{bmatrix}
\end{align}
relative to the symmetric tilt (twin) boundary.
A square inclusion is considered that is aligned with the twin boundary ($\theta=0^\circ$) and rotated ($\theta=45^\circ$).
Nondimensionalized phase field parameters are used: $\gamma=10$, diffuse thickness $\ell_{gb}=0.05$, grain boundary energy $\sigma_0=0.05$.
The grains are pseudolinear cubic with elastic moduli $C_{11}=1.71,C_{12}=1.22,C_{44}=0.75$.
The domain is $[4,4]^2$ and the inclusion is initialized to $[2,2]^2$.
On the four boundaries, Dirichlet (zero-displacement) conditions are applied to the displacement field in the normal direction, and Neumann conditions are applied in the tangential direction.

Two grain boundary orientations are considered: $\theta=0^\circ$ (so that the sides of the cube are twin boundaries) and  $\theta=45^\circ$ (\cref{fig:results/Stabilization}).
For each of these, cases, the relative magnitude of the elastic driving force is changed.
The value fac corresponds to the multiplying factor in the elastic contribution to the phase field driving force, with values of $1,5,10$.
For fac=1, the boundary evolution is primarily curvature-driven.
However, as the elastic energy increases, the sides of the cube are significantly less mobile.
Slightly more motion of the sides occurs for $\theta=0^\circ$ due to their compatibility with the shear coupling matrix.
For $\theta=45^\circ$, there is almost no motion along the edges, as they are not compatible and cause back-stress with any amount of migration.
For $\theta=45^\circ$, motion along the edges is nearly suppressed, reflecting incompatibility with the shear coupled deformation.
The resulting beveling of the corners indicates preferential migration along compatible directions.

\subsection{Shear coupling of symmetric and asymmetric grain boundaries}\label{sec:atgb}

\begin{figure}
  \begin{subfigure}{0.48\linewidth}
    \includegraphics[height=6cm]
{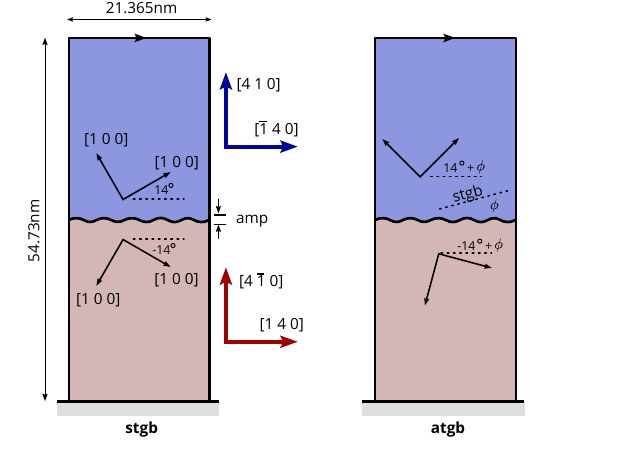}
    \caption{Schematic illustrating the configuration for the symmetric tilt boundary (left) and family of asymmetric tilt boundaries, in which the top and bottom grains are rotated by the same ATGB tilt angle $\phi$.}
    \label{fig:results/Lamination/drawing}
  \end{subfigure}%
  \hfill%
  \begin{subfigure}{0.5\linewidth}

    \includegraphics[height=6cm]
{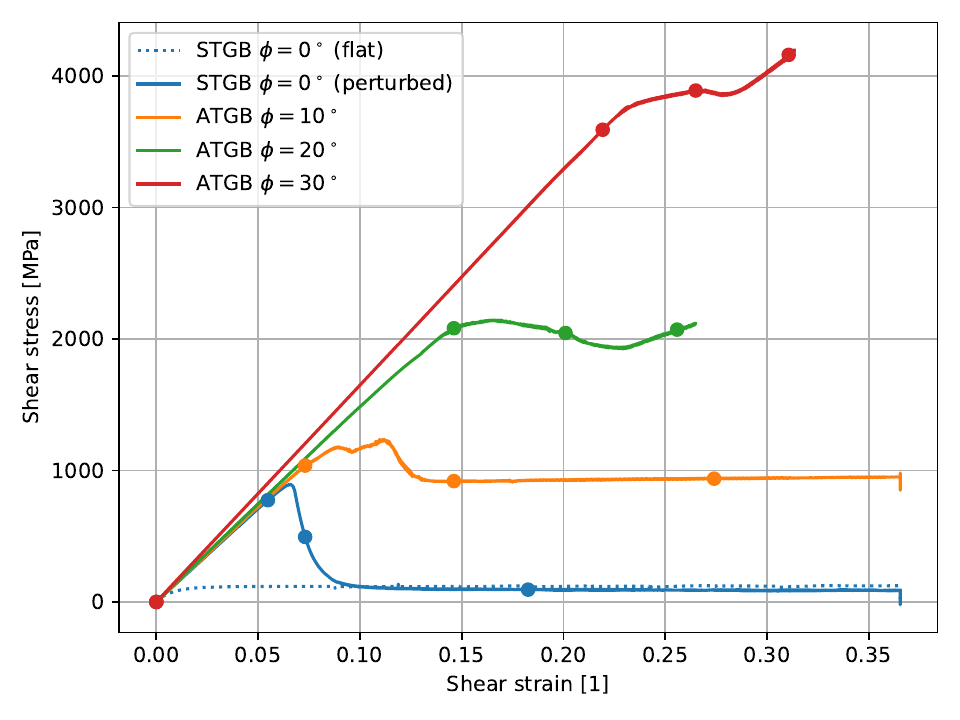}
    \caption{
    Stress strain curves for symmetric tilt boundary ($\phi=0^\circ$, with and without initial perturbation) and asymmetric tilt boundaries.
    Markers correspond to states of the boundary in snapshots presented in
    \cref{fig:results/Lamination/output_0deg_nothreshold_iso,fig:results/Lamination/output_10deg_nothreshold_iso,fig:results/Lamination/output_20deg_nothreshold_iso,fig:results/Lamination/output_30deg_nothreshold_iso}
    }
    \label{fig:results/Lamination/stress_strain}
  \end{subfigure}
  \caption{
  Shear coupling of symmetric and asymmetric tilt boundaries.
  }
\end{figure}

Shear coupling is the canonical example of grain boundary-mediated deformation, whereby an applied shear load causes motion of a boundary, sweeping a region that undergoes permanent shear deformation determined by $\Delta\mathbf{F}^{gb}$.
Shear coupling of STGBs is reasonably well-studied and understood; it is possible to determine the shear coupling factor $\beta$ crystallographically \cite{han2018grainboundary}, and driven boundary simulations show consistent and repeatable results.
The same is not true for ATGBs.
While it is possible to determine disconnection modes for ATGBs \cite{admal2022interface}, atomistic investigations into ATGB migration fail to reveal the same predictable behavior observed in \replaced[id=R1,comment={1.4}]{STGBs}{STGBS}.
In this section, we apply the grain boundary flow rule to understand the effect of asymmetry on boundary migration.

For the cases considered here, phase field parameter values are $\gamma=10$, $\ell_{gb}=1nm$, $\sigma_0=0.443\frac{J}{m^2}$.
A finite deformation pseudolinear (technically, pseudo\textit{affine}) model for cubic elasticity is used, though most deformations fall within the linearized regime about $\mathbf{I}+\Delta\mathbf{F}^{gb}$.
Elastic constants are $C_{11}=108.2\mathrm{GPa}, C_{12}=61.3\mathrm{GPa}, C_{44}=28.3\mathrm{GPa}$, (aluminum), with proper rotations applied depending on the configuration of the boundary to account for elastic anisotropy. 
No explicit thresholding is considered, which is a deliberate choice to accentuate the natural thresholding induced by incompatibility.

The boundary that is selected for consideration is the $\Sigma17$ $\hkl<100>$ tilt boundary.
This boundary is known to have a shear coupling factor of $\beta=0.5$ \cite{cahn2006coupling}.
This symmetric tilt boundary is indicated as $\phi=0^\circ$, where $\phi$ is an asymmetric inclination of the boundary away from the STGB plane (\cref{fig:results/Lamination/drawing}).
ATGBs corresponding to $\phi=10^\circ,20^\circ,30^\circ$ are also considered.
Similar behaviors are observed for a wide range of boundary characters; here, only this representative boundary is presented for the sake of brevity.

The domain size is $21.365\textrm{nm}$ by $54.73\textrm{nm}$.
Dirichlet conditions for the order parameters $\eta_1,\eta_2$ and the displacement $\bm{u}$ are applied at the top and bottom.
Periodic boundary conditions are applied on the sides.
The domain is evenly divided with the boundary plane along the $y=0$ line.
Except for the first case, a slight perturbation with amplitude $1\textrm{nm}$ is applied to the interface (discussion below).
The boundary is loaded through a prescribed displacement along the top, beginning at $0$ and increasing to $10\textrm{nm}$ over $20\mathrm{\mu s}$., resulting in a shear strain rate of approximately $100s^{-1}$, with a maximum overall shear strain of $35\%$.
After loading, for $\phi\le10^\circ$, the boundary is held and allowed to relax.
Five boundaries are considered: a planar STGB, a perturbed STGB, and perturbed ATGBs at 10$^\circ$ increments up to 30$^\circ$.
For each case, the engineering stress-strain response is measured (\cref{fig:results/Lamination/stress_strain}).

\textbf{STGB (unperturbed)}: 
Unsurprisingly, the unperturbed STGB presented a negligible yield/flow stress.
The very small stress that is present in the unperturbed STGB is the result of rate effects, as shown by the drop in stress to zero after the load ceases to increase.
(Snapshots for the unperturbed STGB are available on request, but are not included here as there is little remarkable about them.)

\begin{figure}
  \centering
  \includegraphics      [height=3cm,clip,trim=1cm 15cm 36cm 0cm]
{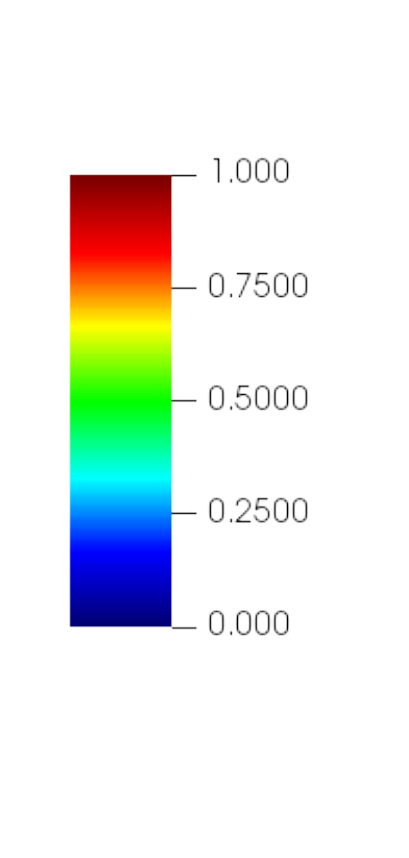}%
  \includegraphics      [height=3cm,clip,trim=8cm 0cm 0cm 0cm]
{9e1fc207f4e64d10.pdf}%
  \hfill\includegraphics[height=3cm,clip,trim=8cm 0cm 0cm 0cm]
{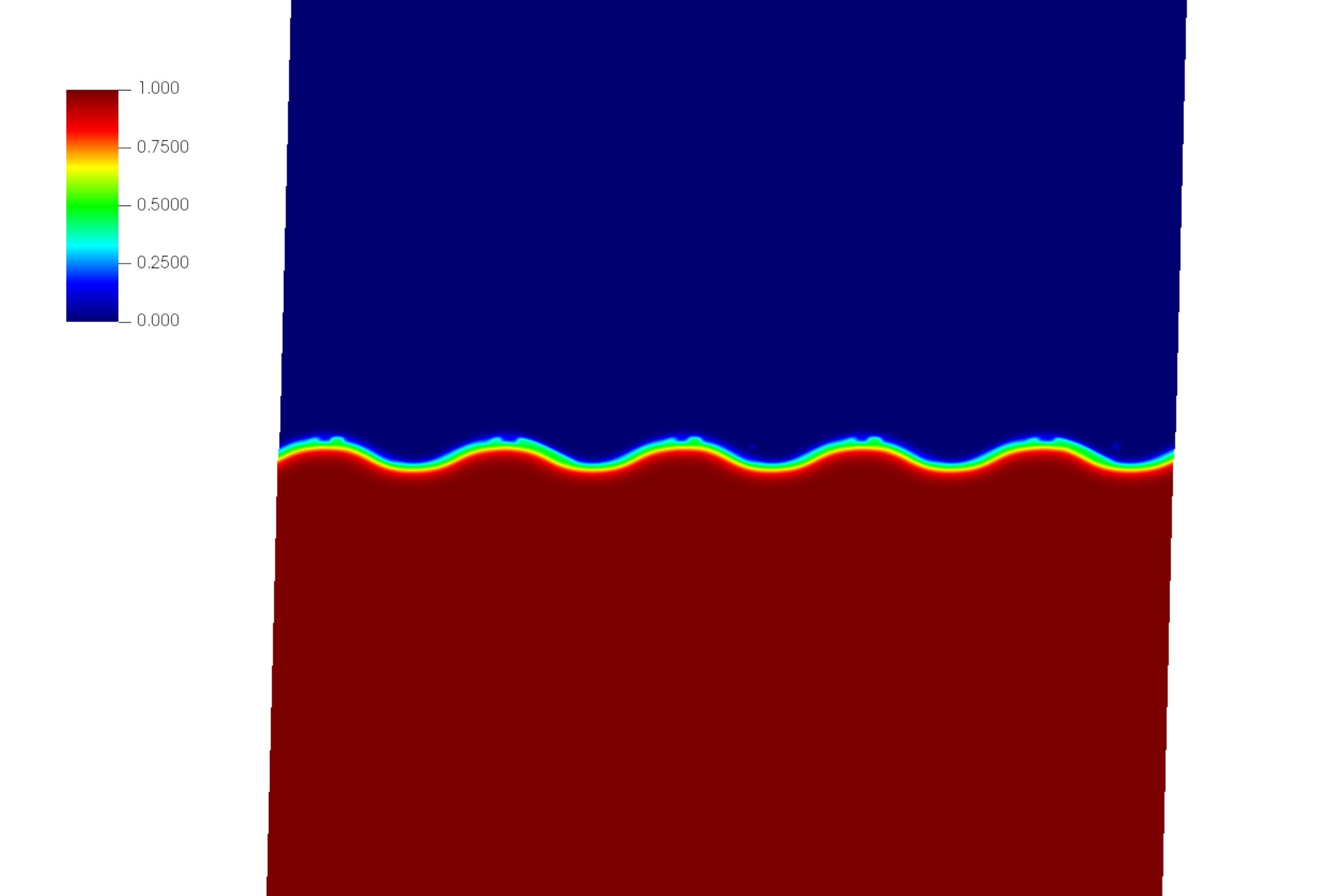}%
  \hfill\includegraphics[height=3cm,clip,trim=8cm 0cm 0cm 0cm]
{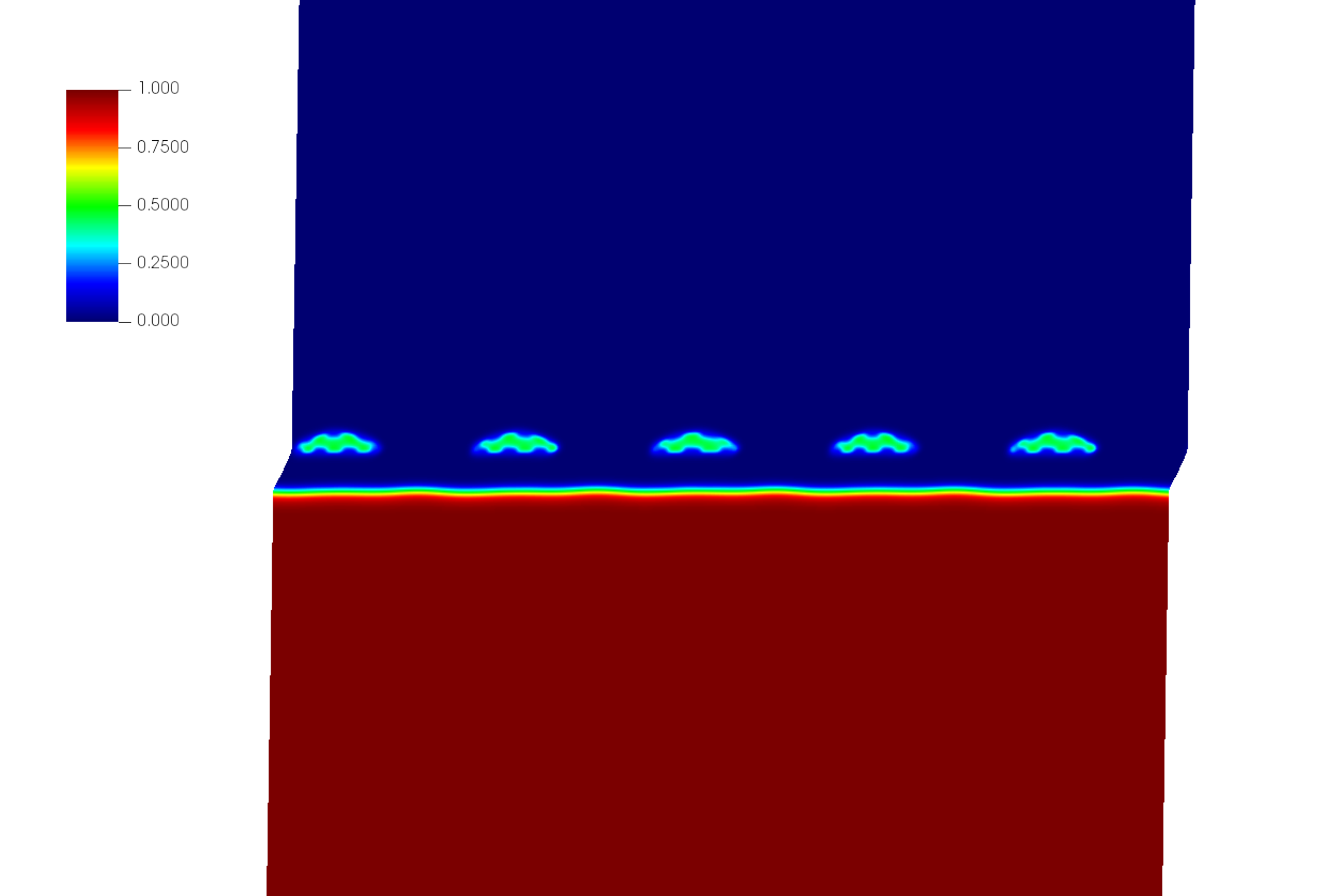}%
  \hfill\includegraphics[height=3cm,clip,trim=8cm 0cm 0cm 0cm]
{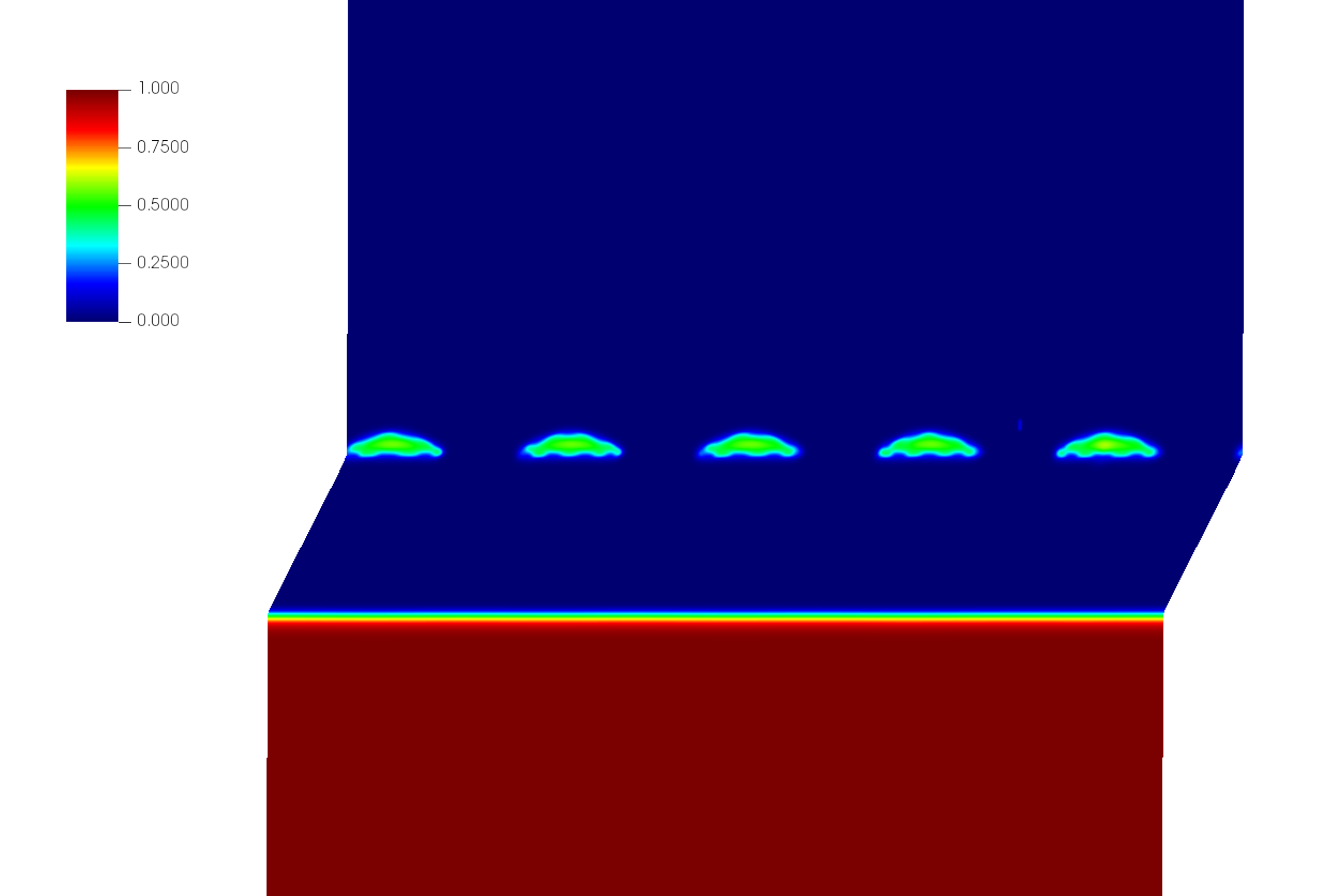}

  \centering
  \includegraphics      [height=3cm,clip,trim=1cm 15cm 36cm 0cm]
{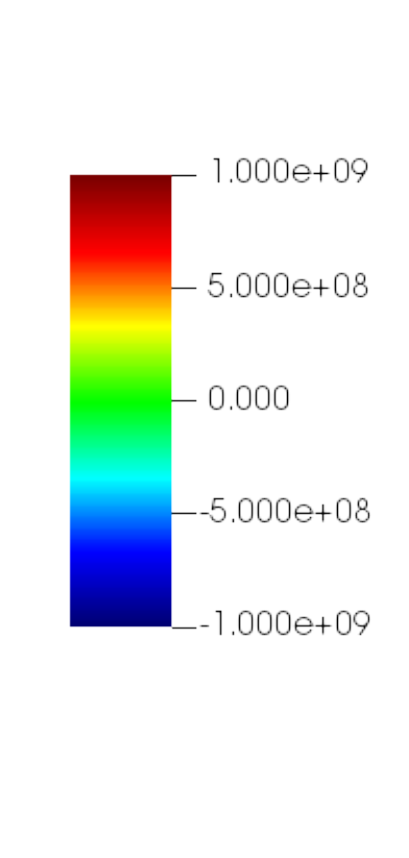}%
  \hfill\includegraphics[height=3cm,clip,trim=8cm  0cm  0cm 0cm]
{ff903e6893968eed.pdf}%
  \hfill\includegraphics[height=3cm,clip,trim=8cm  0cm  0cm 0cm]
{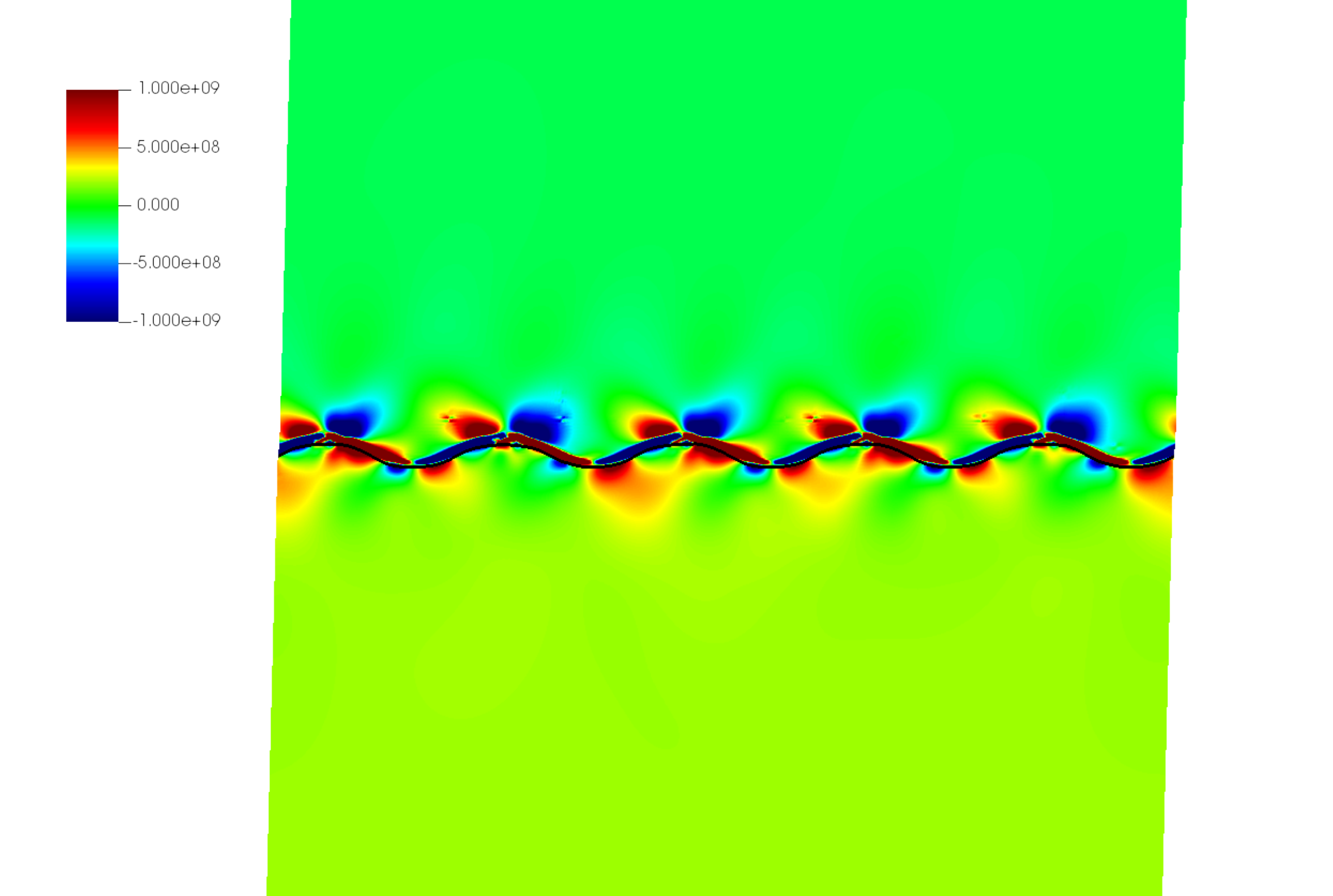}%
  \hfill\includegraphics[height=3cm,clip,trim=8cm  0cm  0cm 0cm]
{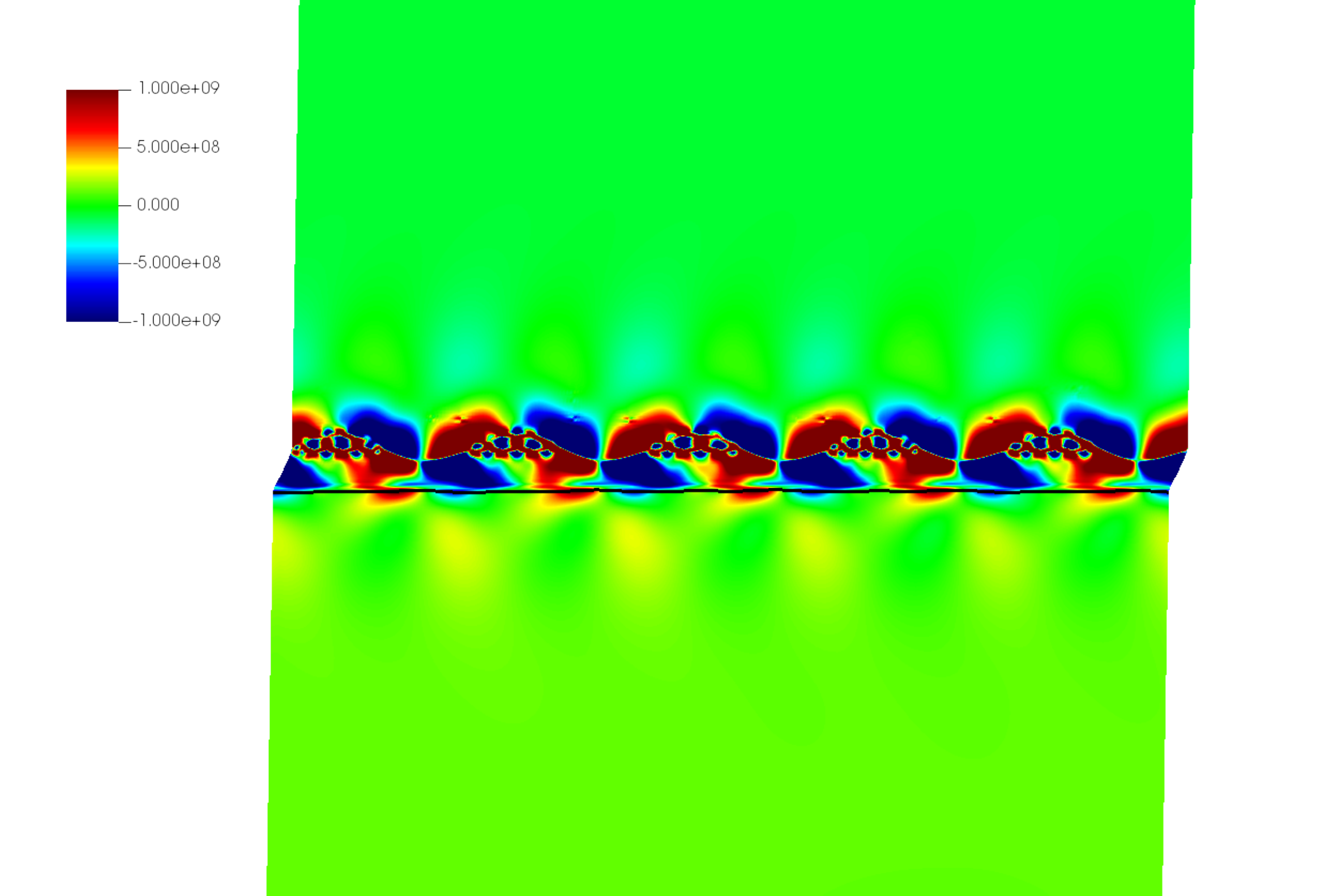}%
  \hfill\includegraphics[height=3cm,clip,trim=8cm  0cm  0cm 0cm]
{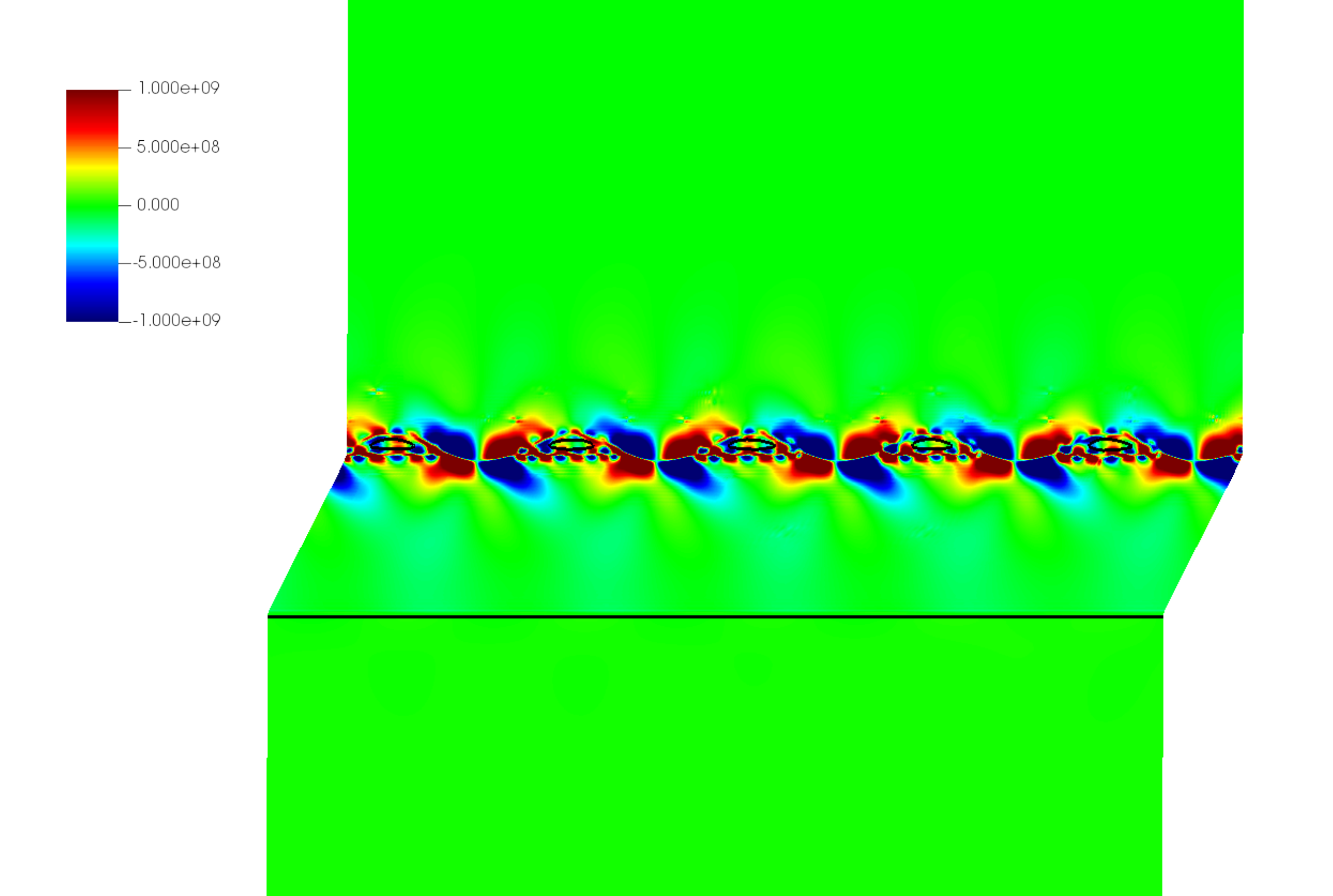}
  \caption{
  Shear-coupled migration of the STGB with a slightly perturbed interface.
  Intervals show $\eta_1 [-]$ and $\sigma_{xx} [\mathrm{MPa}]$ at intervals corresponding to the markers in \cref{fig:results/Lamination/stress_strain}.
  }
  \label{fig:results/Lamination/output_0deg_nothreshold_iso}
\end{figure}

\textbf{STGB (perturbed)}: 
The perturbed STGB (\cref{fig:results/Lamination/output_0deg_nothreshold_iso}) exhibits a clear threshold, with the boundary remaining pinned up to about 7\% strain, even though no rate-independent term is used.
Once the stress exceeds this level, the boundary migrates and the flow stress matches the unperturbed case.
Migration of the (now) flat boundary leaves behind small residual inter-phase regions and stress doublets reminiscent of defect dipoles, consistent with back-stress generated by locally incompatible segments of the perturbed interface.
A planar STGB can migrate with negligible back-stress, whereas a perturbed interface contains both compatible and incompatible segments.
Compatible segments migrate readily; incompatible segments accumulate elastic mismatch, producing residual stress and defect-like remnants.

Some boundaries exhibit an initial resistance to motion, followed by steady subsequent motion \cite{chesser2020continuum}.
This is often explained in terms of the boundary's atomistic structure: the particular state of a boundary may not be conducive to motion, but once a certain driving force is achieved, it may transition to a different structure that is able to move more easily \cite{chesser2021optimal}.
Such behavior can sometimes result in defects - dislocation loops or vacancies/interstitials - from the boundary's structural transition.
The present results provide an analogous, mechanistic explanation for this effect.
However, the present work considers grain boundary migration, with no other damage mechanisms present.
For robust interpretation of residual stresses and left-behind defects, a coupled plasticity model is essential and will be considered in future work.

\begin{figure}
  \centering
  \includegraphics      [height=3cm,clip,trim=1cm 15cm 36cm 0cm]
{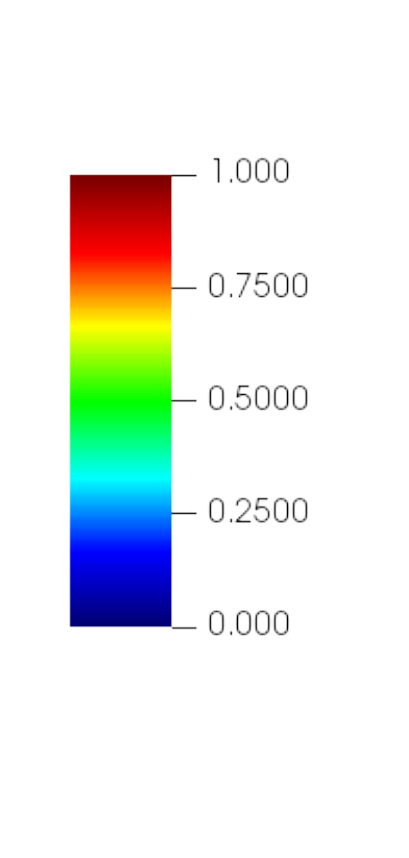}%
  \includegraphics      [height=3cm,clip,trim=8cm  0cm  0cm 0cm]
{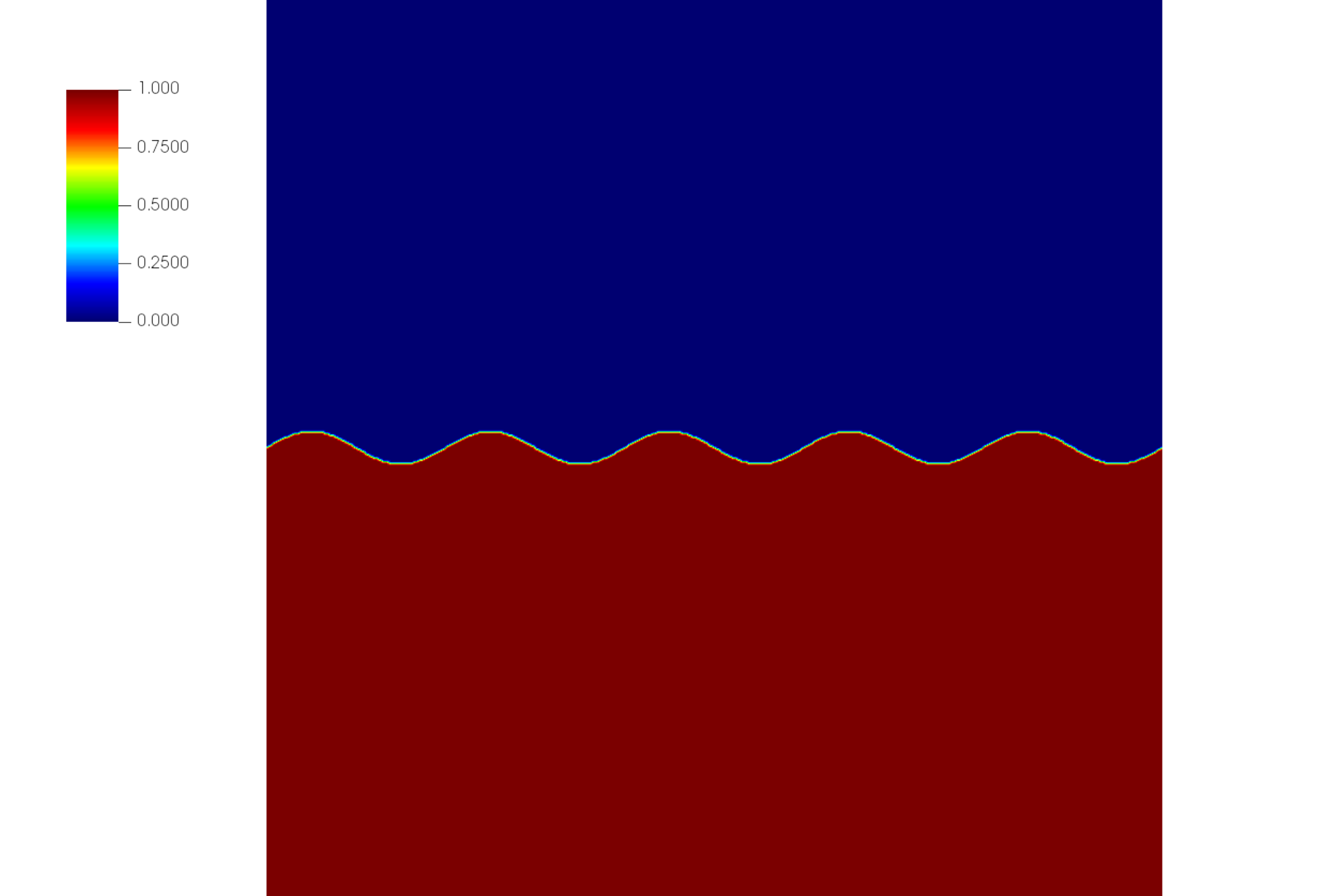}%
  \hfill\includegraphics[height=3cm,clip,trim=8cm  0cm  0cm 0cm]
{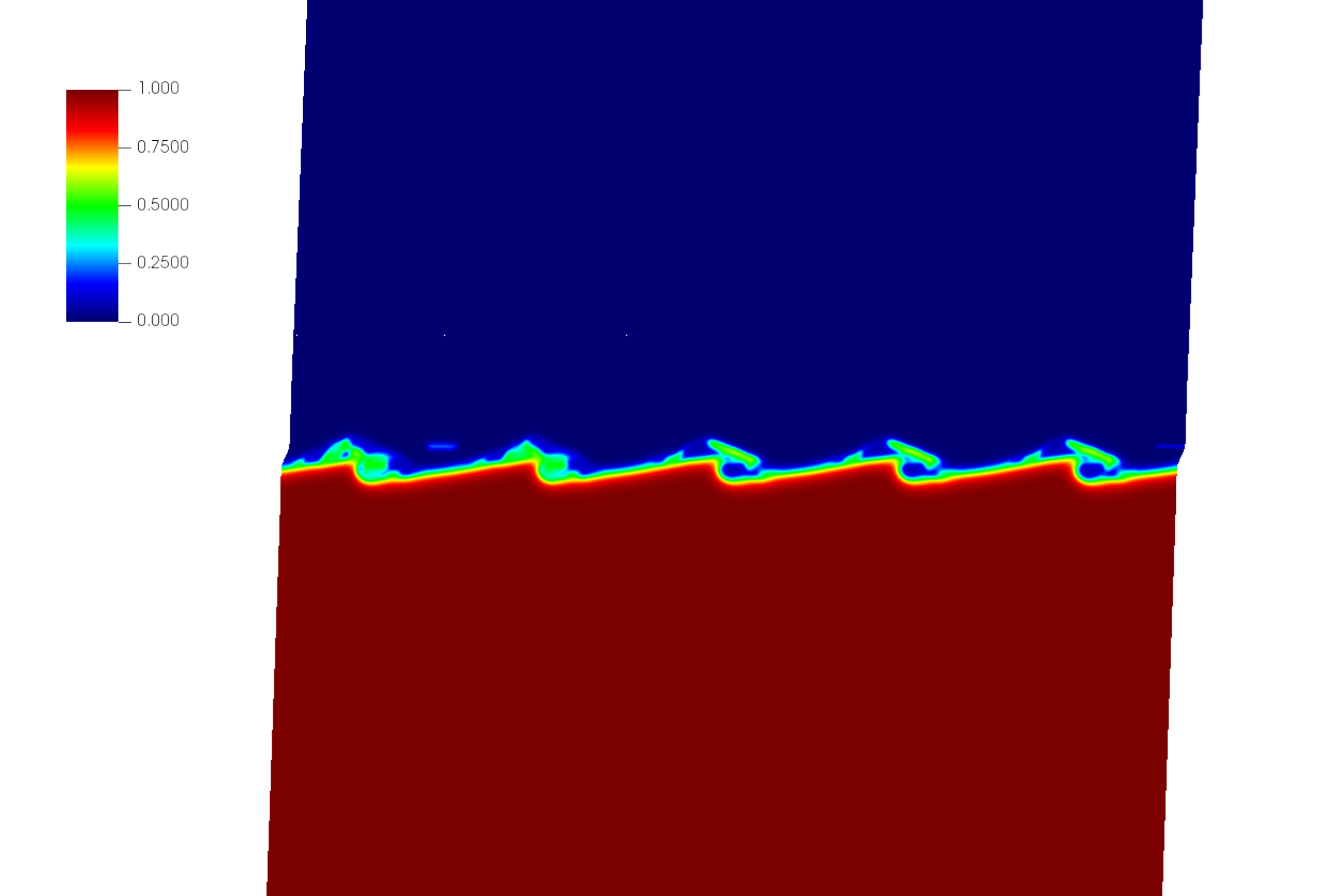}%
  \hfill\includegraphics[height=3cm,clip,trim=8cm  0cm  0cm 0cm]
{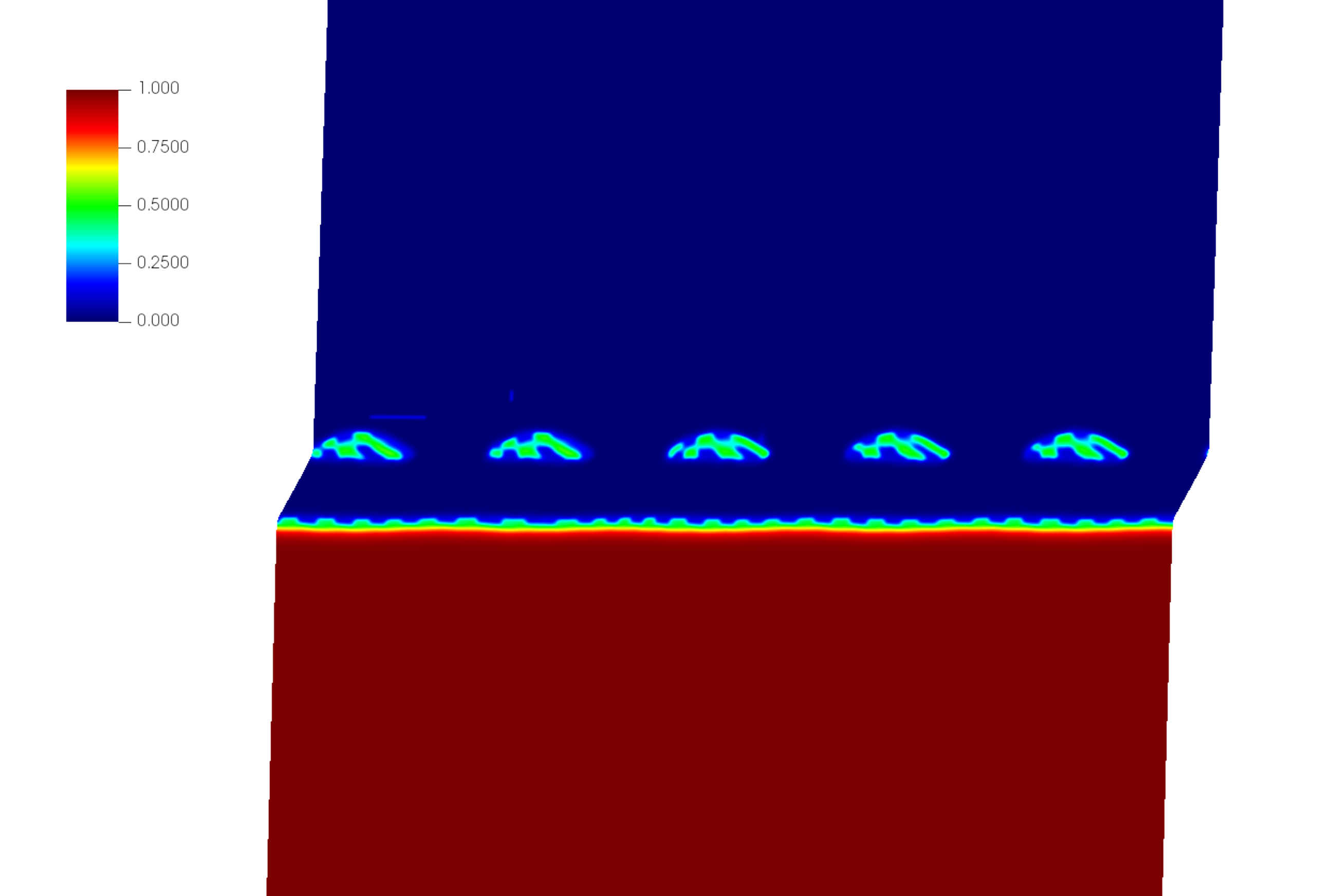}%
  \hfill\includegraphics[height=3cm,clip,trim=8cm  0cm  0cm 0cm]
{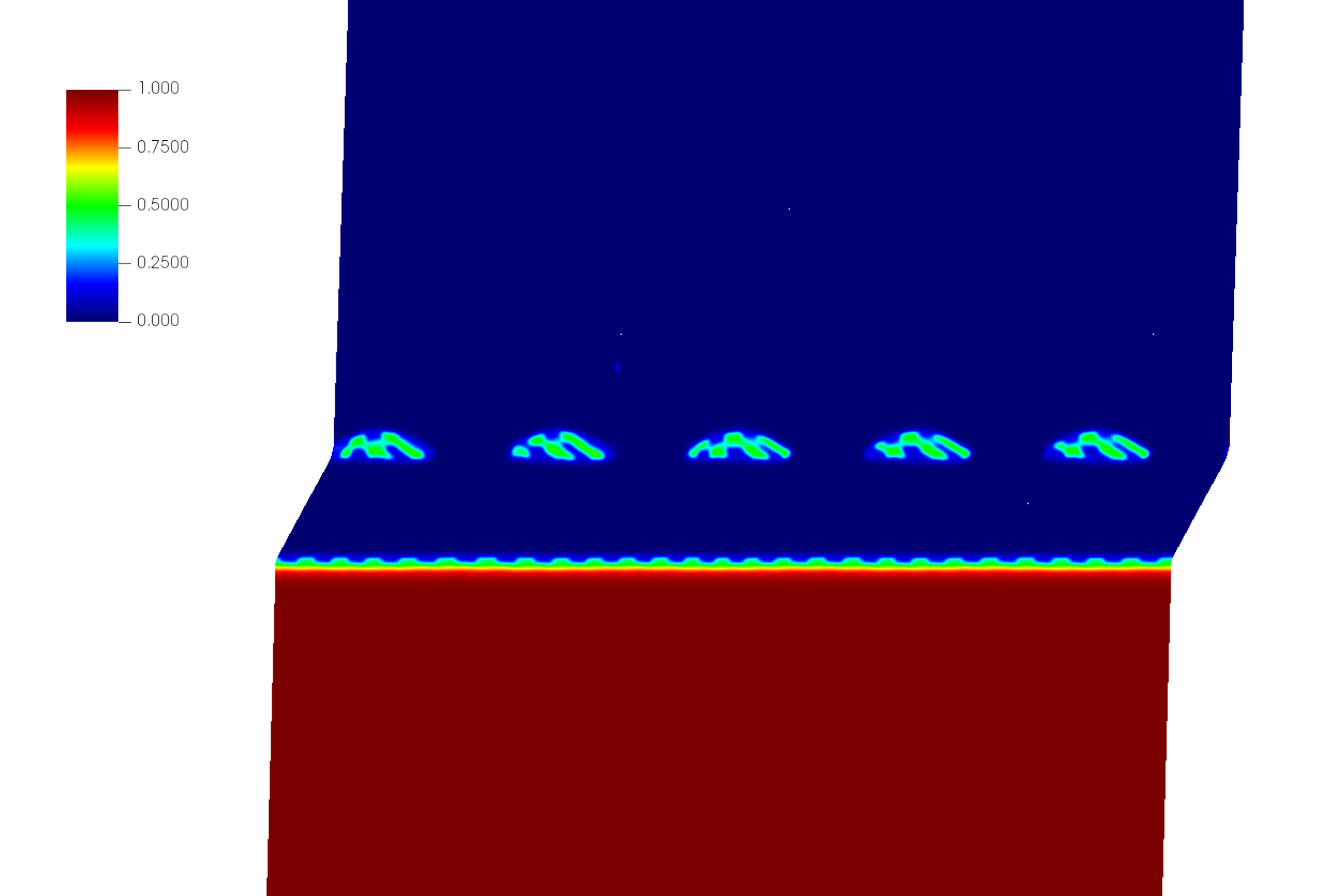}
  \centering
  \includegraphics      [height=3cm,clip,trim=1cm 15cm 36cm 0cm]
{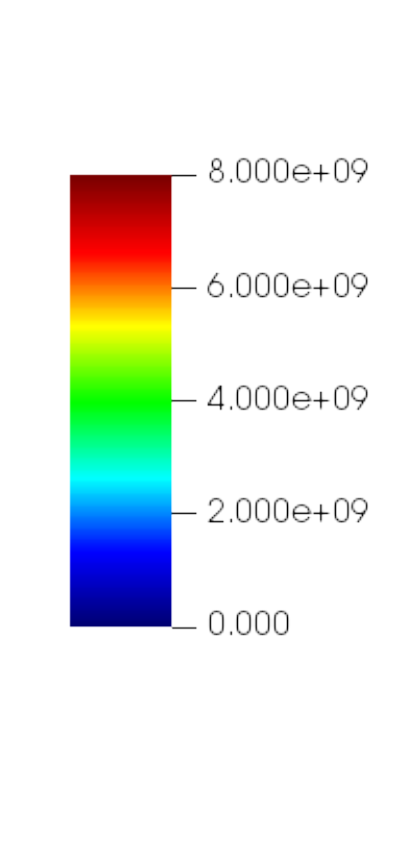}%
  \includegraphics      [height=3cm,clip,trim=8cm  0cm  0cm 0cm]
{2da8192b746a064a.pdf}%
  \hfill\includegraphics[height=3cm,clip,trim=8cm  0cm  0cm 0cm]
{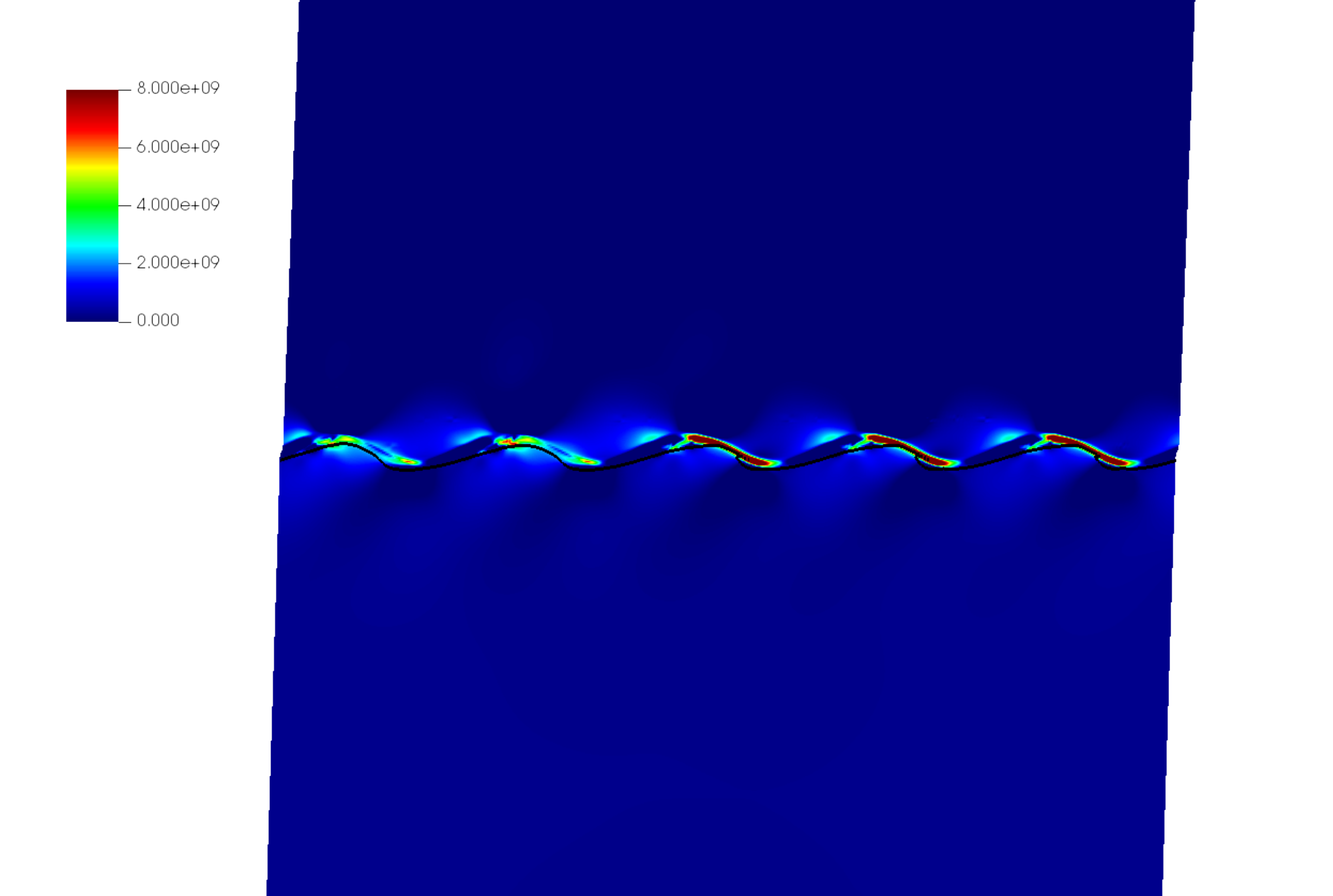}%
  \hfill\includegraphics[height=3cm,clip,trim=8cm  0cm  0cm 0cm]
{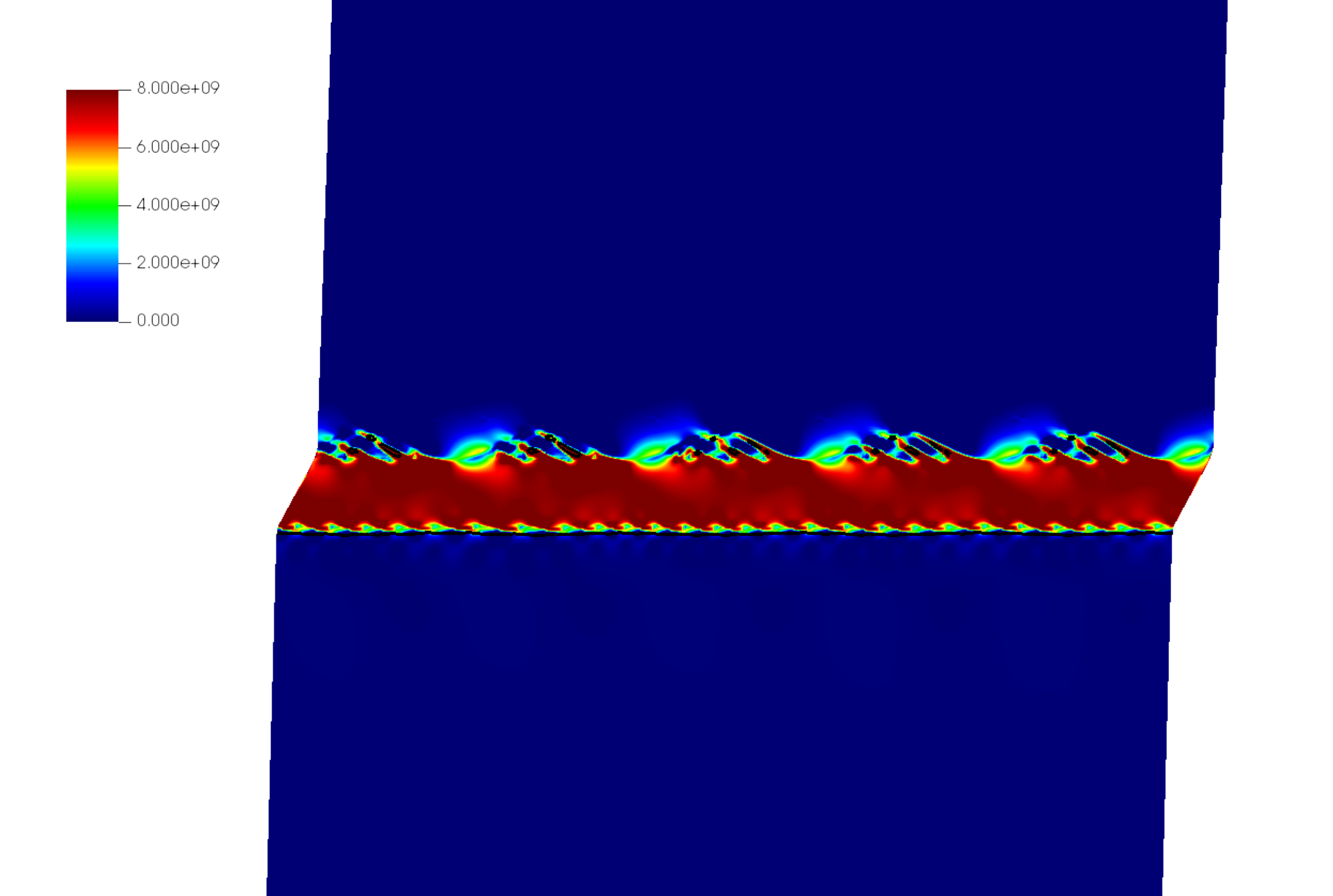}%
  \hfill\includegraphics[height=3cm,clip,trim=8cm  0cm  0cm 0cm]
{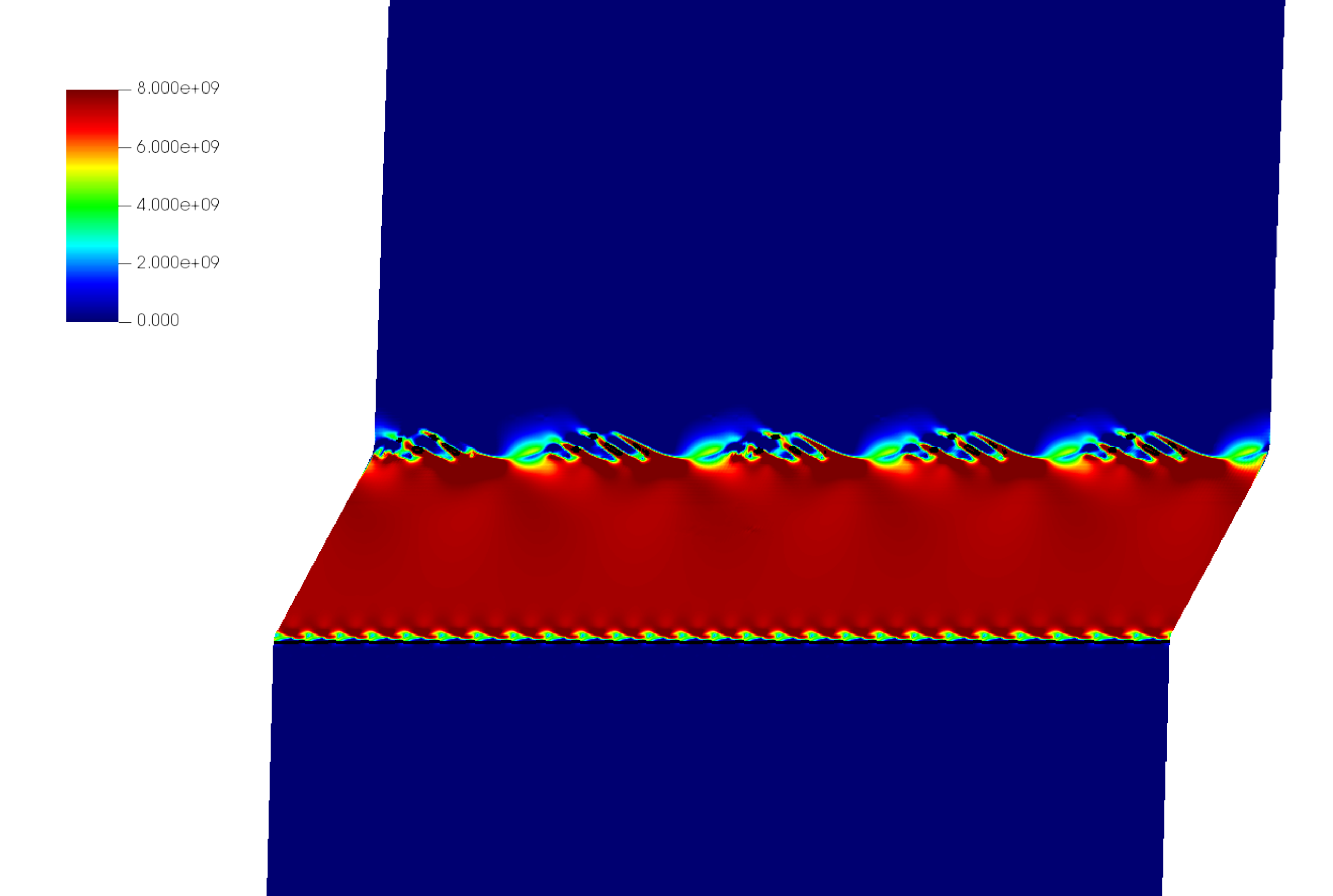}
  \caption{Shear-coupled migration of the ATGB $\phi=10^\circ$ boundary with a slightly perturbed interface. Intervals show $\eta_1[-]$ and $\sigma_{xx}[\mathrm{MPa}]$ at intervals corresponding to the markers in \cref{fig:results/Lamination/stress_strain}.}
  \label{fig:results/Lamination/output_10deg_nothreshold_iso}
\end{figure}

\textbf{10$^\circ$ ATGB}: 
This ATGB (\cref{fig:results/Lamination/output_10deg_nothreshold_iso}) behaves similarly to the perturbed STGB: it exhibits a finite threshold and a higher steady stress.
Unlike the STGB, however, the boundary accumulates a much larger residual stress, because the rotated shear-coupling deformation is incompatible with the nominally planar interface.
Just prior to motion, the ATGB begins to take on a saw-tooth pattern, forming small local STGB segments.
However, as the applied stress continues to drive the boundary, it is still impossible to move in a compatible fashion, even when forming facets (as in \cite{chesser2018understanding}).
As the boundary is driven downward, there is a continual increase in the stress in the x direction.
This is because, after rotation, $\Delta \mathbf{F}^{gb}$, contains components in the xx direction as well as the xy and yy directions.
The stresses exceed typical yield stresses, indicating that dislocation activity would likely accompany migration.
Nevertheless, even in the absence of plasticity, incompatibility alone produces residual back-stress and defect-like structures.

\begin{figure}
  \centering
  \includegraphics      [height=3cm,clip,trim=1cm 15cm 36cm 0cm]
{d5f2f29f1e9f6361.pdf}%
  \includegraphics      [height=3cm,clip,trim=8cm 0cm 0cm 0cm]
{d5f2f29f1e9f6361.pdf}%
  \hfill\includegraphics[height=3cm,clip,trim=8cm 0cm 0cm 0cm]
{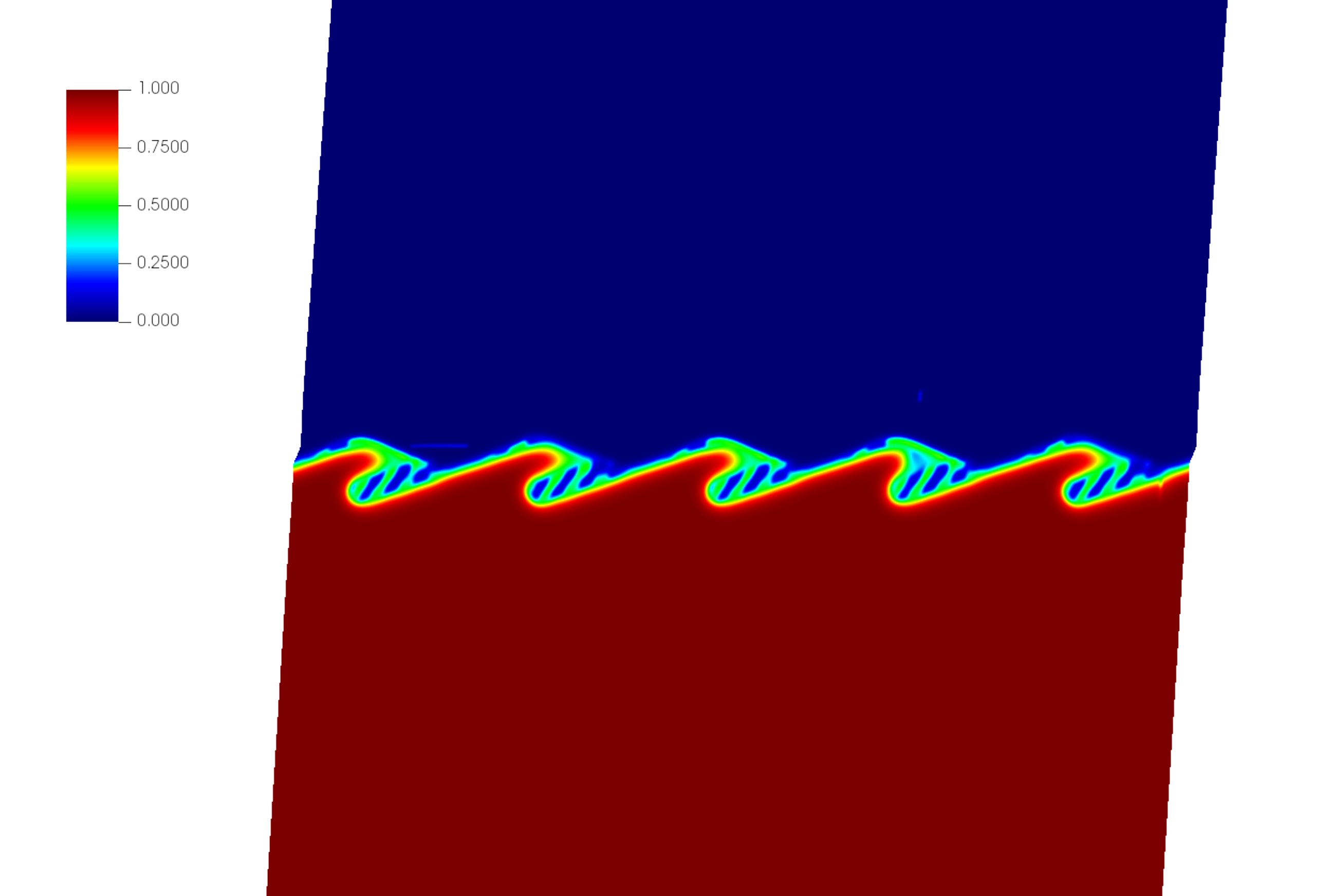}%
  \hfill\includegraphics[height=3cm,clip,trim=8cm 0cm 0cm 0cm]
{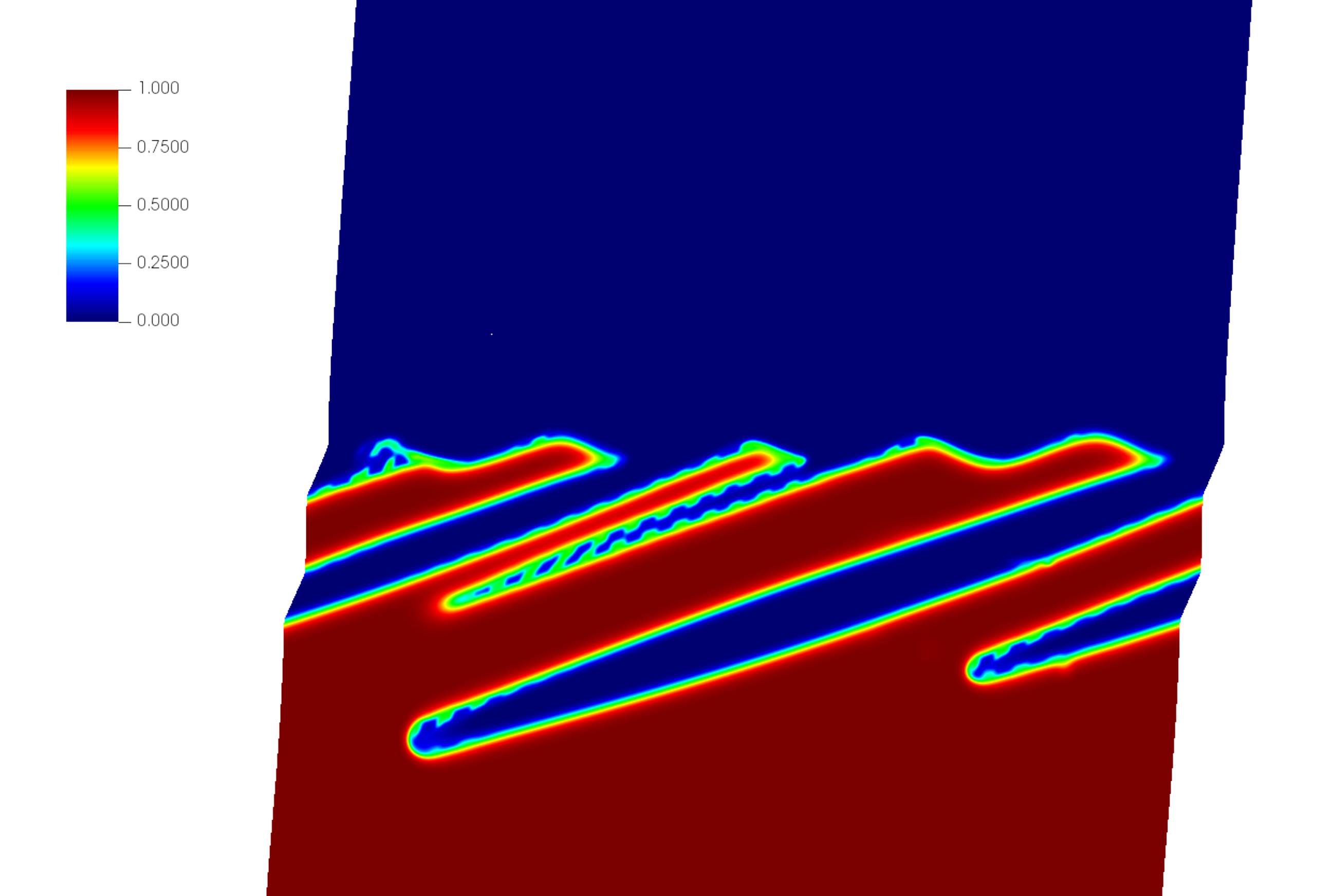}%
  \hfill\includegraphics[height=3cm,clip,trim=8cm 0cm 0cm 0cm]
{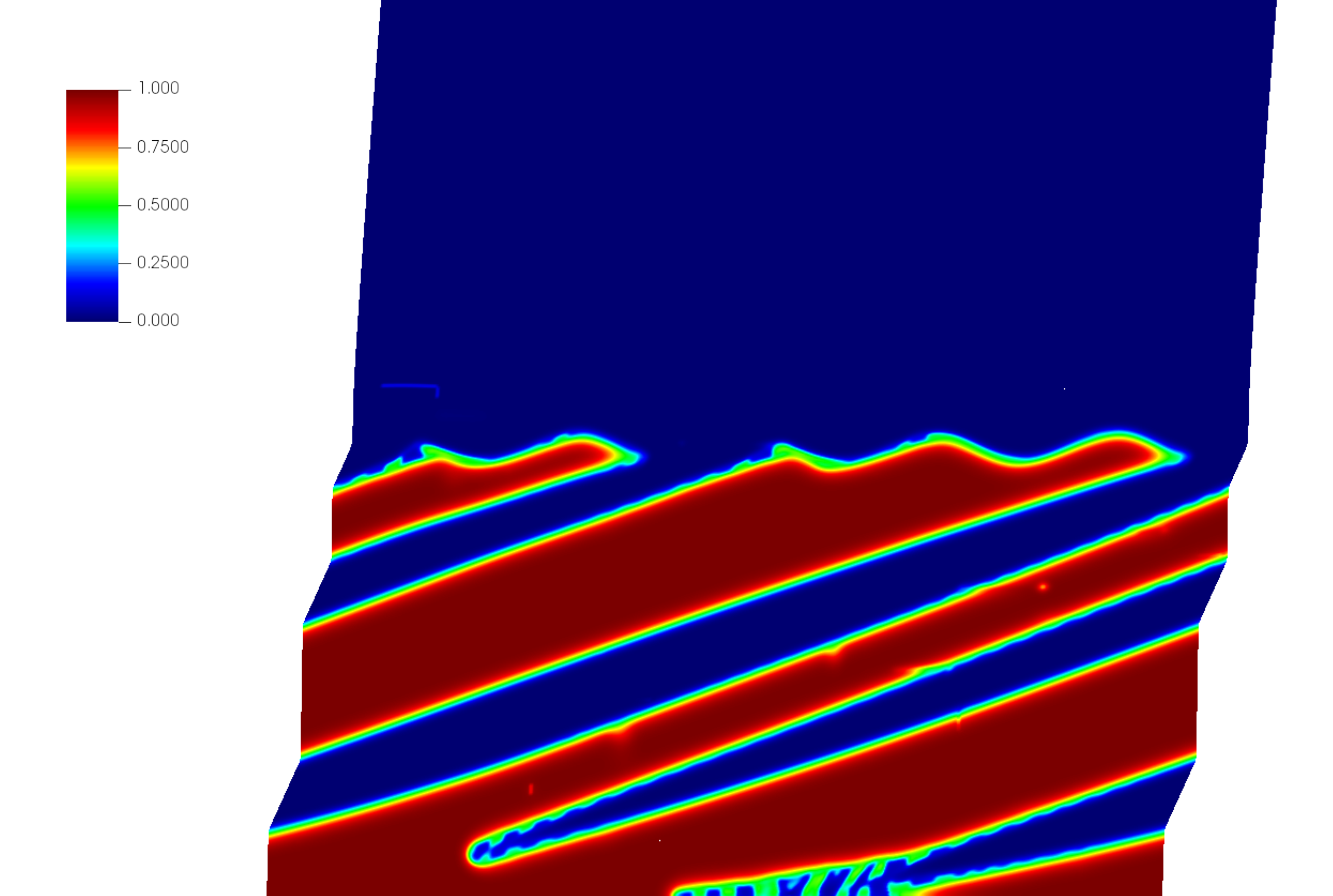}
  \centering
  \includegraphics      [height=3cm,clip,trim=1cm 15cm 36cm 0cm]
{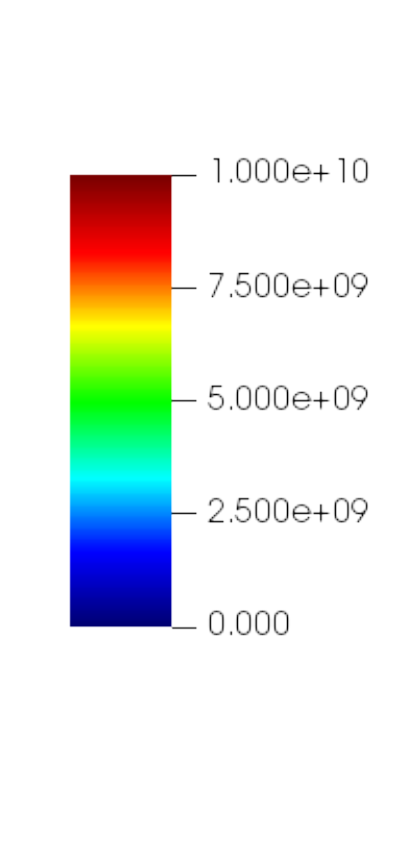}%
  \includegraphics      [height=3cm,clip,trim=8cm 0cm 0cm 0cm]
{28fa18d4d4e0a418.pdf}%
  \hfill\includegraphics[height=3cm,clip,trim=8cm 0cm 0cm 0cm]
{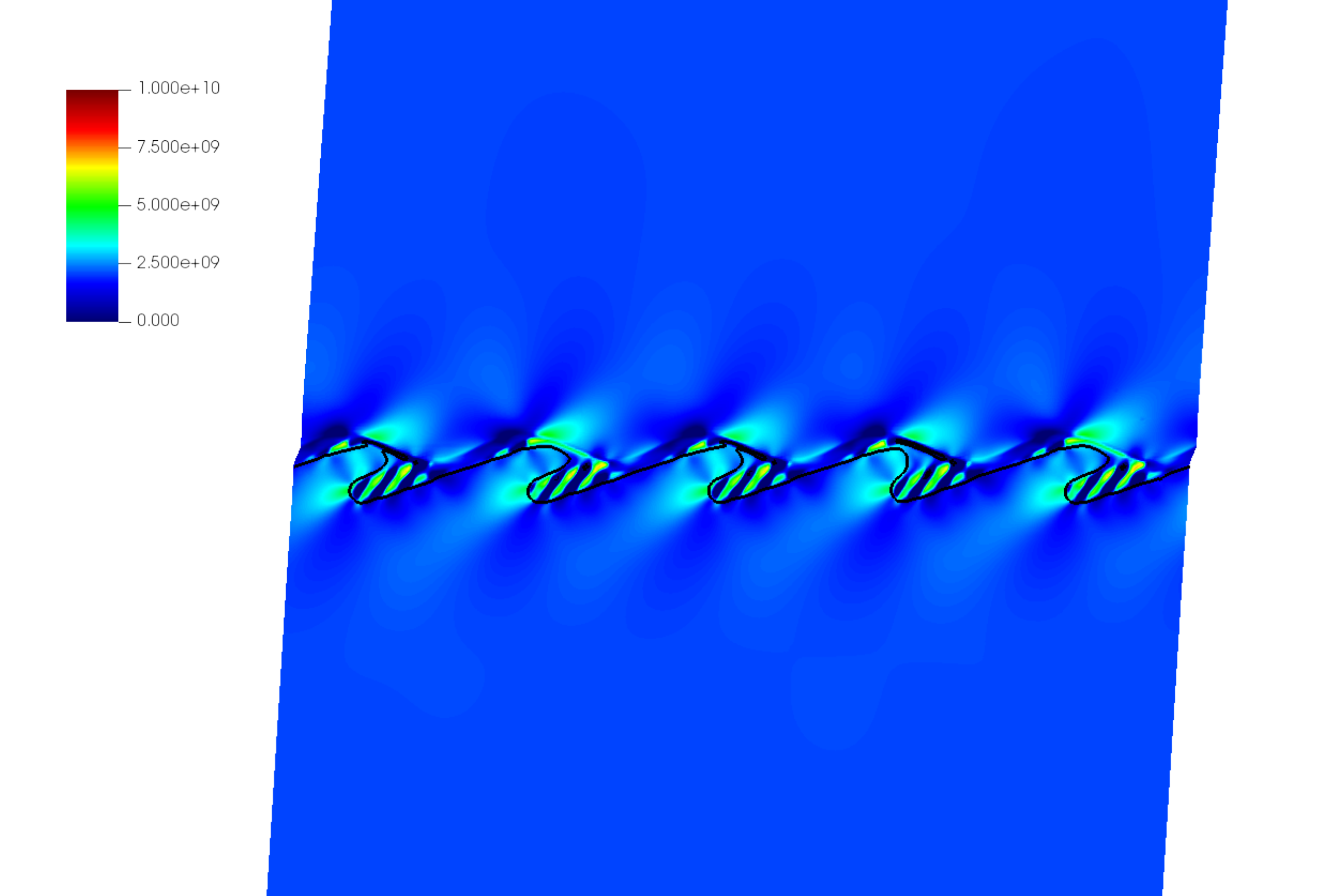}%
  \hfill\includegraphics[height=3cm,clip,trim=8cm 0cm 0cm 0cm]
{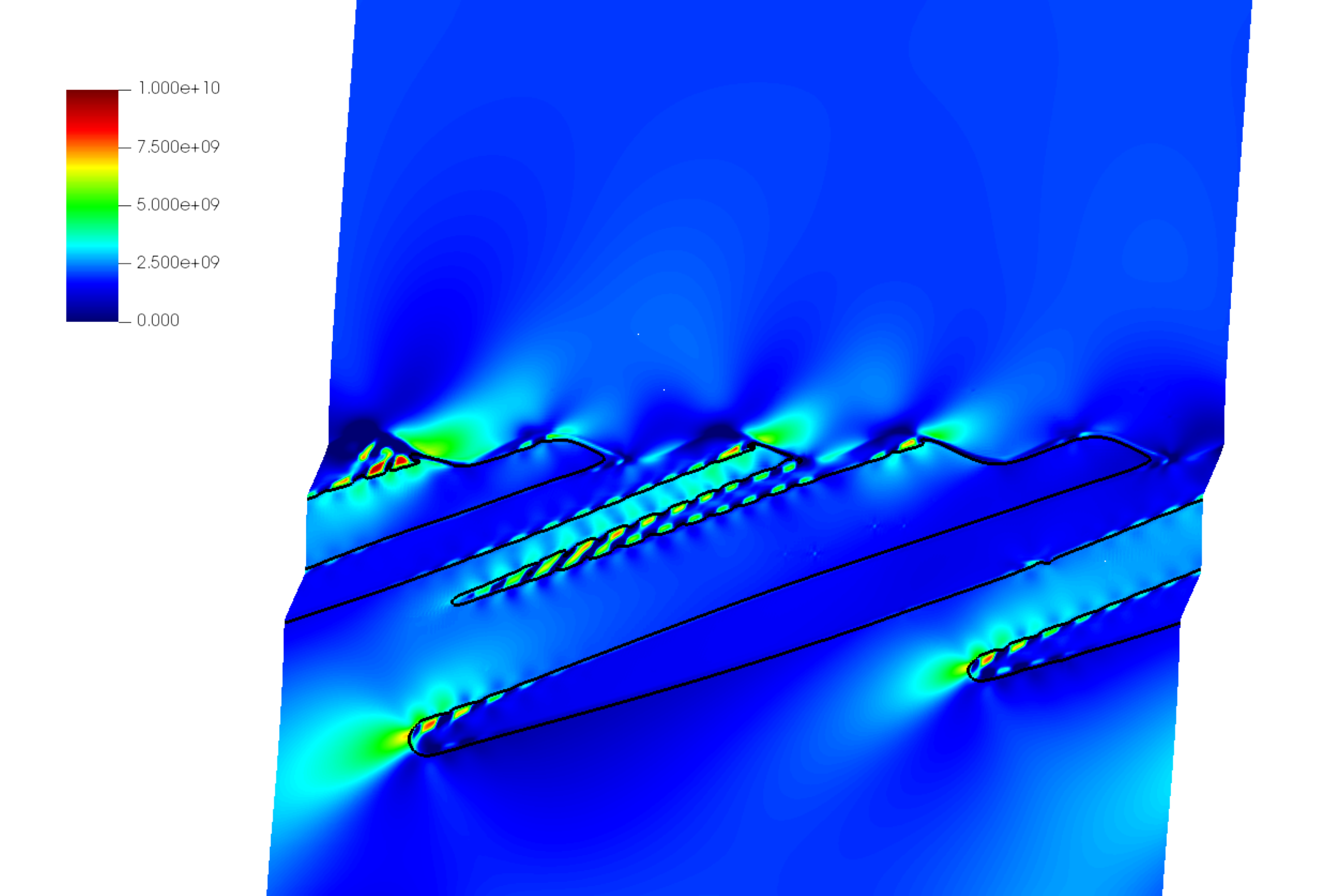}%
  \hfill\includegraphics[height=3cm,clip,trim=8cm 0cm 0cm 0cm]
{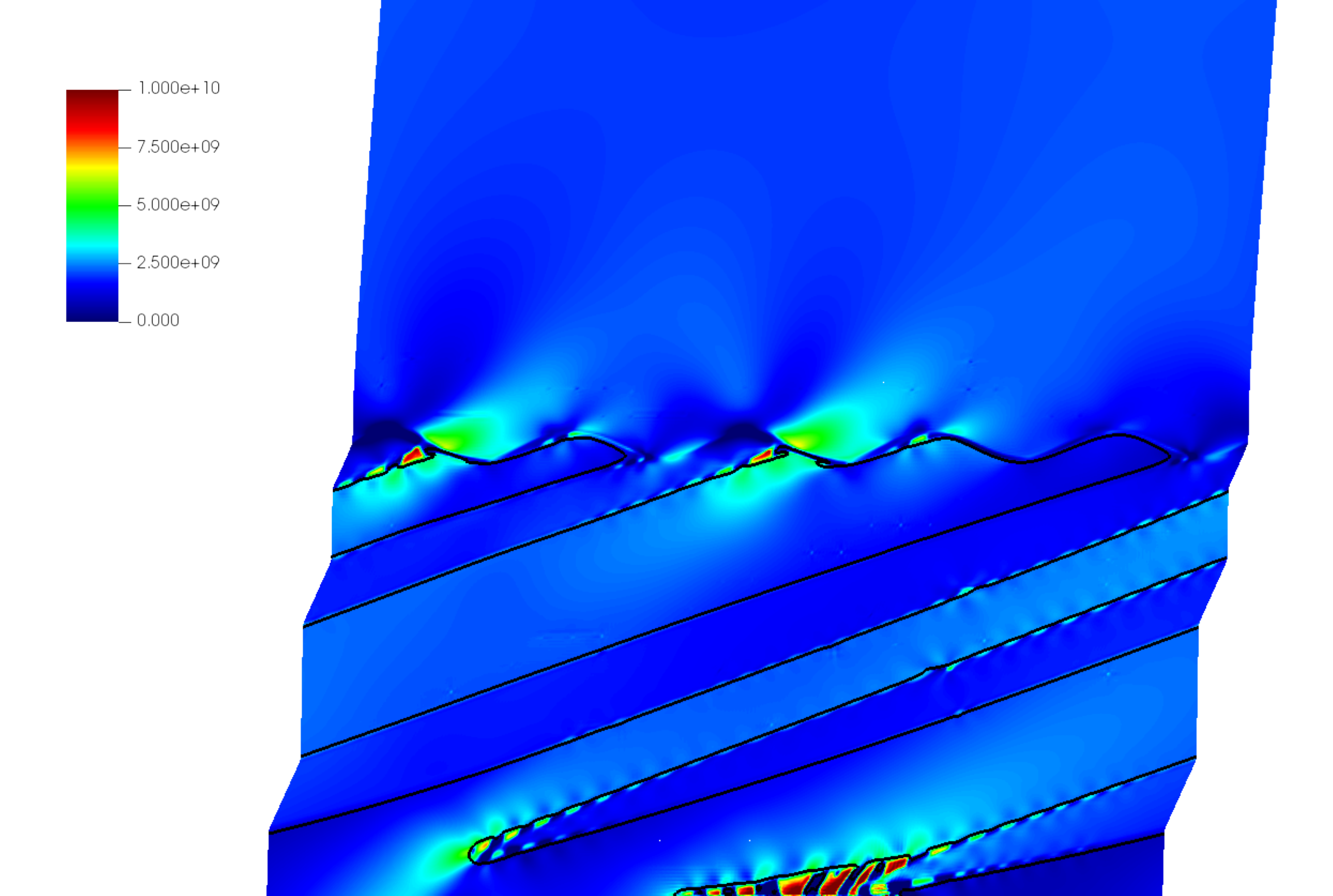}
  \caption{Shear-coupled migration of the ATGB $\phi=20^\circ$ boundary with a slightly perturbed interface. Intervals show $\eta_1[-]$ and $\sigma_{xy}[\mathrm{MPa}]$ at intervals corresponding to the markers in \cref{fig:results/Lamination/stress_strain}.}
  \label{fig:results/Lamination/output_20deg_nothreshold_iso}
\end{figure}

\textbf{20$^\circ$ ATGB}:
As the ATGB inclination angle is increased, a significant change in boundary migration behavior is observed.
Whereas the lower $\phi$ boundaries continued to move in an effectively planar fashion, larger-angle ATGBs accommodated migration by a different migration mechanism altogether.
The trend of increasing yield stress is continued with the $20^\circ$ ATGB, which is approximately twice that of the low-angle ATGB and the perturbed STGB.
However, there is little to no difference between the yield and flow stresses, and the boundary motion never stabilizes to a steady state.
The difference is seen in the morphology of the boundary and the stresses during migration (\cref{fig:results/Lamination/output_20deg_nothreshold_iso}), in which no planar motion occurs at all.

This shows that, once incompatibility (i.e. the requisite strain energy resulting from back-stress) exceeds a critical level, planar migration becomes mechanically inaccessible.
Instead, the system accommodates the applied deformation by introducing new interfaces, resulting in lamination, with laminar boundaries oriented towards the STGB inclination.
Indeed, at this point the system more closely resembles deformation by twinning rather than boundary migration. 
In fact, the inclined boundaries are technically twins as they are symmetric across the inclined plane; however, these are generally not considered true ``twins'' in FCC metals as the boundary is incoherent and high-energy.
This twinning behavior is entirely spontaneous and the natural consequence of forced incompatible motion using the grain boundary flow rule.

\begin{figure}
  \centering
  \includegraphics      [height=3cm,clip,trim=1cm 15cm 36cm 0cm]
{d5f2f29f1e9f6361.pdf}%
  \includegraphics      [height=3cm,clip,trim=8cm 0cm 0cm 0cm]
{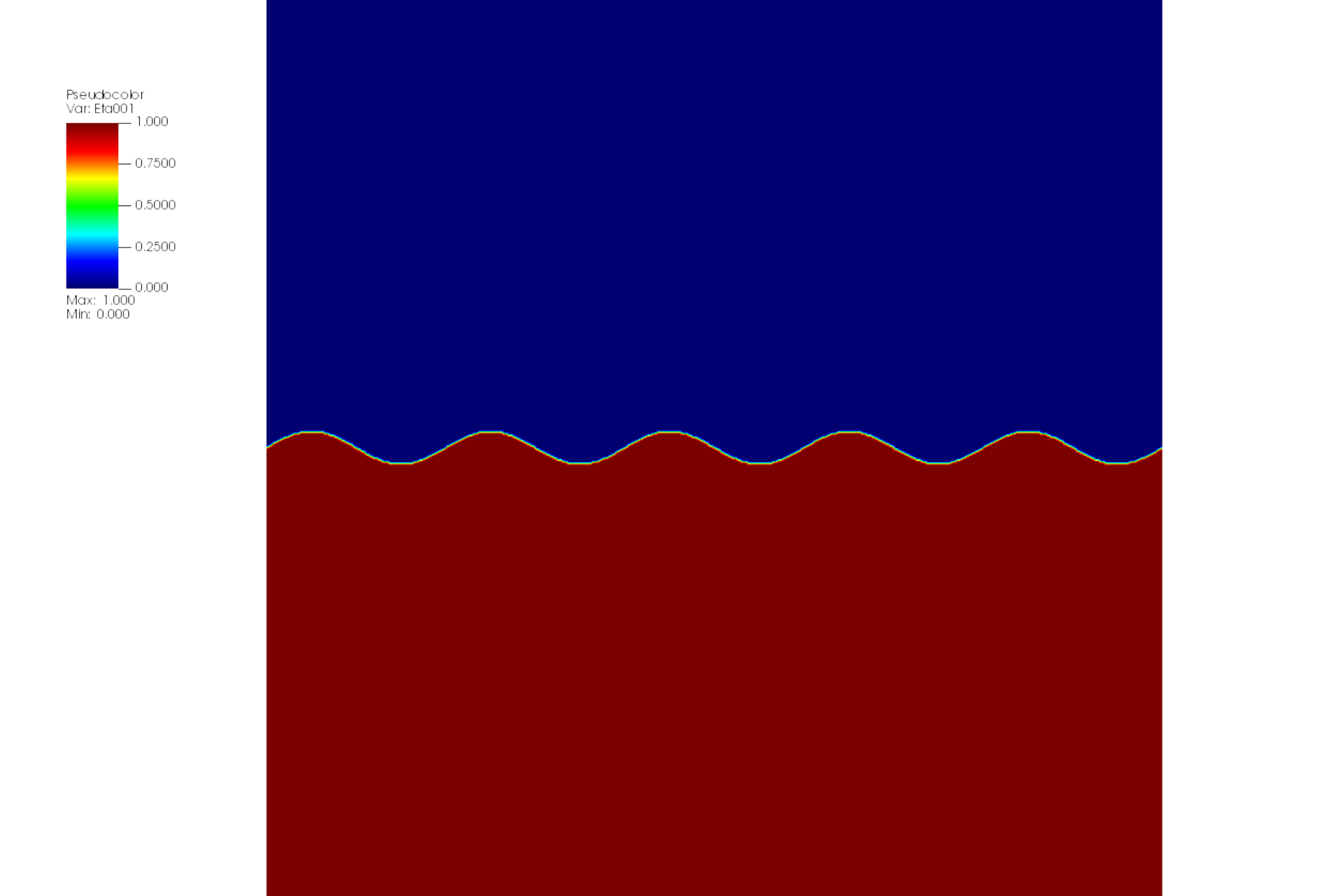}%
  \hfill\includegraphics[height=3cm,clip,trim=8cm 0cm 0cm 0cm]
{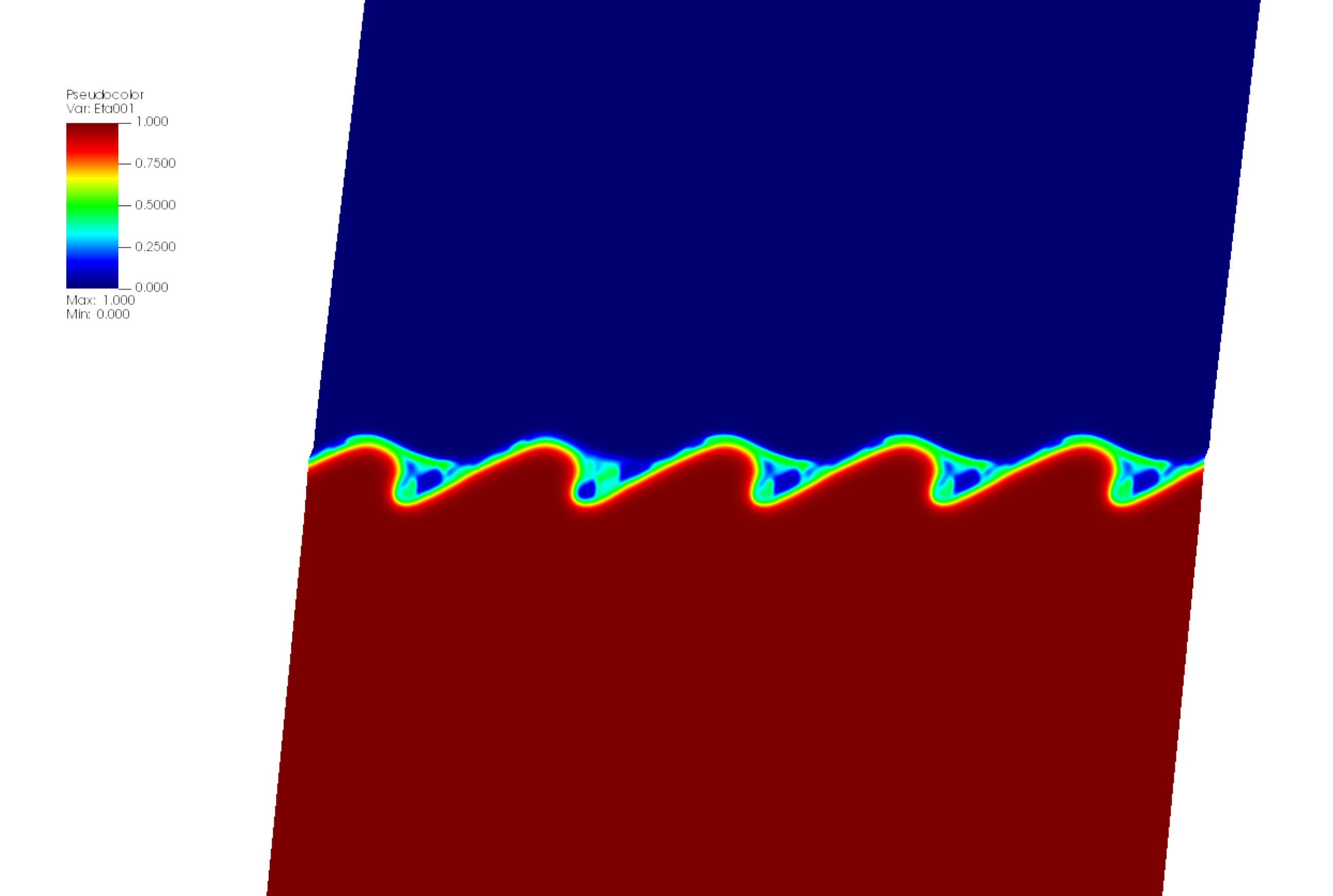}%
  \hfill\includegraphics[height=3cm,clip,trim=8cm 0cm 0cm 0cm]
{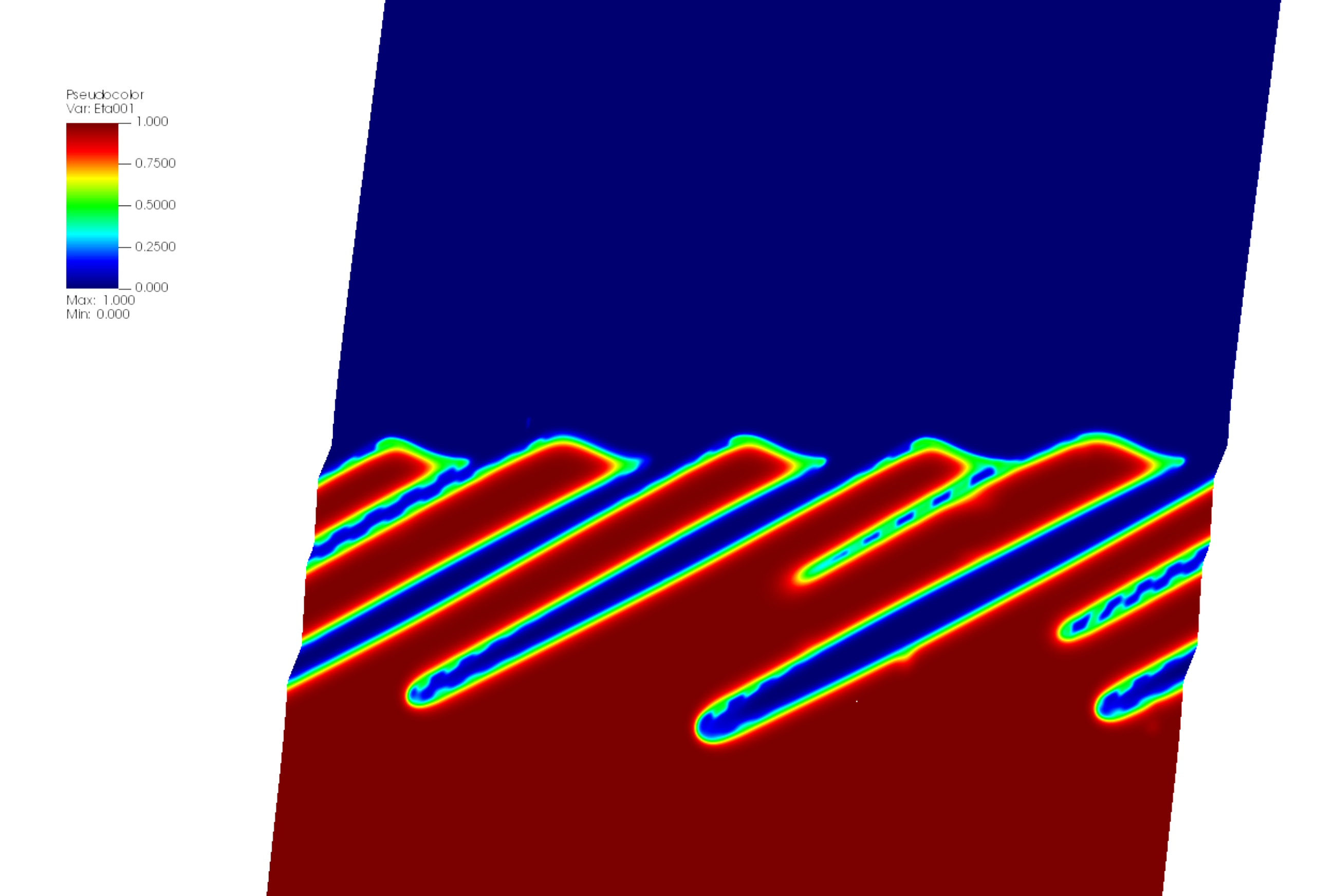}%
  \hfill\includegraphics[height=3cm,clip,trim=8cm 0cm 0cm 0cm]
{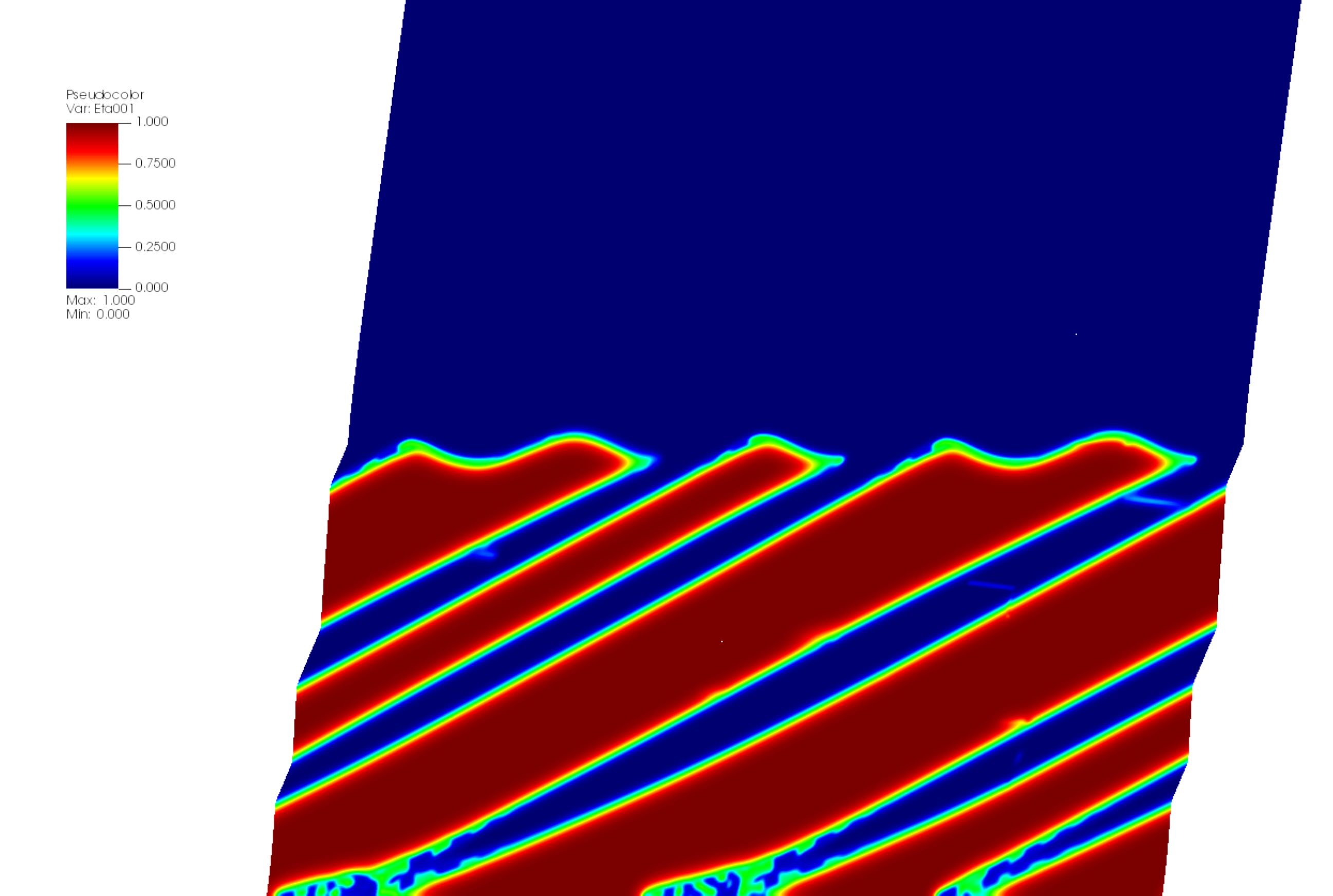}
  \centering
  \includegraphics      [height=3cm,clip,trim=1cm 15cm 36cm 0cm]
{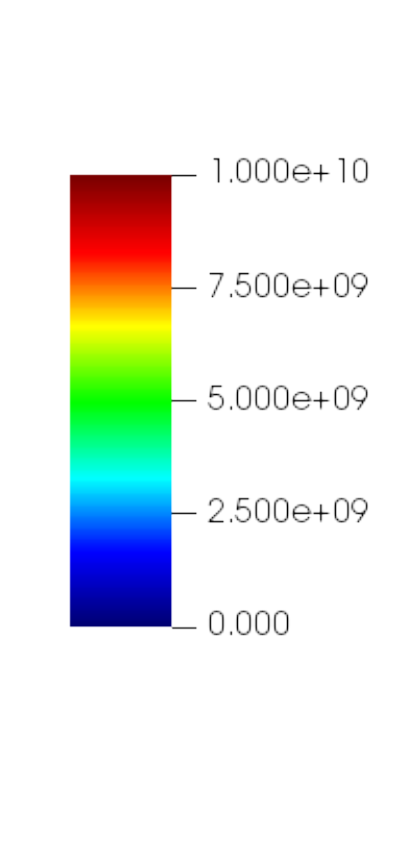}%
  \includegraphics      [height=3cm,clip,trim=8cm 0cm 0cm 0cm]
{1f49d6d2f3a22339.pdf}%
  \hfill\includegraphics[height=3cm,clip,trim=8cm 0cm 0cm 0cm]
{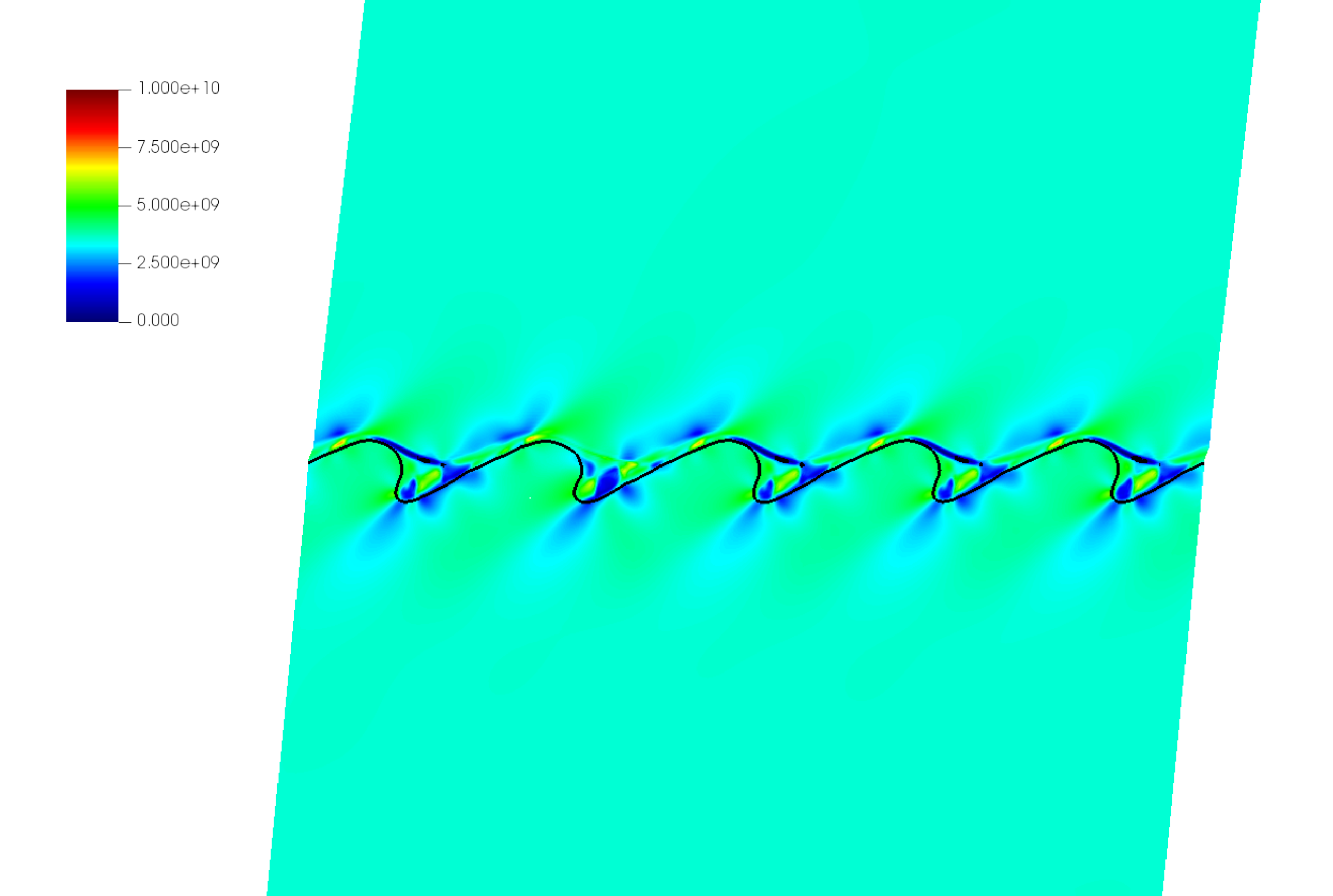}%
  \hfill\includegraphics[height=3cm,clip,trim=8cm 0cm 0cm 0cm]
{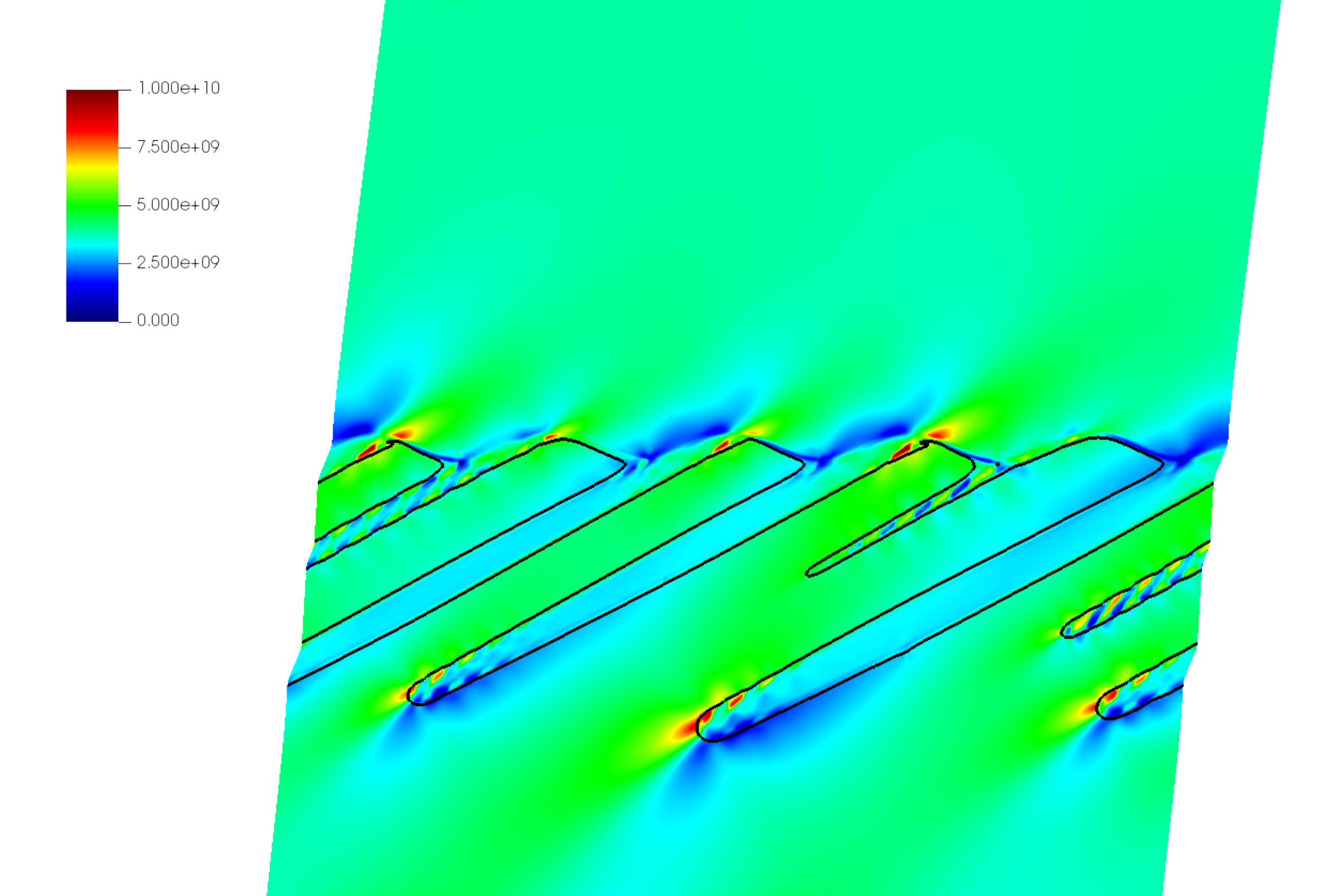}%
  \hfill\includegraphics[height=3cm,clip,trim=8cm 0cm 0cm 0cm]
{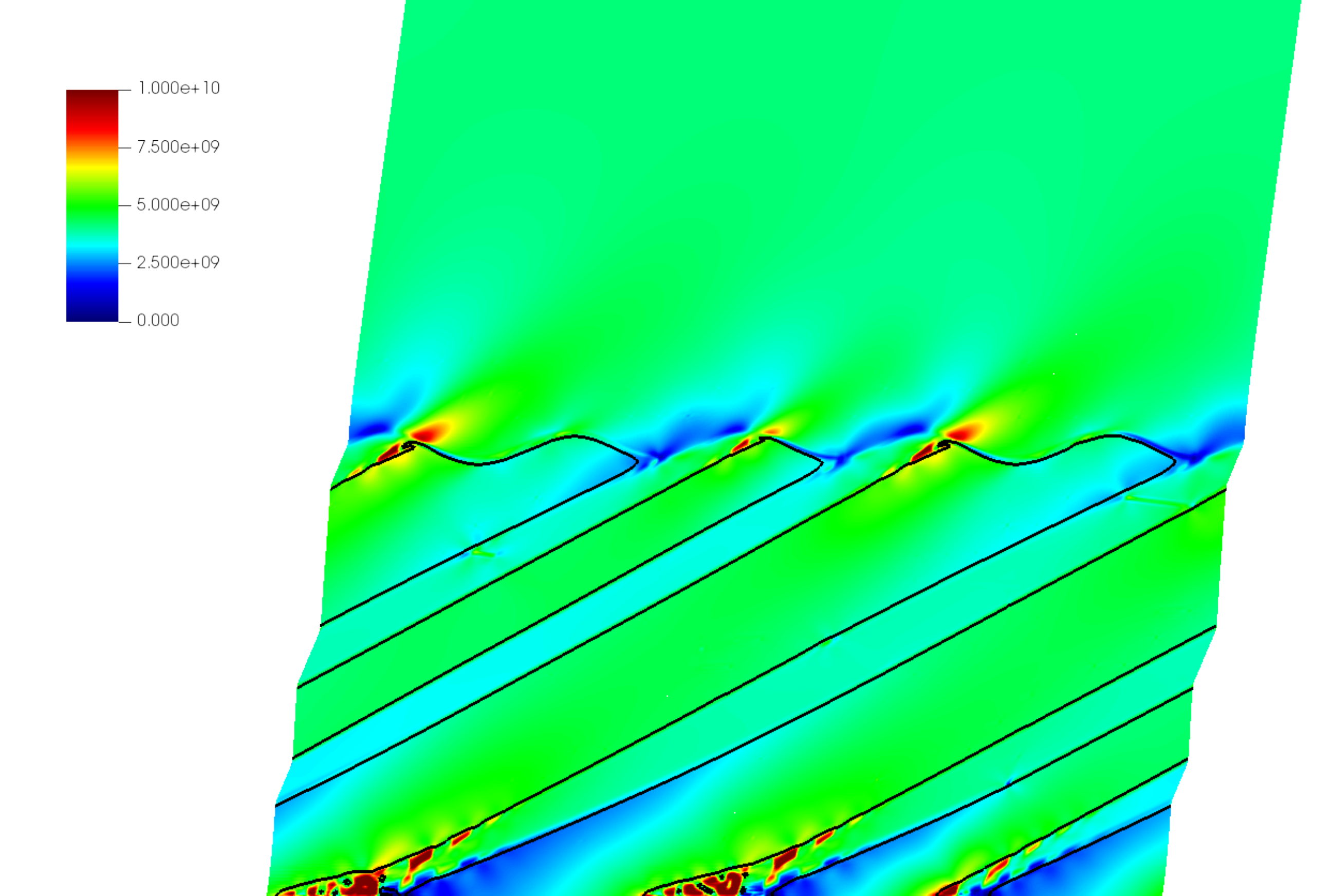}
  \caption{Shear-coupled migration of the ATGB $\phi=30^\circ$ boundary with a slightly perturbed interface. Intervals show $\eta_1[-]$ and $\sigma_{xy}[\mathrm{MPa}]$ at intervals corresponding to the markers in \cref{fig:results/Lamination/stress_strain}.}
  \label{fig:results/Lamination/output_30deg_nothreshold_iso}
\end{figure}

\textbf{30$^\circ$ ATGB}:
Similar behavior for this boundary is observed as for the 20$^\circ$ ATGB (\cref{fig:results/Lamination/output_30deg_nothreshold_iso}).
The main difference is the unsurprising increase in yield and flow stress, along with the increase in the inclination angle of the laminate structures to 30$^\circ$ from 20$^\circ$.
In both the 20 and the 30$^\circ$ ATGB cases, the simulation is ended once the laminate growth is arrested by the domain boundary.
(Larger simulations can certainly be run to examine long-time flow behavior, although with additional cost due to the increased amount of boundary present).
Investigation into larger ATGB angle values, including 35$^\circ$, 40$^\circ$, and 45$^\circ$, showed that these boundaries proved to be entirely immobile.

The laminate structures, and indeed even some of the left-behind defect structures, exhibit some interesting features that are discussed here.
A common occurrence in the initial formation of laminate structures is the emergence of what appear to be \textit{sub-laminates} (especially apparent in the second snapshots of both the 20$^\circ$ and 30$^\circ$ cases).
However, as the larger laminates form, the sub-laminate structures generally dissipate, as can be seen by the smooth boundaries of the larger laminates and the slightly corrugated edges of some of the smaller laminates.
The existence, size, and morphology of these sub-laminates is driven by many factors:
\begin{enumerate}[nosep]
\item The reason for the initial emergence of sub-laminates is the very high driving force that exists just as the boundary begins to move.
  However, once the larger laminates form, the structure relaxes and is able to remove some of the excess boundary energy resulting from the sub-laminates.
\item Since the length scale of the laminate structures is governed by the ratio of driving force to boundary energy, ATGBs with lower overall boundary energy are able to more economically form complex structures to relieve stress.
  Indeed, this is exactly why twins are so much more likely to form with $\Sigma3$ coherent twins.
  (It should also be mentioned that grain boundary anisotropy, though not considered here, has a profound effect on the formation of boundary structures \cite{runnels2016relaxation}).
\item The diffuse length scale also affects what substructures are resolvable.
  Here, the choice of $\ell_{gb}$  is too large to efficiently resolve all boundary structures (though, in the case of 0 and 10 boundaries, structures exist that are apparently below the diffuse length scale).
\item Boundary substructures can be significantly influenced by thresholding/dissipation energy, which encourages the creation of structures under high driving force, and the persistence of such structures under low driving force.
  This effect can also be artificially induced by the frequency of elastic solves; it was determined in some cases that less frequent solves can lead to more prevalent, stable substructure formation.
\end{enumerate}

The solution is surprisingly impervious to numerical errors.
Interestingly, small perturbations that can sometimes arise in numerical solves were shown to have little to no effect on the final solution; indeed, perturbations were observed to stabilize quickly.
This bolsters confidence in the robustness of the method, as well as the local optimality of the phase field result.

\subsection{Asymmetric grain boundary migration}\label{sec:ratcheting}

\begin{table}
  \begin{addedbox}{C0}{1.2}
    \caption{Crystallographic and geometric parameters for synthetic driving force and mechanically driven migration \cite{qiu2024grain}}
    \label{tab:asymmetry_crystallography}
    \begin{tabularx}{\linewidth}{Xllllllllll}
      \toprule
      Type & Tilt axis & CSL & $\phi$ & $\bm{X}_1$ axis & $\bm{X}_2$ axis  & $\bm{Y}_1$ axis & $\bm{Y}_2$ axis & $\beta$ & W & H\\
      \midrule
      STGB &$[111]$ & $\Sigma39$ & $0.0^\circ$  &  $\hkl[-5 7 -2]$ & $\hkl[-7 5 2]$ & $\hkl[-3 -1 4]$ & $\hkl[-1 -3 4]$ & 0.5 & 26.96nm & 72.06nm\\
      ATGB &$[111]$ & $\Sigma39$ & $9.8^\circ$  &  $\hkl[-35  -3 38]$ & $\hkl[-17 -25 42]$ & $\hkl[41 -73 32]$ & $\hkl[67  -59 -8]$ & 0.5 & 21.07nm & 54.74nm\\
      ATGB &$[110]$ & $\Sigma11$ & $46.7^\circ$ &  $\hkl[-5   5 -18]$ & $\hkl[-13 13  -6]$ & $\hkl[-9   9   5]$ & $\hkl[-3   3  13]$ &  0.35 & 23.83nm & 56.18nm\\
      ATGB &$[110]$ & $\Sigma25$ & $26.6^\circ$ &  $\hkl[0  3  1]$ & $\hkl[0 13  9]$ & $\hkl[0 -1  3]$ & $\hkl[0 -9 13]$ & -0.25 & 32.47nm & 58.46nm\\
      \bottomrule
    \end{tabularx}
  \end{addedbox}
\end{table}

Asymmetric tilt boundaries sometimes seem to exhibit a strong dependence of migration response on the sign of the driving force.
This has been interpreted as evidence for direction-dependent kinetic coefficients (leading to grain boundary ``ratcheting'' \cite{qiu2024grain}).
Here, an alternative hypothesis is tested: that apparent asymmetry can arise from incompatibility-induced back-stress, even when mobility is strictly symmetric.

Representative simulations from \cite{qiu2024grain} are replicated as exactly as possible using the present method.
Parameters for the present model are chosen to be consistent with the reference results, and are taken from literature or direct calculation whenever possible.
The multiphase field barrier energy parameter is $\gamma=10$; the diffuse boundary length is $1nm$.
Grain boundary energy is highly anisotropic; however, for the boundaries considered here, it is reasonable to approximate as not depending on orientation.
Therefore a value of $\sigma=0.443 \frac{J}{m^2}$ (approximated from \cite{homer2022examination}) is used.
A constant value of $M=1\times 10^{-11} \frac{m}{s\cdot Pa}$ is used for the mobility; there is no directionality dependence or orientation dependence.

For the mechanics solver, a pseudo-linear finite deformation material model is used, with elastic constant for aluminum taken to be $C_{11}$=108.2 GPa, $C_{12}$=61.3 GPa, $C_{44}$=28.3 GPa.
The elasticity tensor is rotated appropriately based on the crystallographic orientation of each grain to account for any crystallographic anisotropy.
The shear coupling factor $\beta$ is accounted for based on \cite{cahn2006coupling}, and the asymmetry of the boundary is accounted for by rotating the $\mathbf{t}$ and $\mathbf{n}$ by the ATGB inclination angle $\phi$.
Specific values are indicated for each of the examples considered in this section.

\subsubsection{Synthetic Driving Force}

The synthetic driving force (SDF) is used in molecular dynamics to drive a microstructure towards a particular state through an artificial energy penalty based on the local microstructure.
This technique is replicated for phase field by adding the following to the free energy functional
\begin{align}
  W_{\text{sdf}}[\eta] = \sum_n \psi_n g_n(\eta)
\end{align}
where $\psi_n$ are the synthetic driving forces applied to grain $n$, and $g_n$ is the same interpolation function defined in \cref{sec:grain_boundary_flow_rule}.

\begin{figure}

  \begin{addedbox}{C0}{1.3}
  \begin{subfigure}{0.5\linewidth}

    \includegraphics[height=5.5cm]
{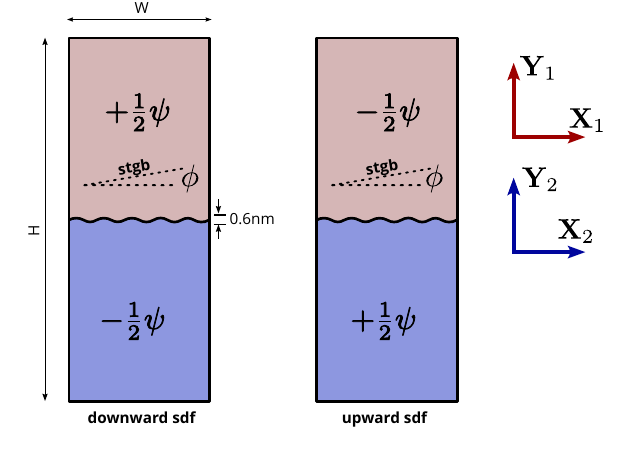}
    \caption{Schematic}
    \label{fig:results/SDF/drawing}
  \end{subfigure}%
  \begin{subfigure}{0.5\linewidth}
    \includegraphics[height=5.5cm]
{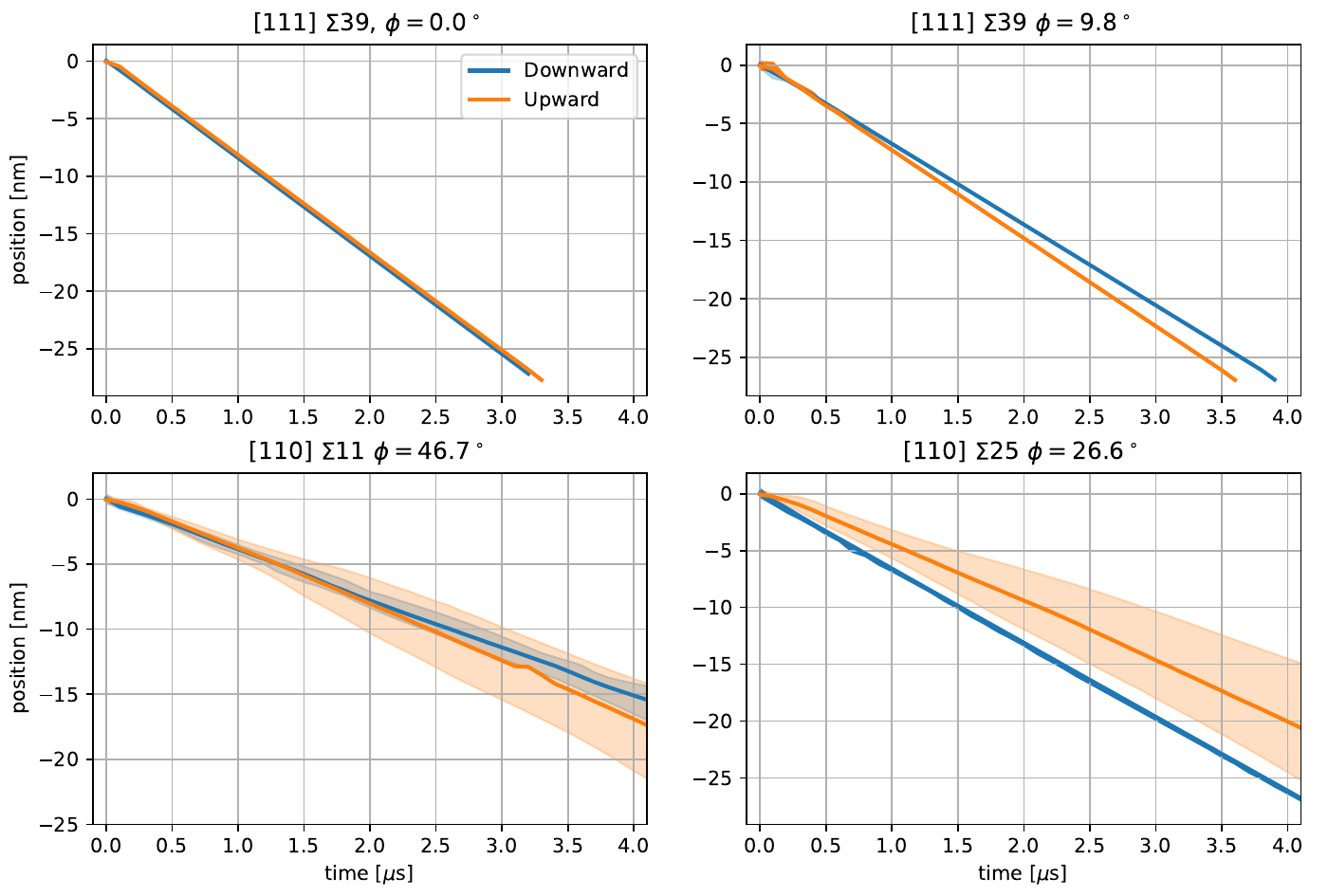}
    \caption{Comparison of upward and downward motion}
    \label{fig:results/SDF/fraction_length}
  \end{subfigure}
  
  \caption{
  Each boundary is subjected to a downward and upward SDF loading.
  (a) The boundary is determined by crystallographic axes $\bm{X}_1,\bm{X}_2,\bm{Y}_1,\bm{Y}_2$, the shear coupling factor $\beta$, and the asymmetric tilt axis angle $\phi$.
  (b) 
  The solid lines represent the average position of the boundary (defined as the $\eta=0.5$ isocontour).
  The shaded regions represent the maximum and minimum boundary location with time.
  The STGB exhibits consistent mobility in both downward and upward motion.
  The other boundaries exhibit 11\%, 17\%, 33\% differences in apparent mobility as evidenced by the differing slopes.
  }
  \end{addedbox}
\end{figure}

\added[id=R1]{Four aluminum grain boundaries are considered: one symmetric tilt boundary and three asymmetric tilt boundaries with \hkl[111] and \hkl[110] tilt axes.}
\added[id=R1]{The dimensions and crystallographic orientations of the grains (\cref{fig:results/SDF/drawing,tab:asymmetry_crystallography}) are specified to match the GB considered in \cite[Table S1, Figure S17A]{qiu2024grain}.}
\added[id=R1]{The vectors $\bm{X}_{1,2}$ $\bm{Y}_{1,2}$ correspond to the crystallographic axis aligned to the respective axis for the respective grain.}
\added[id=R1,comment={1.3}]{The character of the boundary ranges from STGB $\phi=0^\circ$ to a large-angle ATGB ($\phi=46.7^\circ$).}
A slight perturbation is applied to the interface with amplitude 0.3nm.
The base grid has 32x64 cells.
Three levels of refinement are used in domains of interest, for a finest grid spacing of 0.04nm.
The phase field evolution equations are explicitly integrated with a base timestep on the course grid of 1ns, 0.0625 on the finest level.
\replaced[id=R1,comment={1.3}]{The elastic solver is run at every interval, and periodic boundary conditions are used in the $x$ direction.}{The elastic solver is run every 10 timesteps; shorter intervals were tested and found to be unnecessary.
To mimic periodicity, Neumann boundary conditions are applied to all boundaries in both directions.}

\added[id=R1]{Two loadings are applied: one in which the boundary is encouraged to move downward (``downward SDF''); the other, where the sign of the SDF is reversed to encourage upward motion (``upward SDF'').}
\added[id=R1]{(\cref{fig:results/SDF/fraction_length}, solid lines)}
The value of the SDF is $\pm80\times 10^{7}\frac{J}{m^3}$.
Nothing else is changed between the simulations except the sign of the driving force.
\added[id=R1]{In all but the STGB boundary, a significant difference is observed between the downward-driven vs the upward-driven motion.}
\added[id=R1]{For the STGB, both downward and upward loads yield the same boundary velocity, indicating no asymmetry.}
\added[id=R1]{The remaining boundaries exhibit differences in velocity of 11\%, 17\%, and 33\%, respectively.}
\added[id=R1]{Since magnitude of driving force is identical in all cases, this results in a total \textit{apparent} mobility difference of the same amount.}
\deleted[id=R1,comment={1.3}]{The original text reported a single-boundary apparent mobility difference of 14\%.}

\begin{figure}
  \begin{addedbox}{C0}{1.3}
    \centering
    \includegraphics[width=0.8\linewidth]
{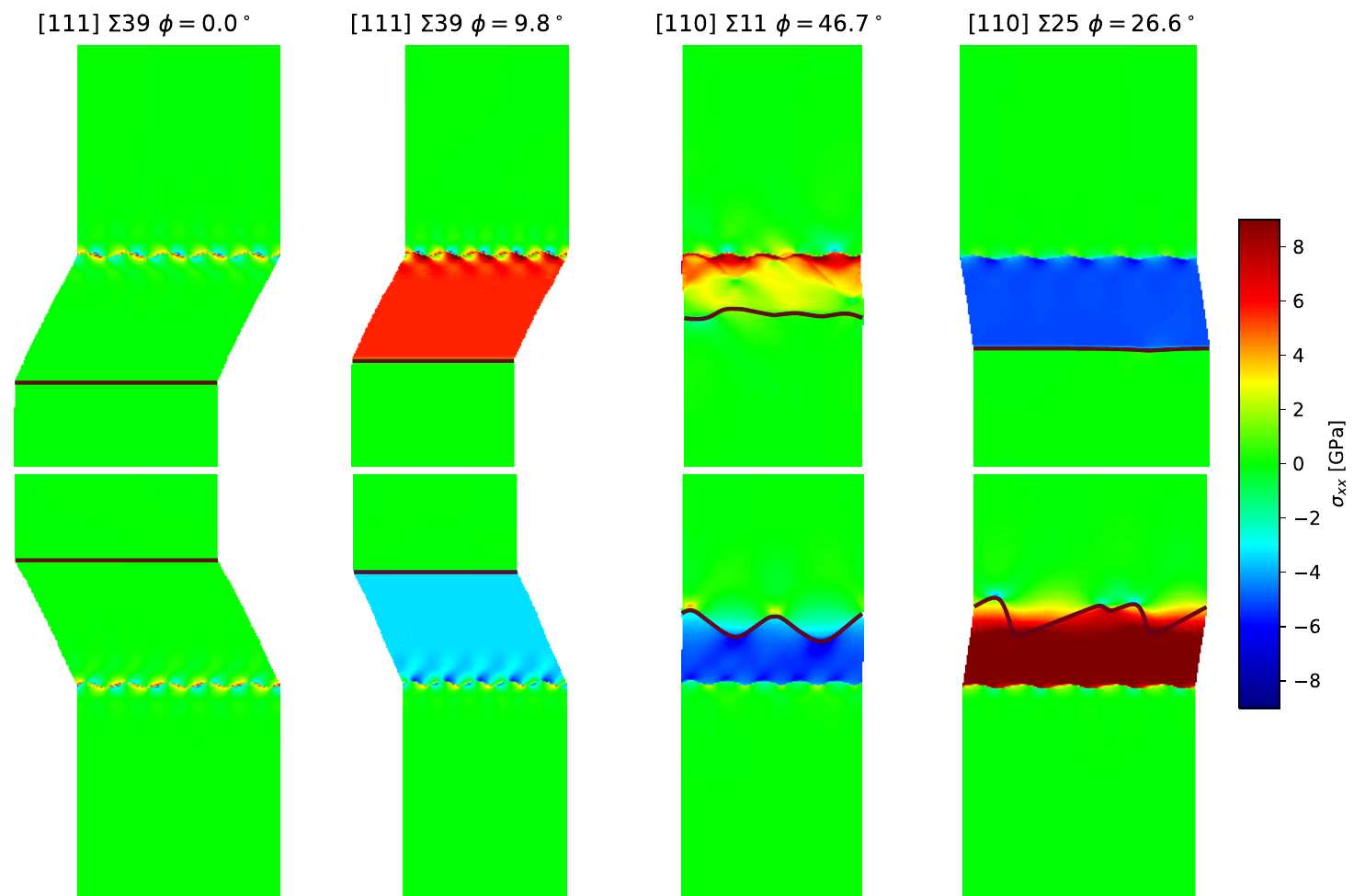}
    \caption{Comparison of $\sigma_{11}$ stress fields (color) and $\eta=0.5$ isocontour for downward-driven SDF (top row) vs upward-driven SDF (bottom row) for the $\Sigma39$ STGB, $\Sigma39$ ATGB, $\Sigma11$ ATGB, and $\Sigma25$ ATGB.}
    \label{fig:results/Comparison/stresses}
  \end{addedbox}
\end{figure}

\added[id=R1,comment={1.3}]{
Since the \textit{actual} mobility of the boundary is constant ($M=10^{-11} \frac{m}{s\cdot Pa}$), the asymmetry arises from the mechanistic incompatibility of the rotated shear coupling matrix compared to the flat ATGB surface.
Since $\Delta\mathbf{F}^{gb}$ alone does not satisfy the Hadamard criterion across the boundary, the deformation gradients on either side $\mathbf{F}=\mathbf{F}^e\mathbf{F}^{gb}$ must contain an elastic deformation $\mathbf{F}^e\ne\mathbf{I}$ to maintain compatibility.
In other words, the boundary incompatibility induces back-stress, which preferentially increases the energetic penalty for boundary motion.
This is observed by considering the components of mechanical stress induced by the boundary's motion (\cref{fig:results/Comparison/stresses}).
The STGB is unsurprisingly identical in both cases.
The boundary's motion induces strikingly different stress fields for the ``upward SDF'' scenario vs the ``downward SDF''.
Most notably, for the $\Sigma11$ and $\Sigma25$ boundaries, there is an obvious difference between the morphologies of the downward-moving vs upward-moving boundaries.
(Interestingly the emergence of faceting was noted by \cite{qiu2024grain} as well).
Moreover, the residual stresses from the initial perturbation are quite distinct, with different back-stress doublets arising for upward and downward motion in all three ATGB cases.
The real difference lies in the $\sigma_{xx}$ back-stress, since upward and downward migration accumulate different amounts of strain energy.
}

\subsubsection{Mechanical Driving Force}

\begin{figure}


  \begin{addedbox}{C0}{1.3}
    \begin{subfigure}{0.48\linewidth}
      \includegraphics[height=5.5cm]
{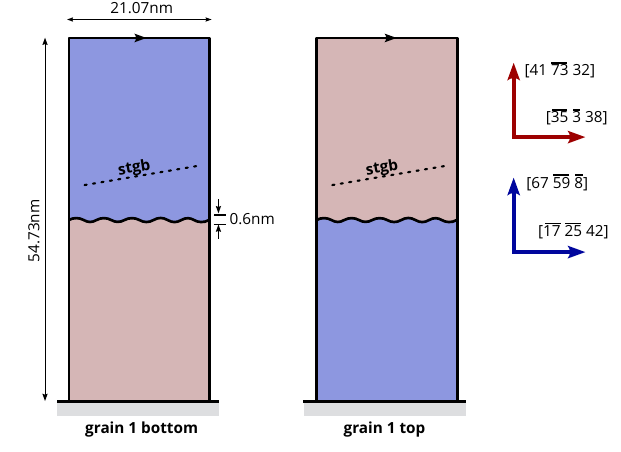}
      \caption{Schematic illustrating exchanged grain configurations}
      \label{fig:results/Mechanical/drawing}
    \end{subfigure}%
    \begin{subfigure}{0.52\linewidth}
      \includegraphics[height=5.5cm]
{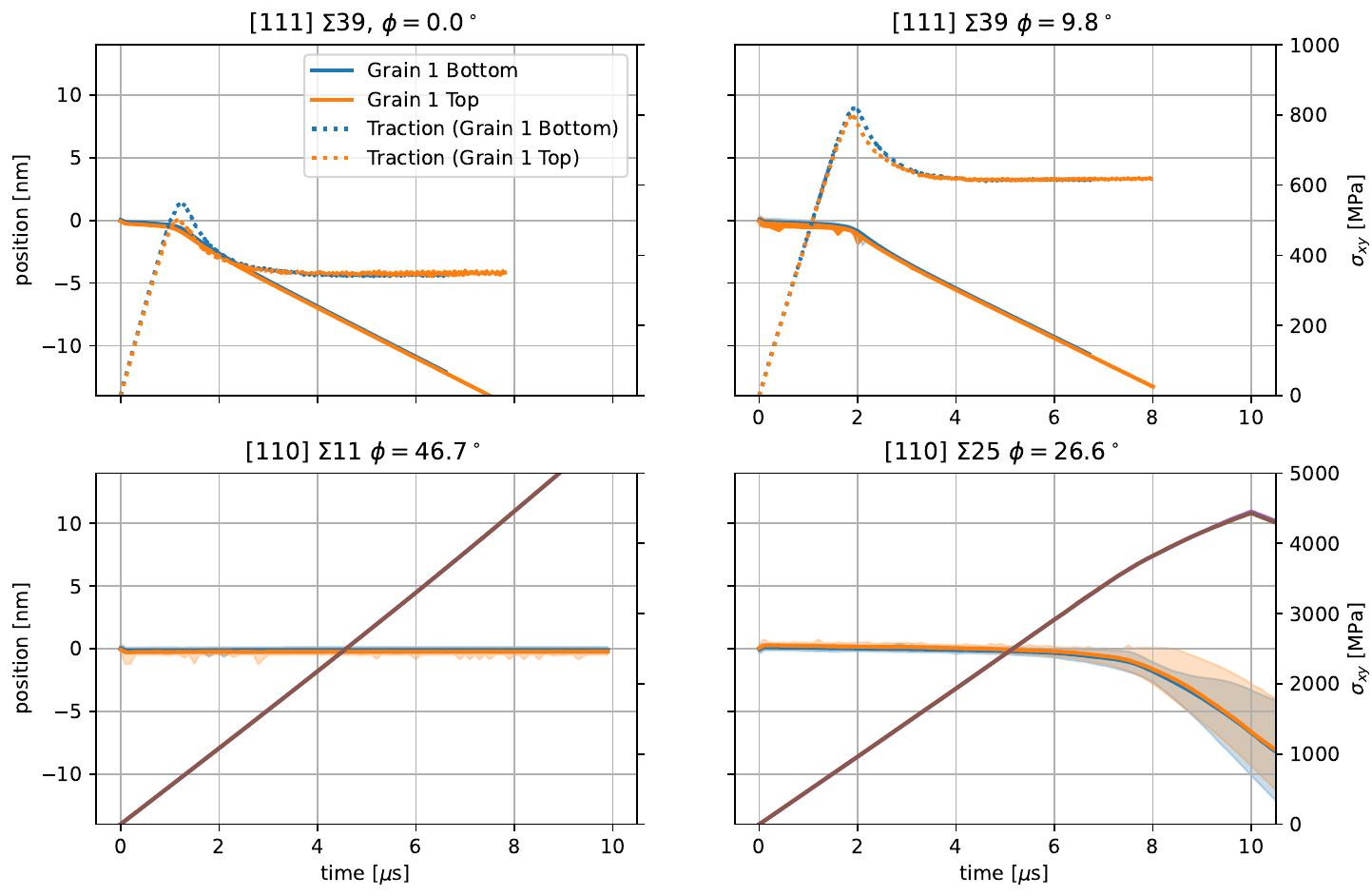}
      \caption{Plot of boundary position (solid) and traction (dashed) vs time}
      \label{fig:results/Mechanical/stress_volfrac}
    \end{subfigure}
    \caption{The mechanical driving force test is similar to the synthetic driving force, except that a mechanical load is applied in the form of horizontal displacement along the top.
    To test for directional dependence, the grains are inverted as shown in (a).
    The interface location (b - solid) indicates the average position of the isocontour $\eta=0.5$ (negative position for the inverted case for comparison).
    The faceting of the interface (b - shaded) is apparent especially in the $[110]$ $\Sigma25$ $\phi=26.6^\circ$ case.
    The traction indicates the mechanical relaxation due to boundary motion (b - dashed).
    }
  \end{addedbox}
\end{figure}

\begin{figure}
  \begin{addedbox}{C0}{1.3}
    \centering
    \includegraphics[width=0.8\linewidth]
{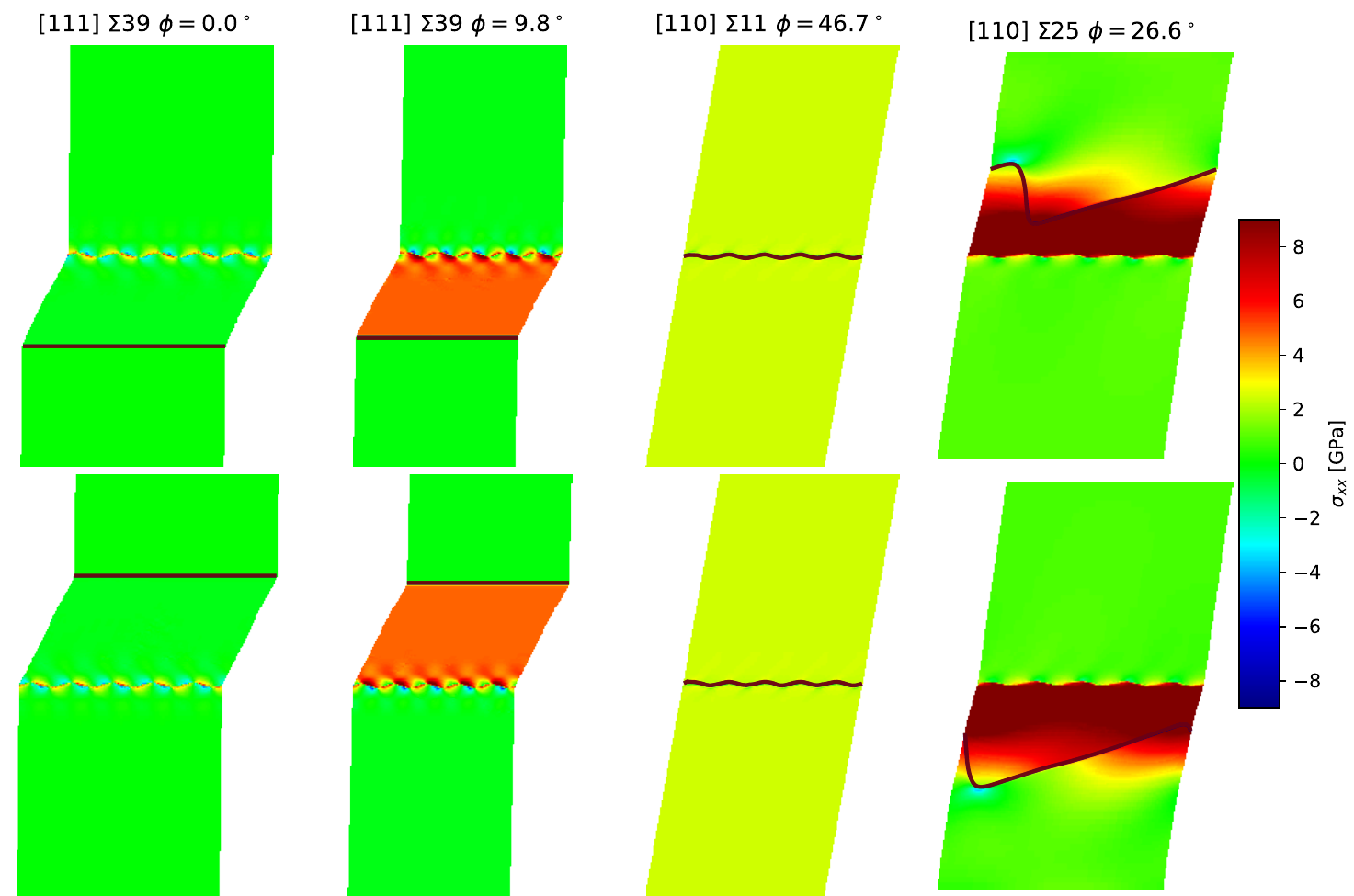}
    \caption{Mechanically driven stress fields}
  \end{addedbox}
\end{figure}

This section considers the same boundaries as above, except subjected to a mechanical rather than synthetic driving force.
Asymmetry is tested by \replaced[id=R1,comment={1.3}]{exchanging}{flipping} the domains occupied by grains 1 and 2, respectively (\cref{fig:results/Mechanical/drawing}).
To apply the mechanical load, the displacement is fixed on the bottom side of the domain.
For time $t\in[0,10\mu s]$, the top load is moved in the positive direction at a rate of $1\frac{nm}{\mu s}$, and then is fixed at a displacement of 10nm for $t>10\mu s$.
This shear rate, which is much slower than accessible by traditional atomistic methods, is beneficial because it allows for suitable relaxation of the boundary with time.
\replaced[id=R1,comment={1.3}]{Tests at higher and lower strain rates produced the same qualitative conclusion: under this mechanical driving protocol, the exchanged configurations do not develop a substantial directional mobility asymmetry.}{Nevertheless, the key asymmetric behavior is observed for both higher and lower strain rates.}

\added[id=R1,comment={1.3}]{The traction response for each boundary indicates the degree of similarity between the grain-1-bottom and grain-1-top configurations (\cref{fig:results/Mechanical/stress_volfrac} - dashed lines).}
\added[id=R1,comment={1.3}]{Because the displacement rate is constant for t<10$\mu s$ (white region), the traction-time response is equivalent to a stress-strain diagram up to scaling in the x direction.}
\added[id=R1,comment={1.3}]{Also, the boundary positions are nearly identical after sign inversion (\cref{fig:results/Mechanical/stress_volfrac} - solid lines).}
\replaced[id=R1,comment={1.3}]{As with symmetric tilt boundaries, the presence of initial ``defects'' - in the form of boundary perturbation - initially frustrates the motion of the boundary.}{Just as with symmetric tilt boundaries, the presence of initial ``defects'' - in the form of boundary perturbation - initially frustrates the motion of the boundary.}
The result is that the boundary is rendered immobile until a critical stress is reached, in this case between 300-350 MPa, despite that the flow of $\eta$ is nominally along the gradient.
\deleted[id=R1,comment={1.3}]{More relevant to this discussion, however, is the difference in flow stress between the grain-1-bottom and grain-1-top cases, which is nearly 40\%.
Since the boundary position is identical, this means that there is also a difference in {\it apparent mobility} of 40\%.
Moreover, the equilibrium stress for the two cases differs by almost a factor of two.}

\added[id=R1,comment={1.3}]{Unlike the application of the synthetic driving force, the mechanically driven cases show almost no directional dependence under the loading considered here.
For the exchanged grain configurations, the boundary position and stress are nearly identical for each boundary, although the \hkl[110] cases behave differently from the \hkl[111] cases.
In the \hkl[110] $\Sigma11$ boundary, the interface does not move at all; in the \hkl[110] $\Sigma25$ case, the motion is very small and accompanied by faceting.
This is anticipated due to the very small value of the effective $\beta$ that can be seen by considering the deformations due to SDF driven migration: since the $\mathbf{P}:\Delta\mathbf{F}^{gb}$ is so small, there is no resulting mechanical driving force on the boundary.}

\subsubsection{Curvature Driving Force}

\begin{figure}
  \begin{subfigure}{0.53\linewidth}
    \centering
    \includegraphics[height=5.5cm]
{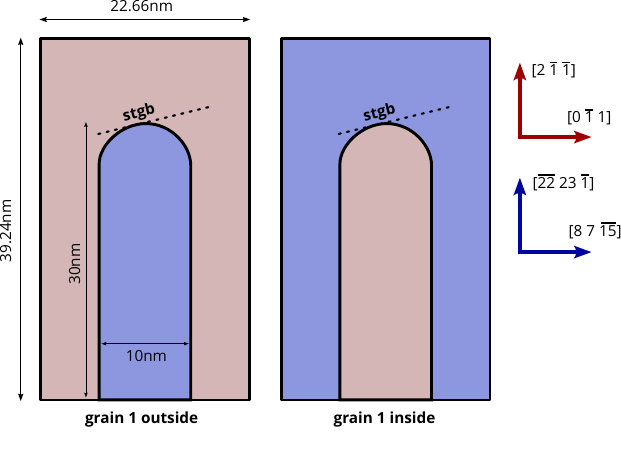}
    \caption{
    Schematic and geometry of half-loop configuration.
    Grain 1 indicated by red, grain 2 indicated by blue.
    The normal axis for both grains is \hkl[111], and the dashed line indicates the orientation of the symmetric tilt grain boundary.
    }
    \label{fig:results/HalfLoop/drawing}
  \end{subfigure}\hfill
  \begin{subfigure}{0.45\linewidth}
    \centering
    \includegraphics[height=5.5cm]
{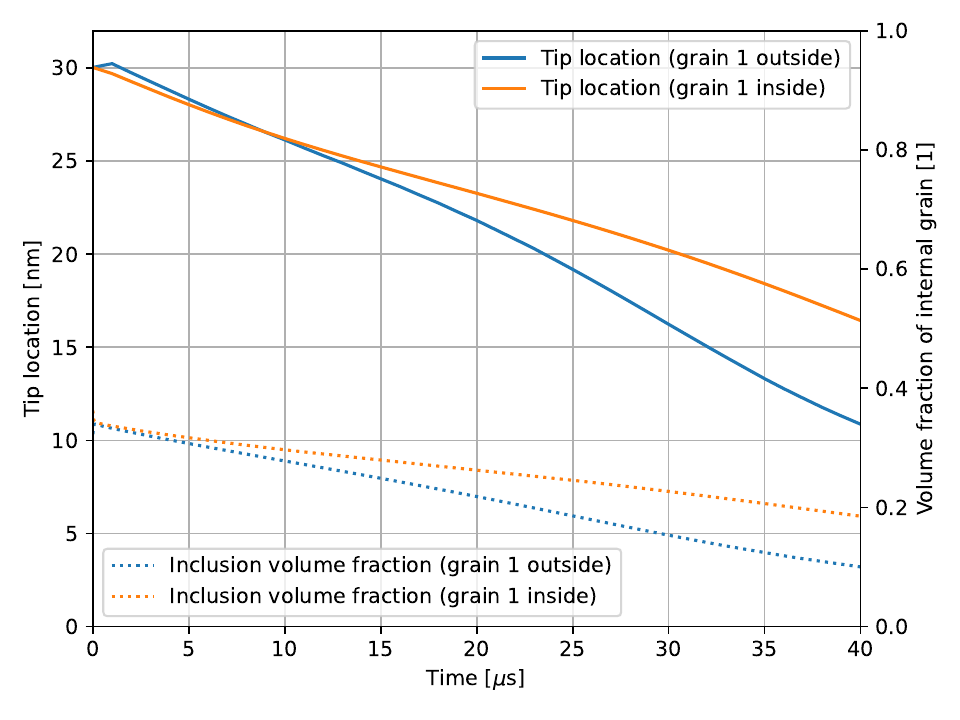}
    \caption{
    (Solid lines) Location of the tip of the half-loop measured from the bottom
    (Dashed lines) Volume of the inclusion (with a small correction added to offset the initial transient)
    }
  \end{subfigure}
  \caption{
  An asymmetric response is observed in the half-loop when the interior and exterior grains are exchanged, as indicated by the significantly faster migration of the boundary in the first case.
  The average velocity of the half-loop tip for the grain 1 inside case is 0.47 $nm/\mu s$; the velocity for the reversed case is 0.34 $nm /\mu s$.
  }
  \label{fig:results/HalfLoop/velocities}
\end{figure}

\begin{figure}
  \begin{subfigure}{0.5\linewidth}
    \includegraphics[width=1cm,clip,trim=0cm 27cm 25cm 3cm]
{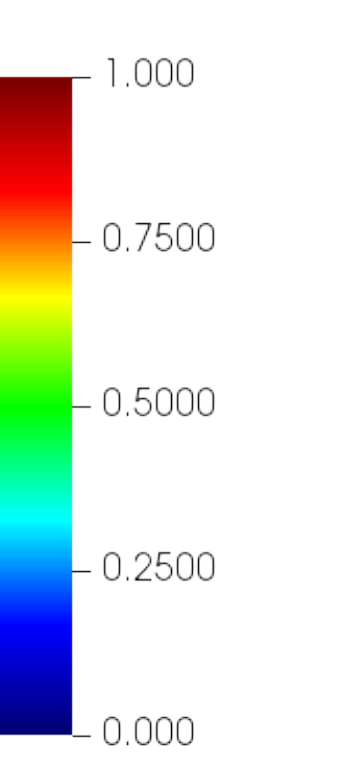}%
    \includegraphics[height=3cm,clip,trim=6cm 0cm 0cm 0cm]
{8721671fd3cb3103.pdf}%
    \includegraphics[height=3cm,clip,trim=6cm 0cm 0cm 0cm]
{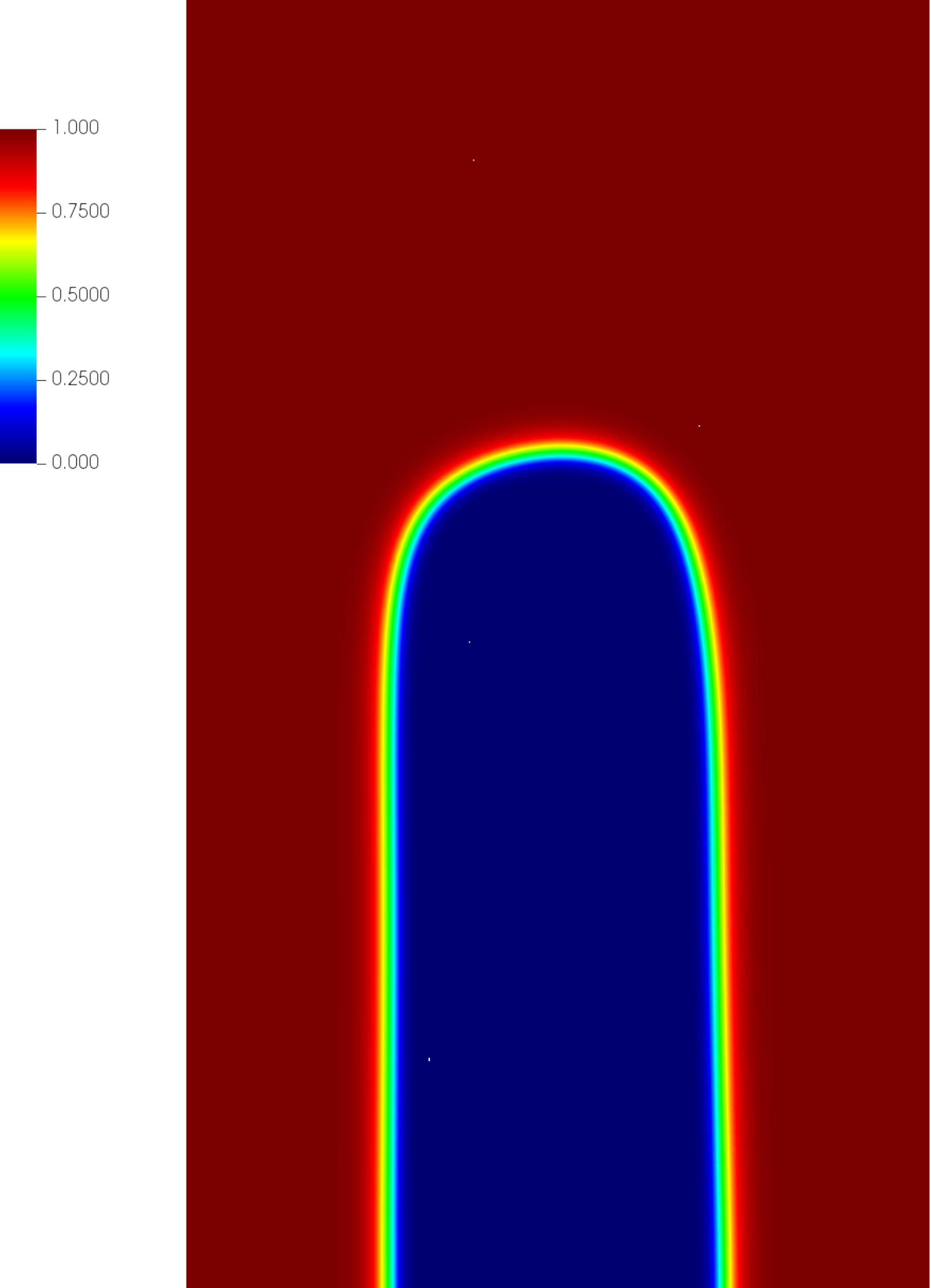}%
    \includegraphics[height=3cm,clip,trim=6cm 0cm 0cm 0cm]
{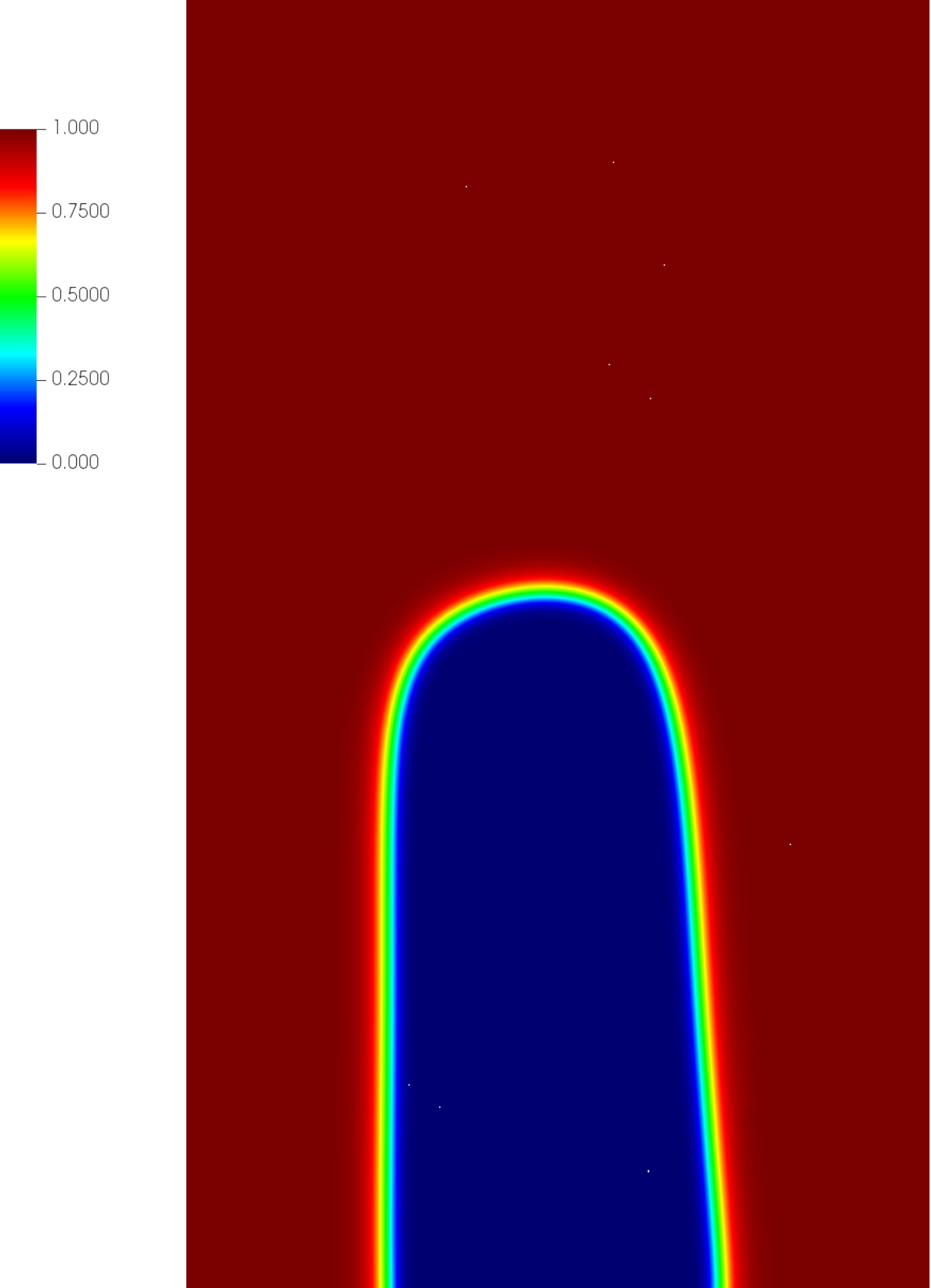}%
    \includegraphics[height=3cm,clip,trim=6cm 0cm 0cm 0cm]
{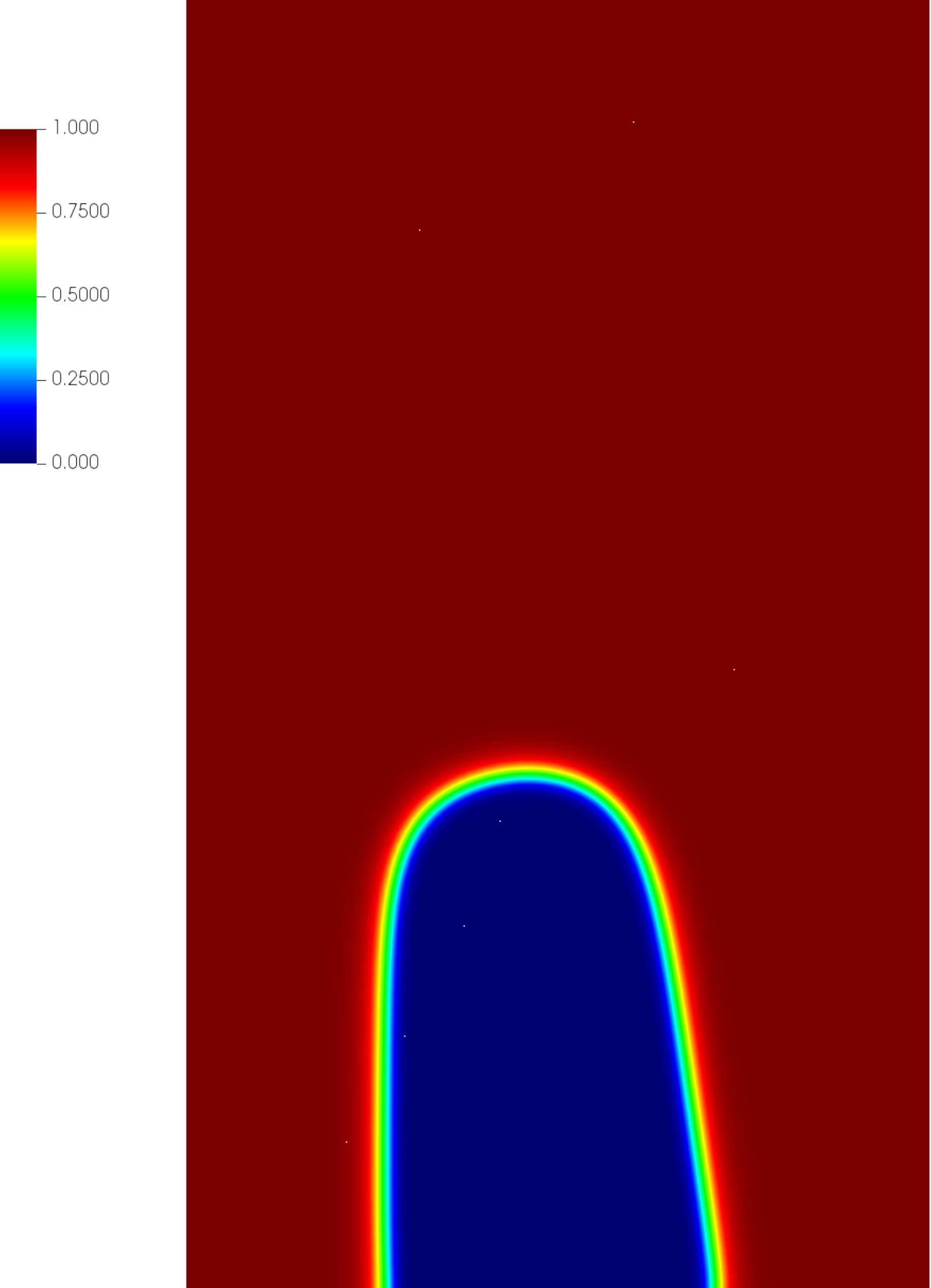}\par
    \hspace{1cm}%
    \includegraphics[height=3cm,clip,trim=6cm 0cm 0cm 0cm]
{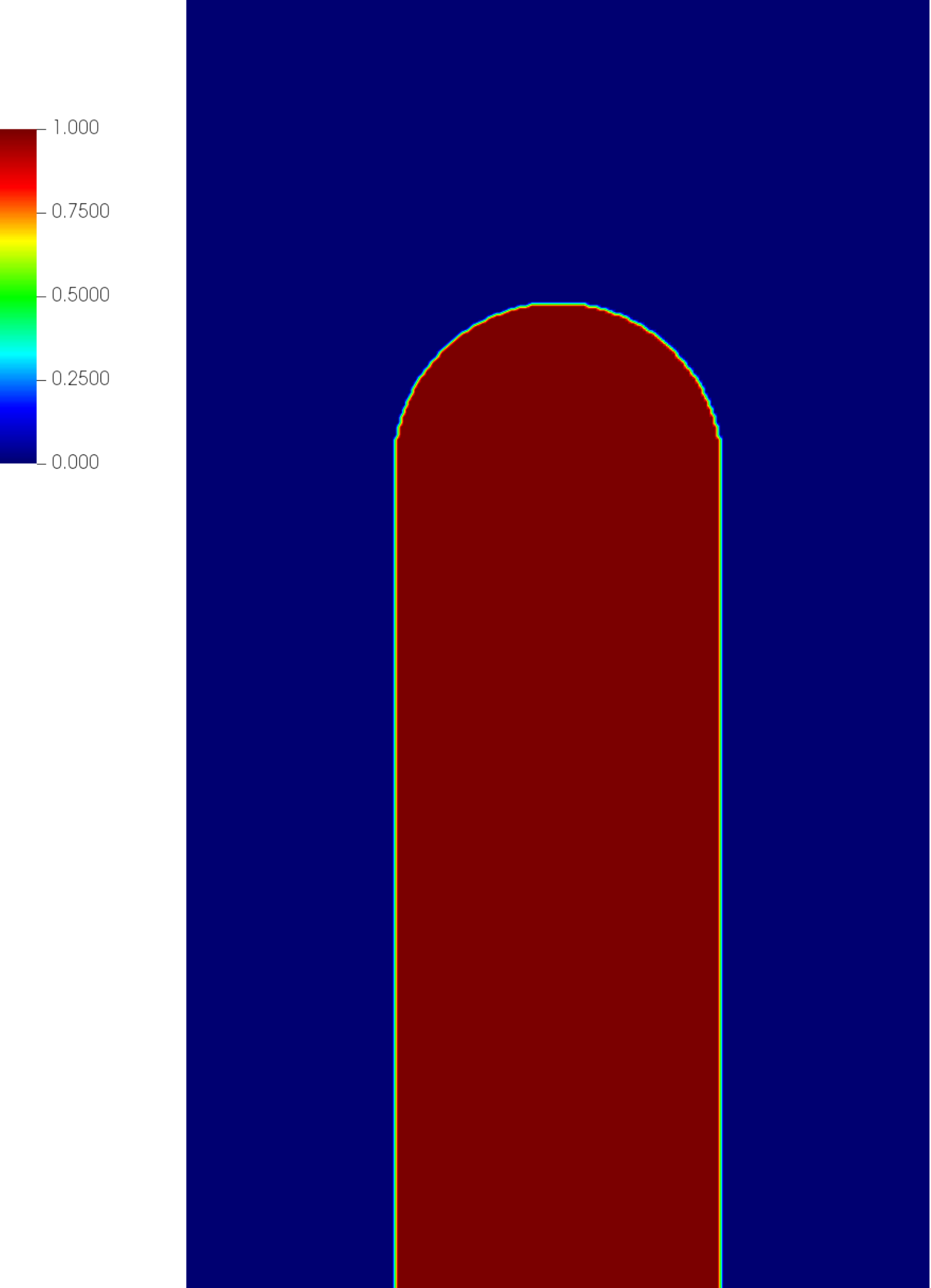}%
    \includegraphics[height=3cm,clip,trim=6cm 0cm 0cm 0cm]
{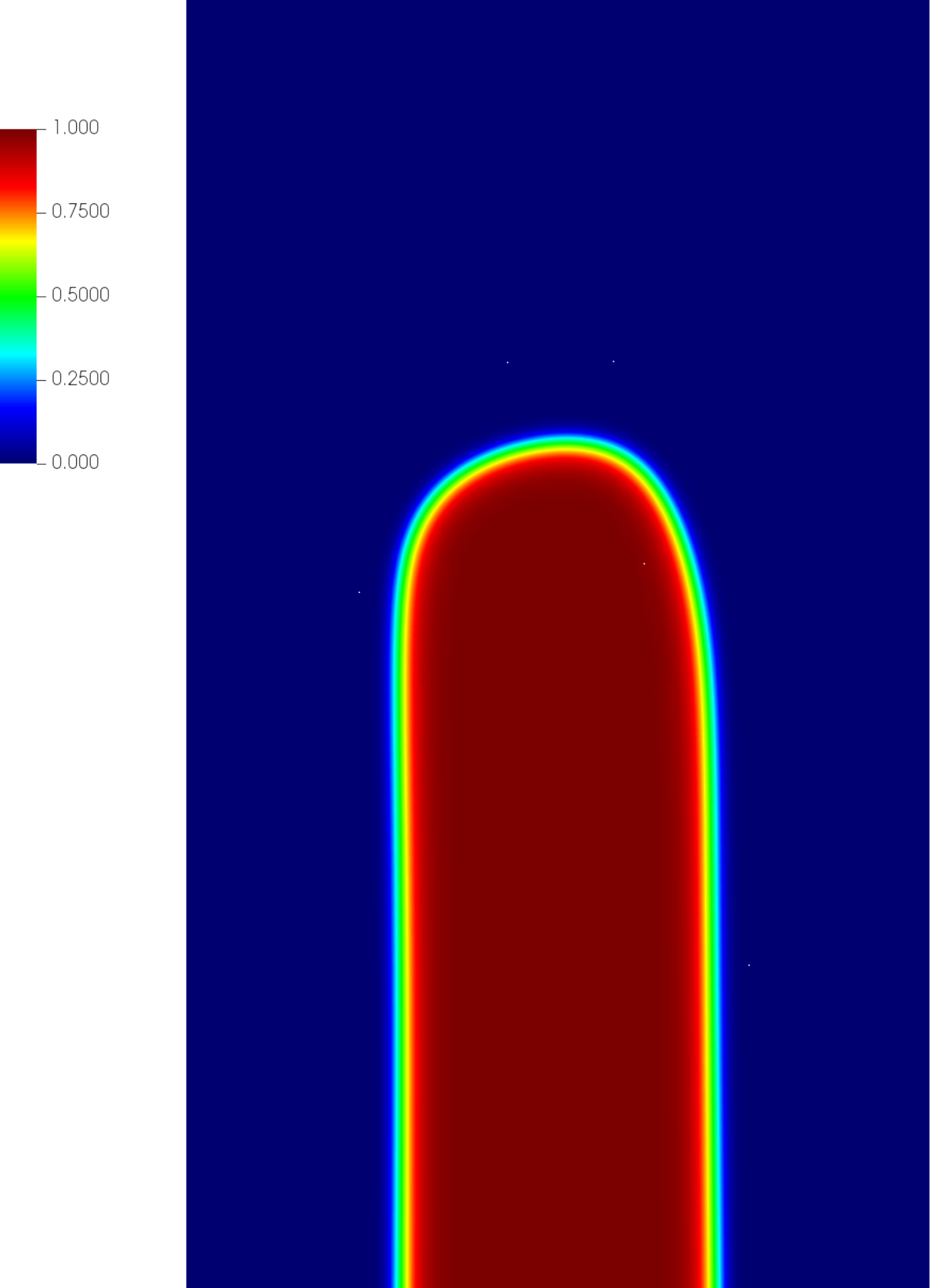}%
    \includegraphics[height=3cm,clip,trim=6cm 0cm 0cm 0cm]
{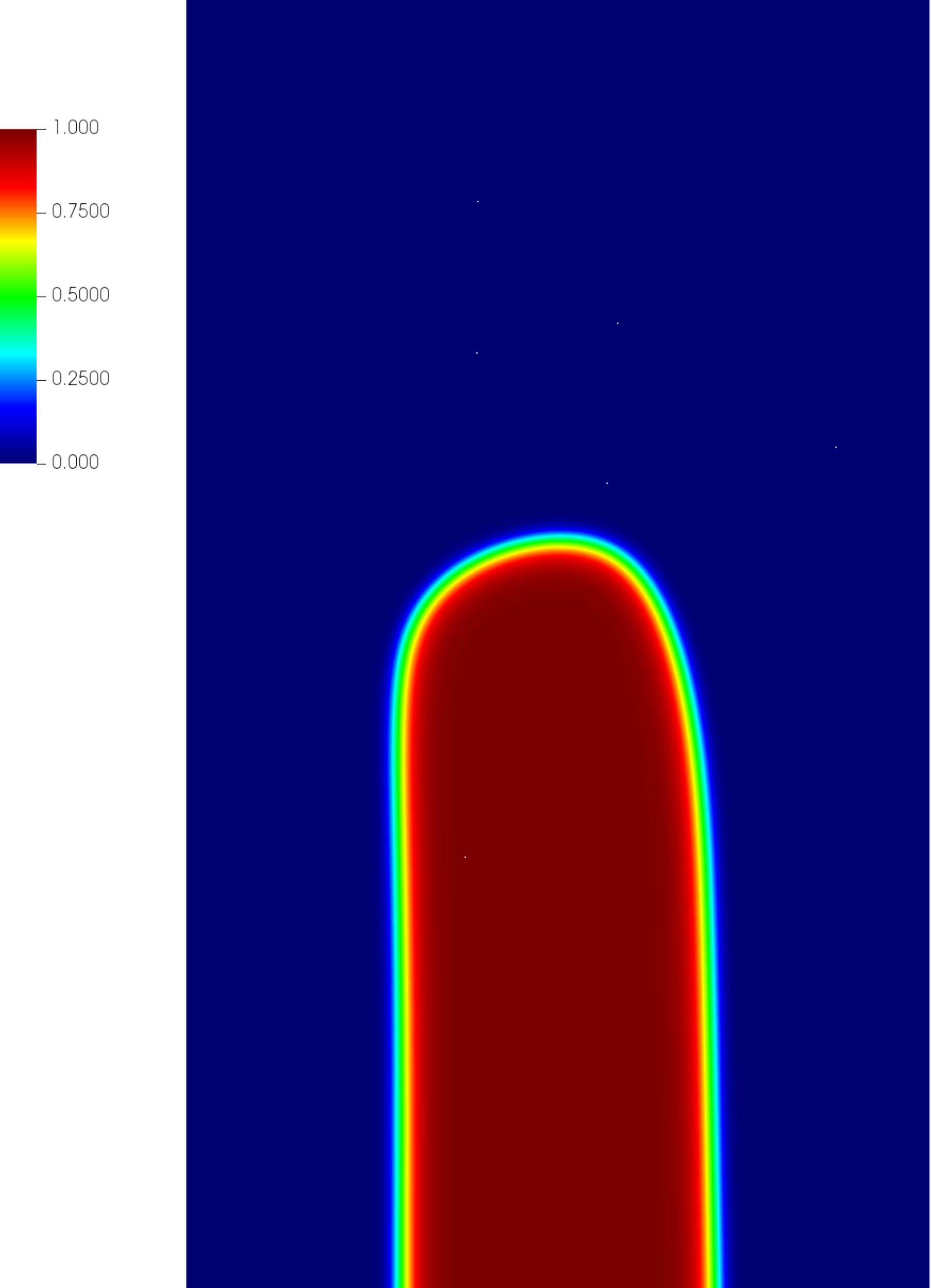}%
    \includegraphics[height=3cm,clip,trim=6cm 0cm 0cm 0cm]
{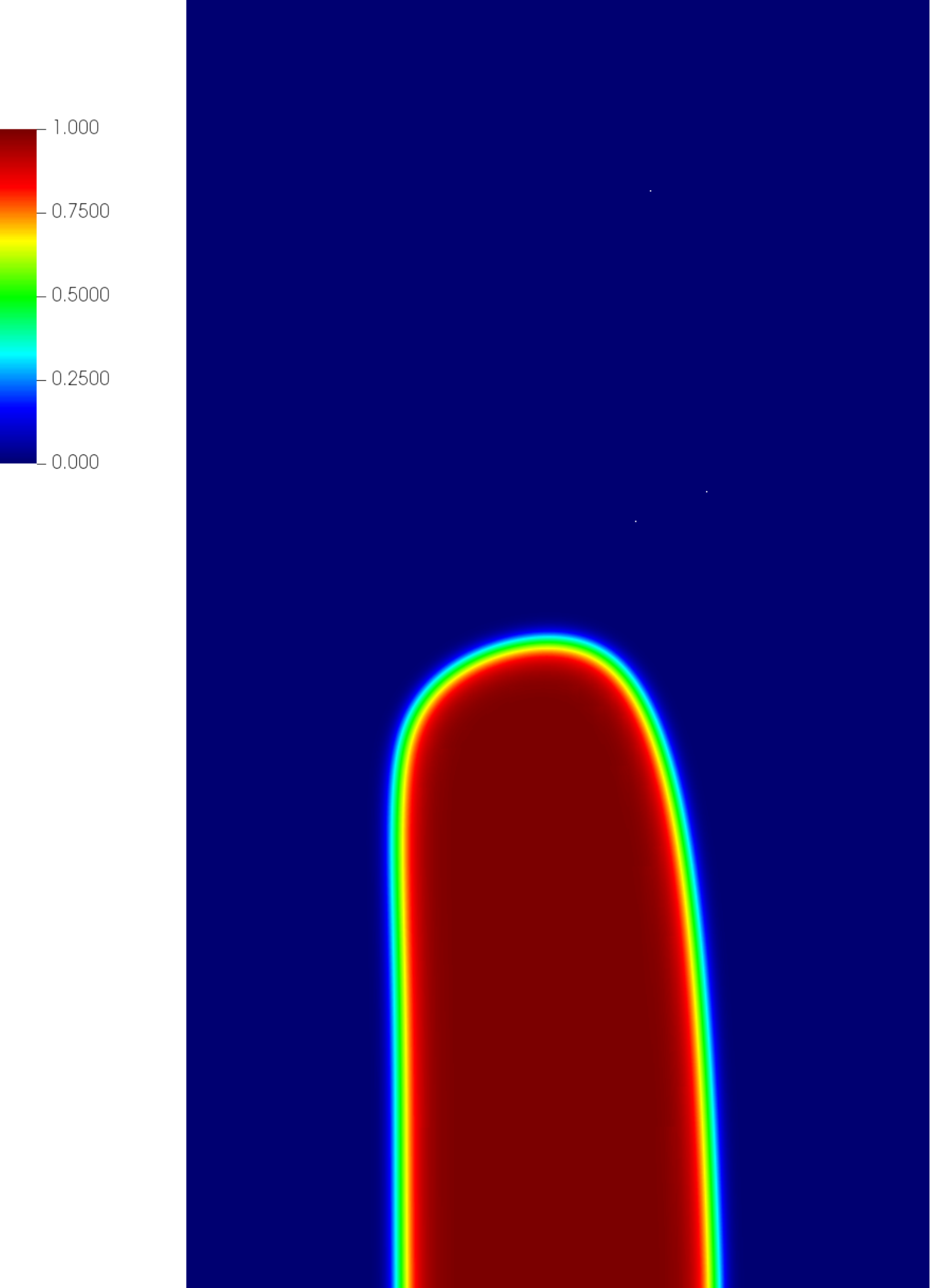}%
    \caption{Order parameter $\eta_1$ [-]}
    \label{fig:results/HalfLoop/output_reverse/eta}
  \end{subfigure}%
  \begin{subfigure}{0.5\linewidth}
    \includegraphics[width=1cm,clip,trim=0cm 27cm 25cm 3cm]
{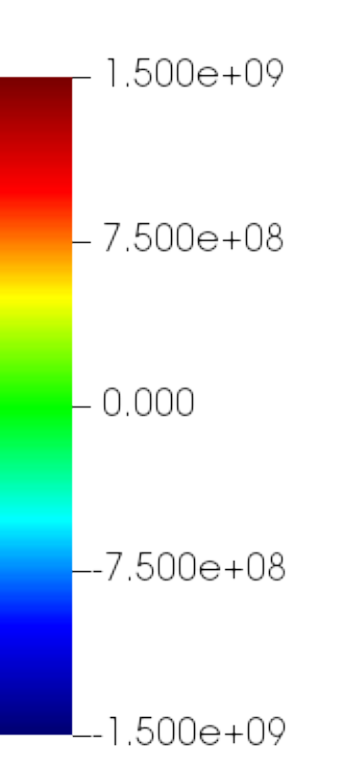}%
    \includegraphics[height=3cm,clip,trim=6cm 0cm 0cm 0cm]
{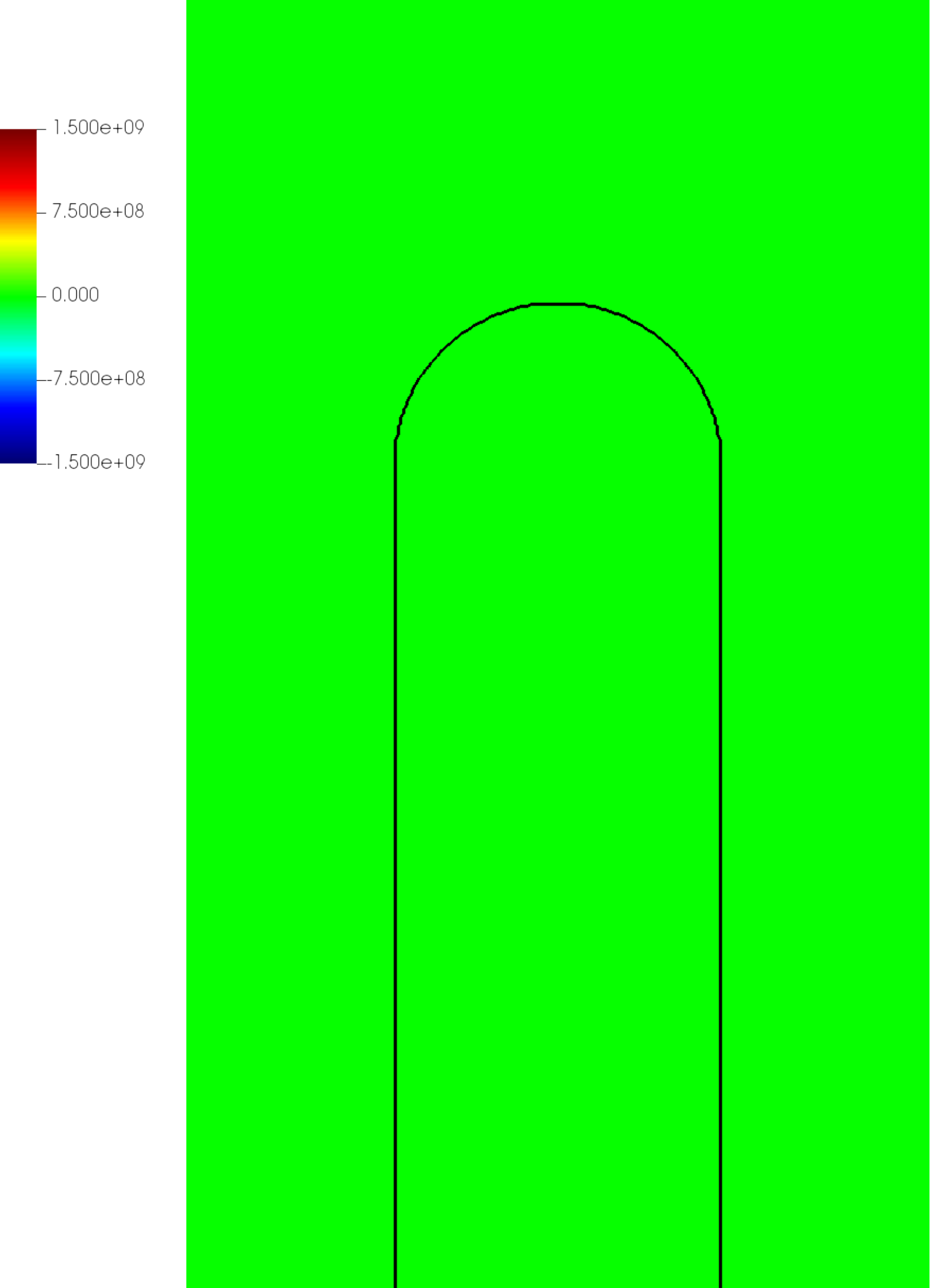}%
    \includegraphics[height=3cm,clip,trim=6cm 0cm 0cm 0cm]
{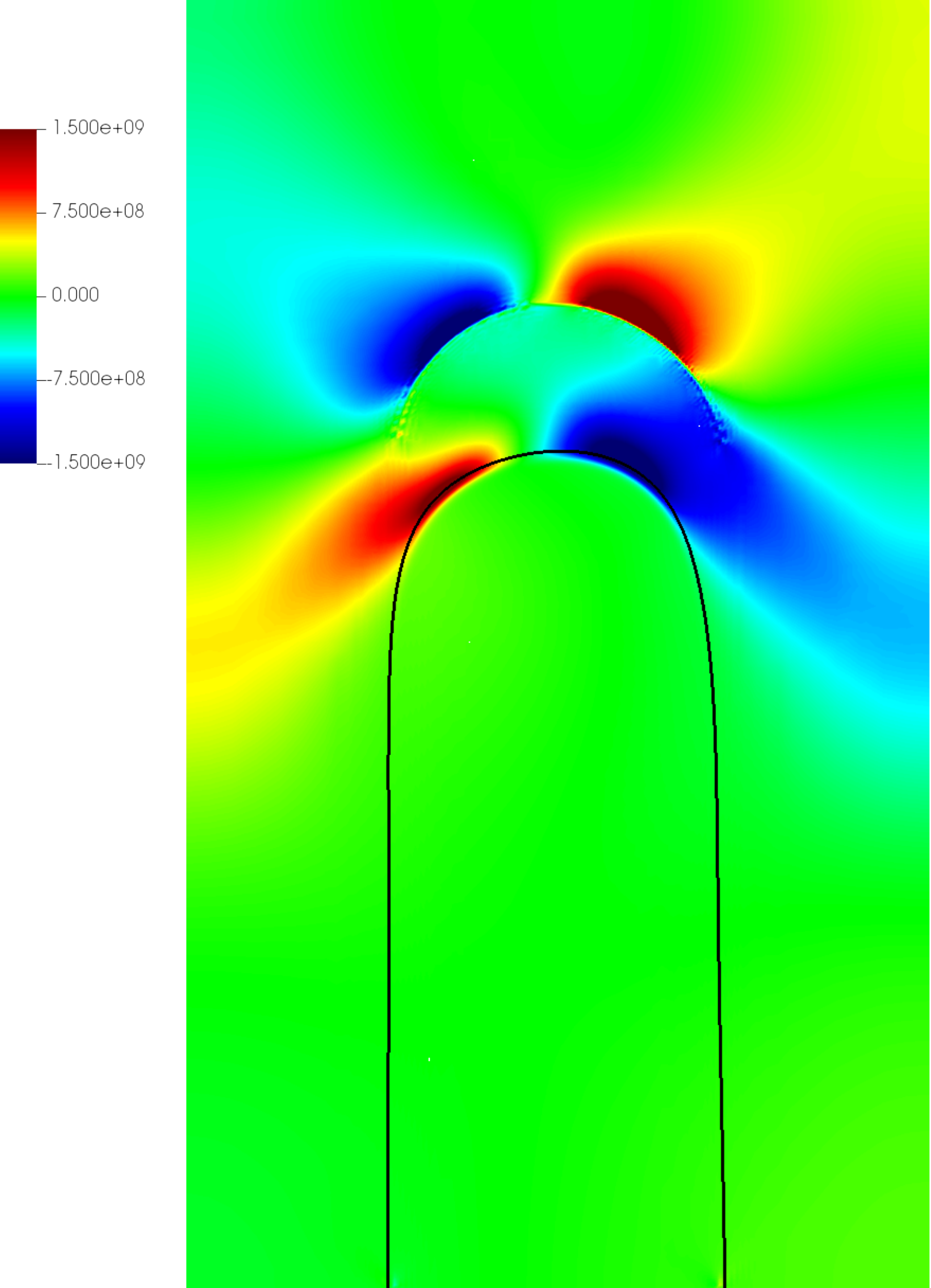}%
    \includegraphics[height=3cm,clip,trim=6cm 0cm 0cm 0cm]
{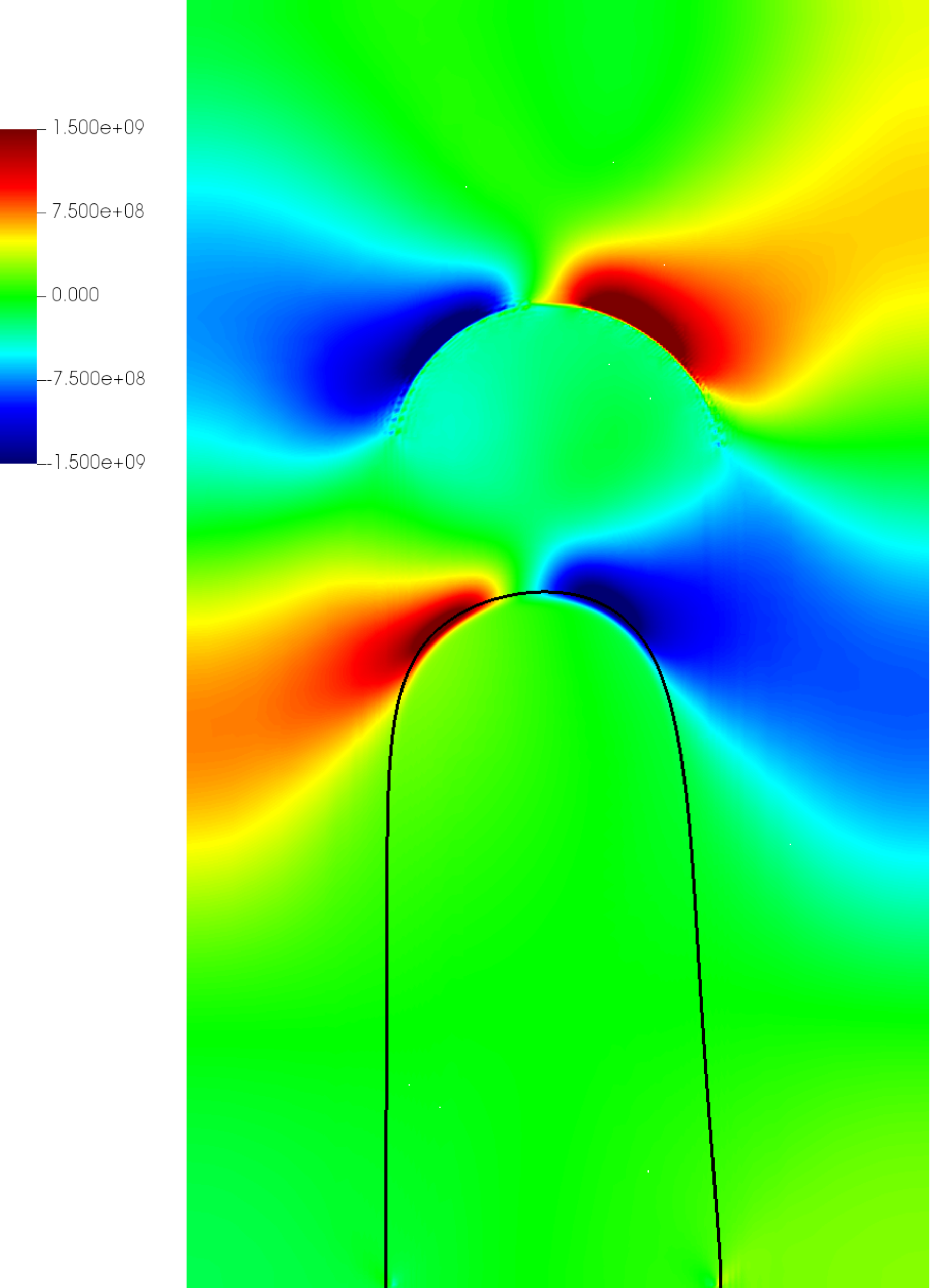}%
    \includegraphics[height=3cm,clip,trim=6cm 0cm 0cm 0cm]
{c47030fa180cc5ef.pdf}\par
    \hspace{1cm}%
    \includegraphics[height=3cm,clip,trim=6cm 0cm 0cm 0cm]
{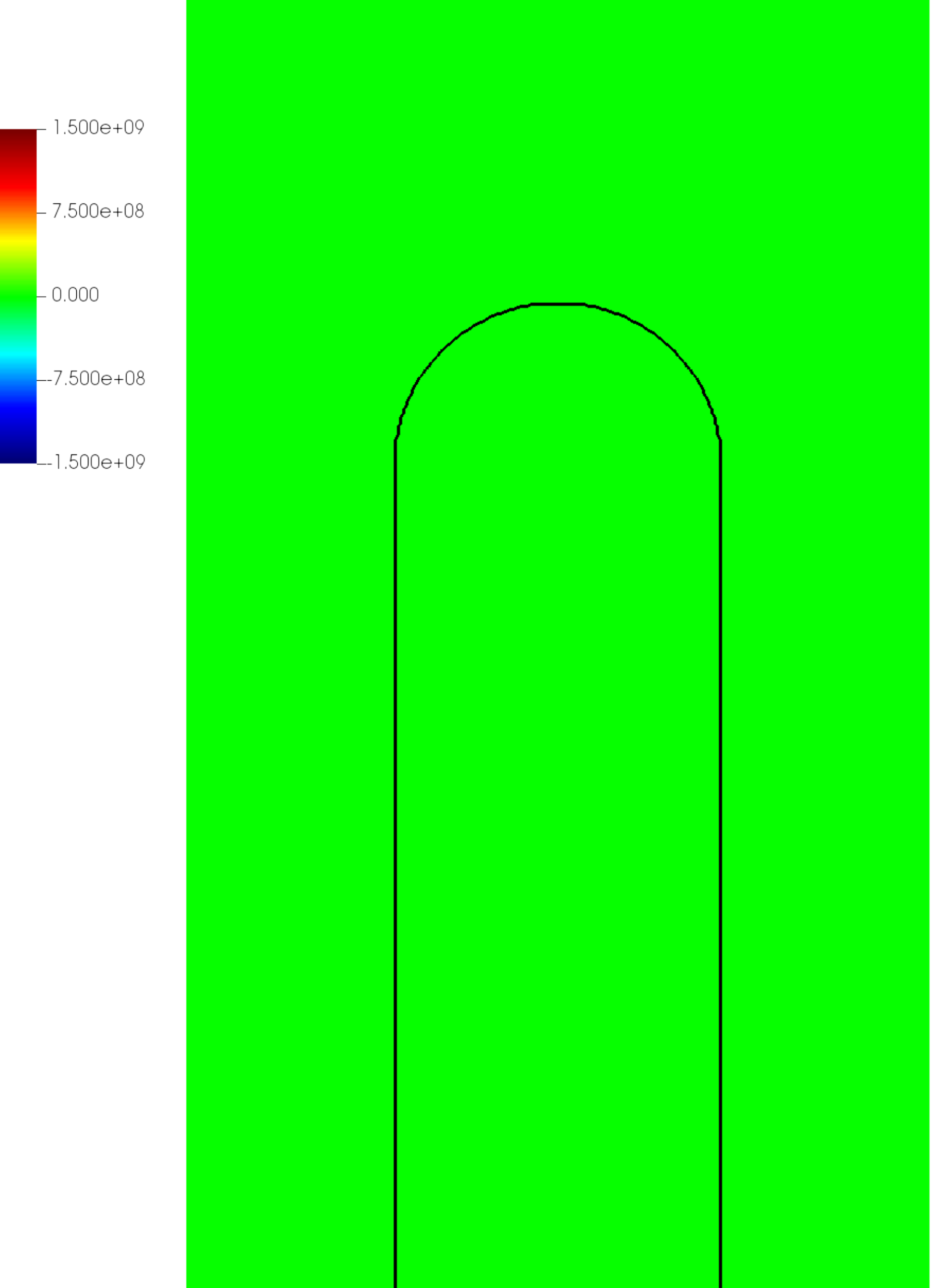}%
    \includegraphics[height=3cm,clip,trim=6cm 0cm 0cm 0cm]
{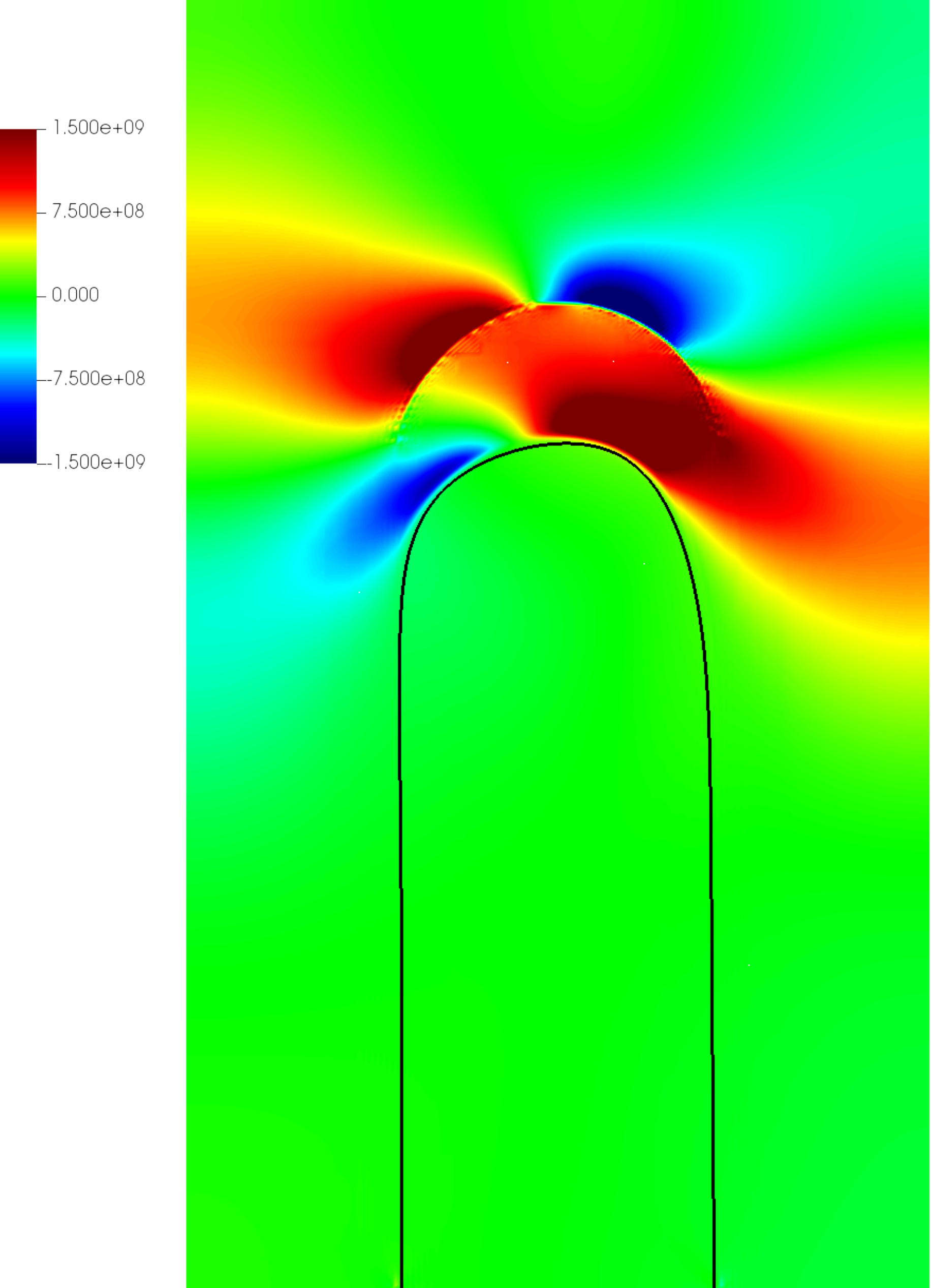}%
    \includegraphics[height=3cm,clip,trim=6cm 0cm 0cm 0cm]
{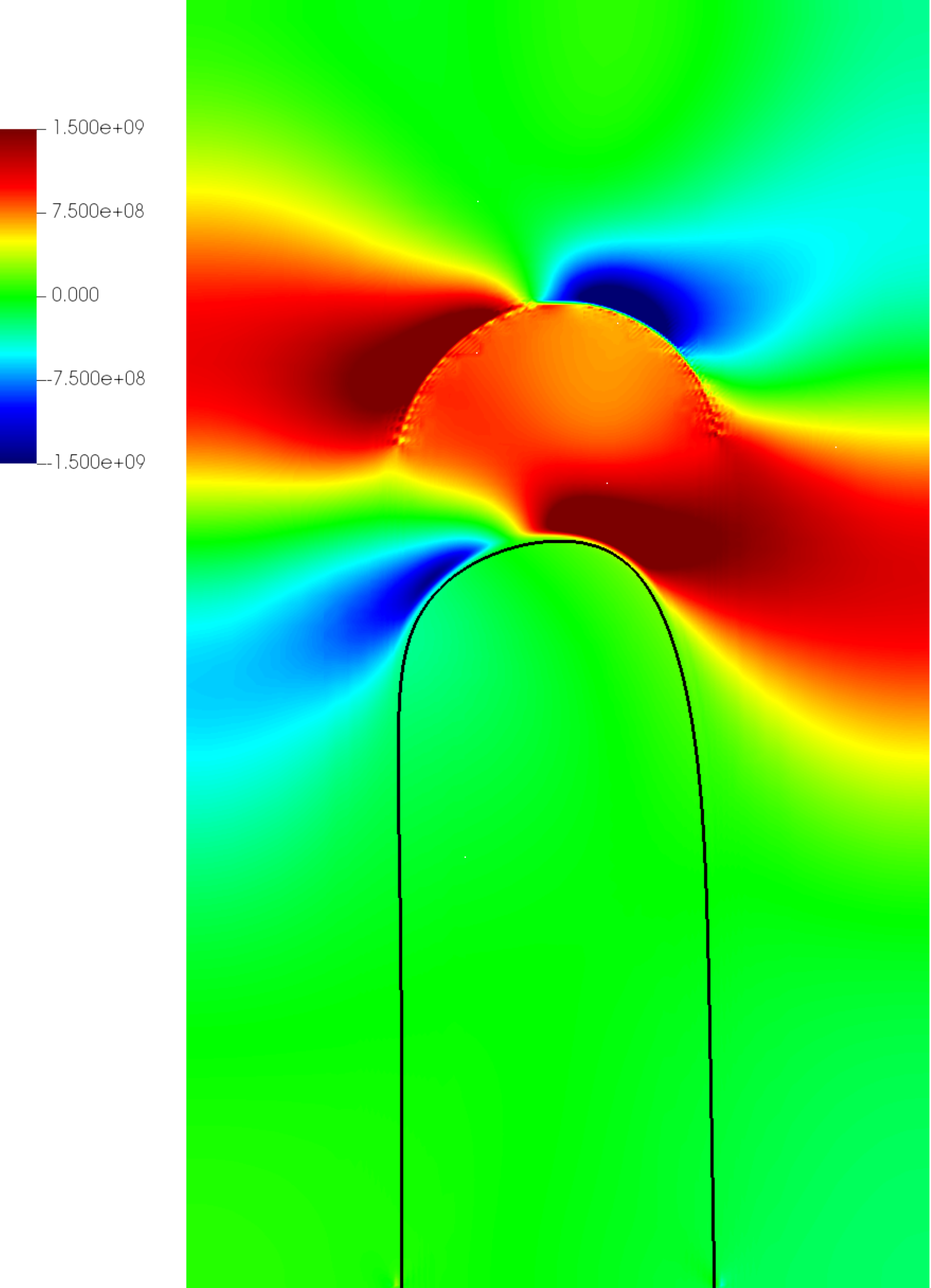}%
    \includegraphics[height=3cm,clip,trim=6cm 0cm 0cm 0cm]
{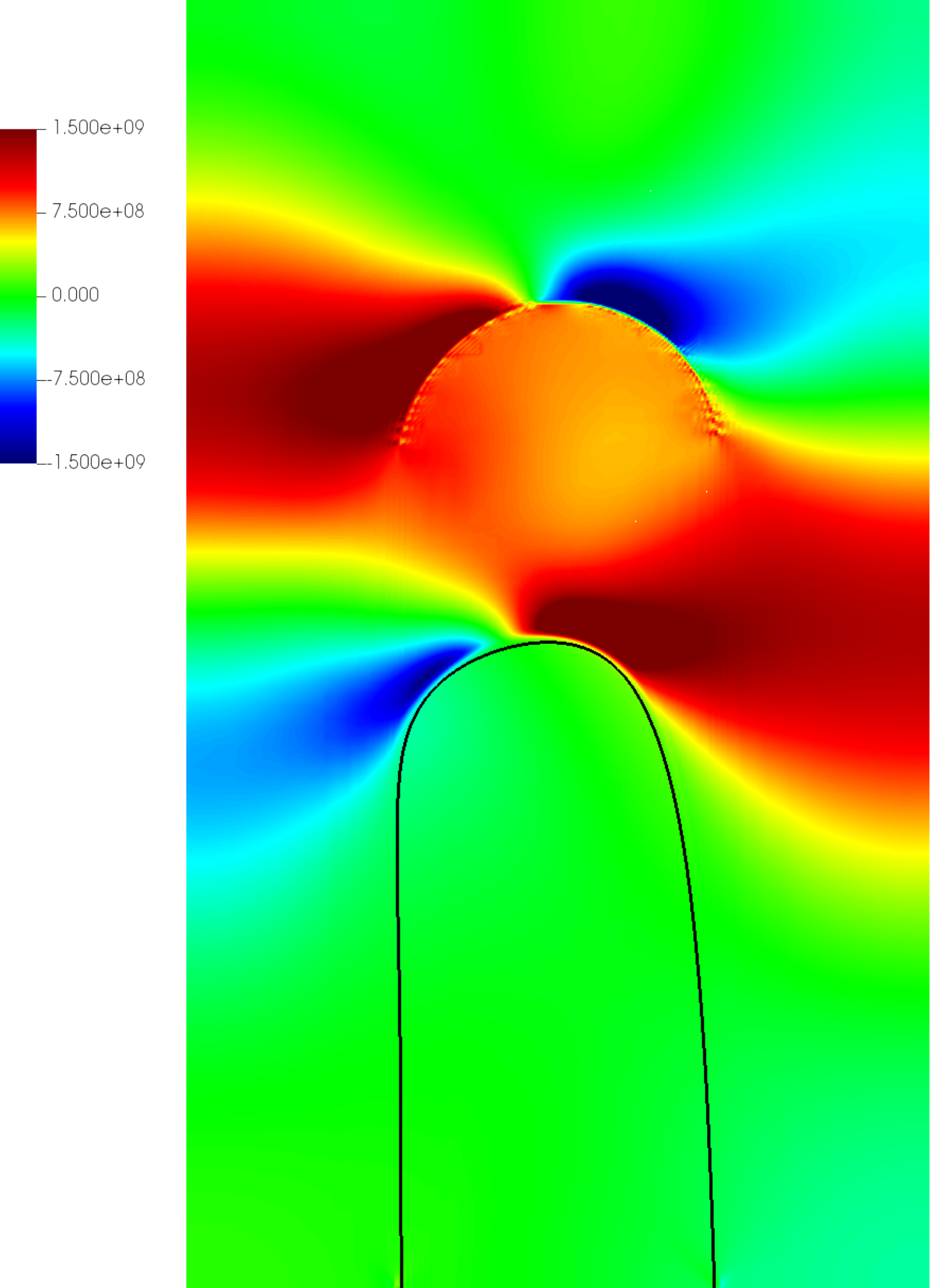}%
    \caption{Normal stress $\sigma_{xx}$ [Pa]}
    \label{fig:results/HalfLoop/output_reverse/sigma_xx}
  \end{subfigure}

  \begin{subfigure}{0.5\linewidth}
    \includegraphics[width=1cm,clip,trim=0cm 27cm 25cm 3cm]
{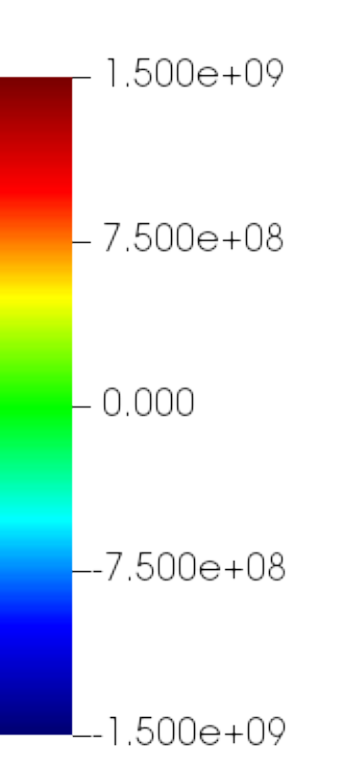}%
    \includegraphics[height=3cm,clip,trim=6cm 0cm 0cm 0cm]
{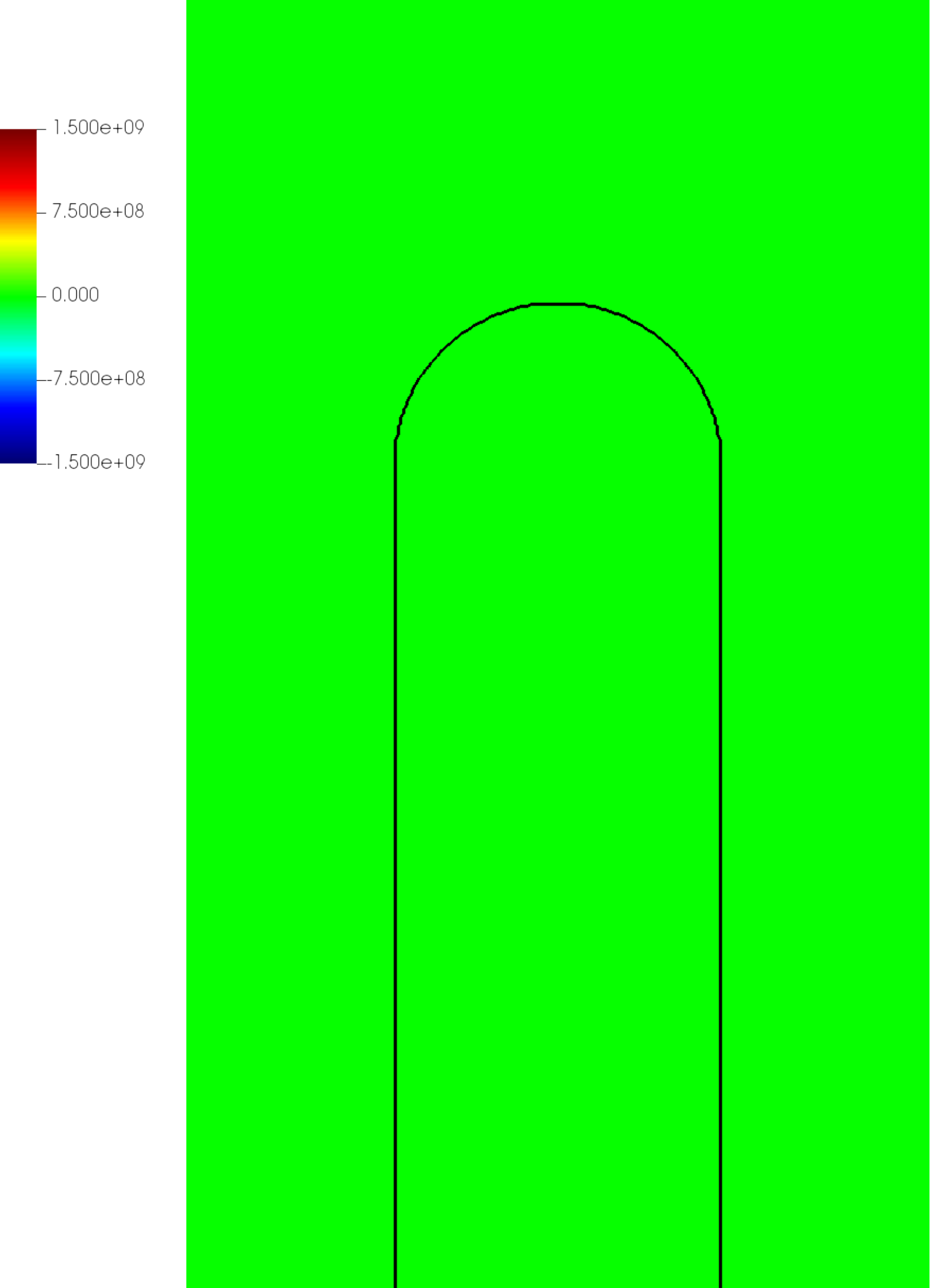}%
    \includegraphics[height=3cm,clip,trim=6cm 0cm 0cm 0cm]
{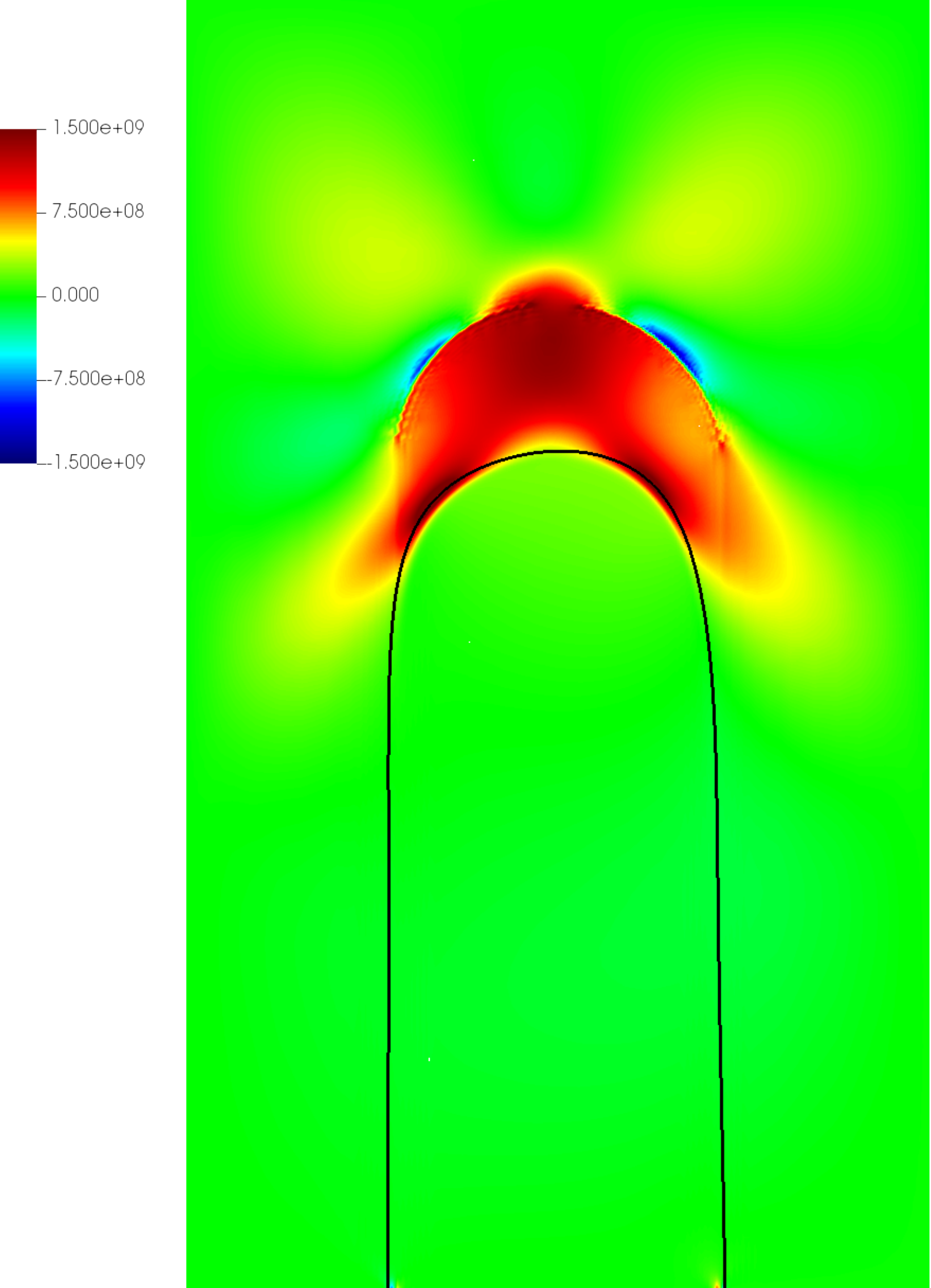}%
    \includegraphics[height=3cm,clip,trim=6cm 0cm 0cm 0cm]
{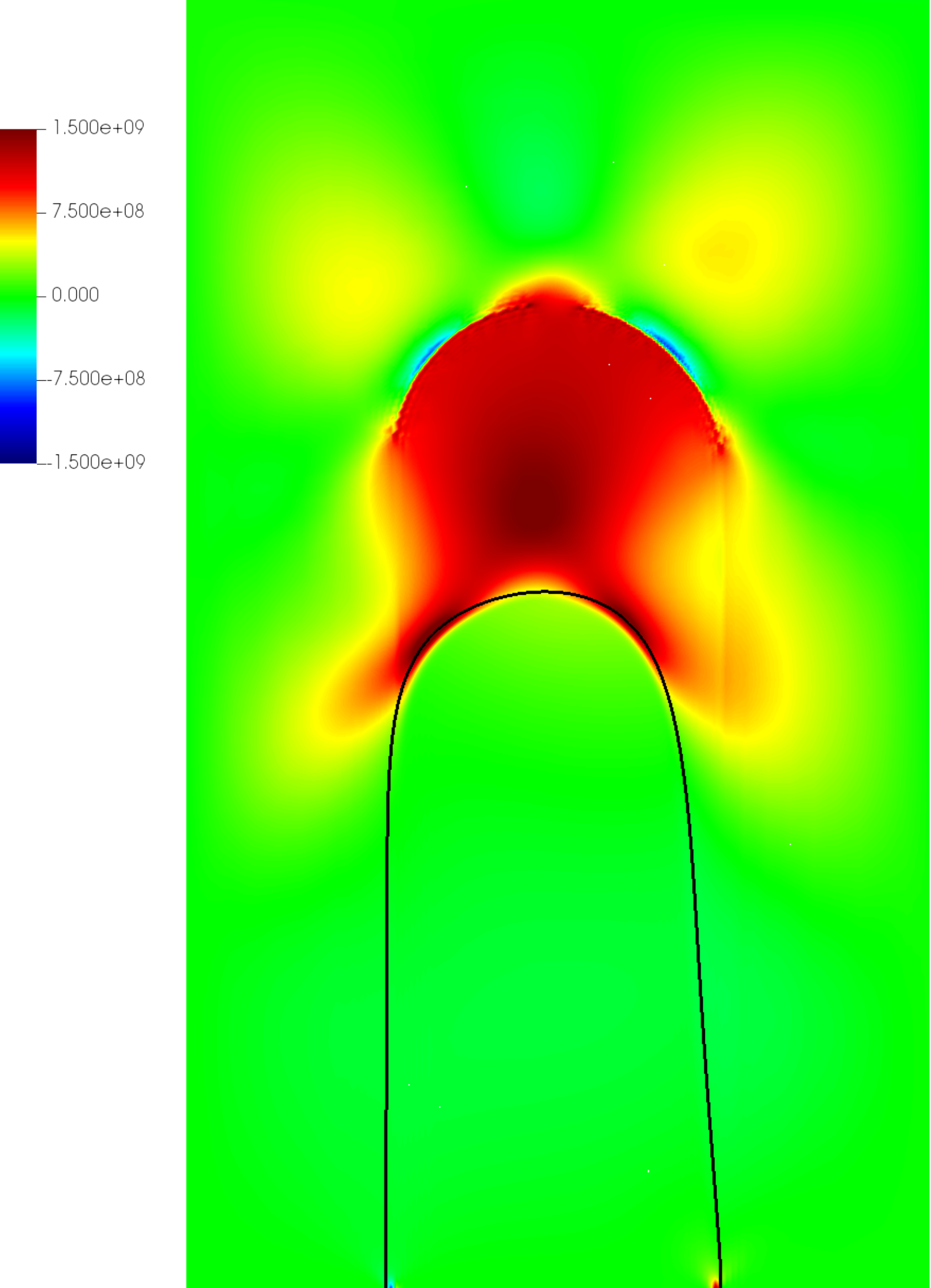}%
    \includegraphics[height=3cm,clip,trim=6cm 0cm 0cm 0cm]
{e944dc53e206b85f.pdf}\par
    \hspace{1cm}%
    \includegraphics[height=3cm,clip,trim=6cm 0cm 0cm 0cm]
{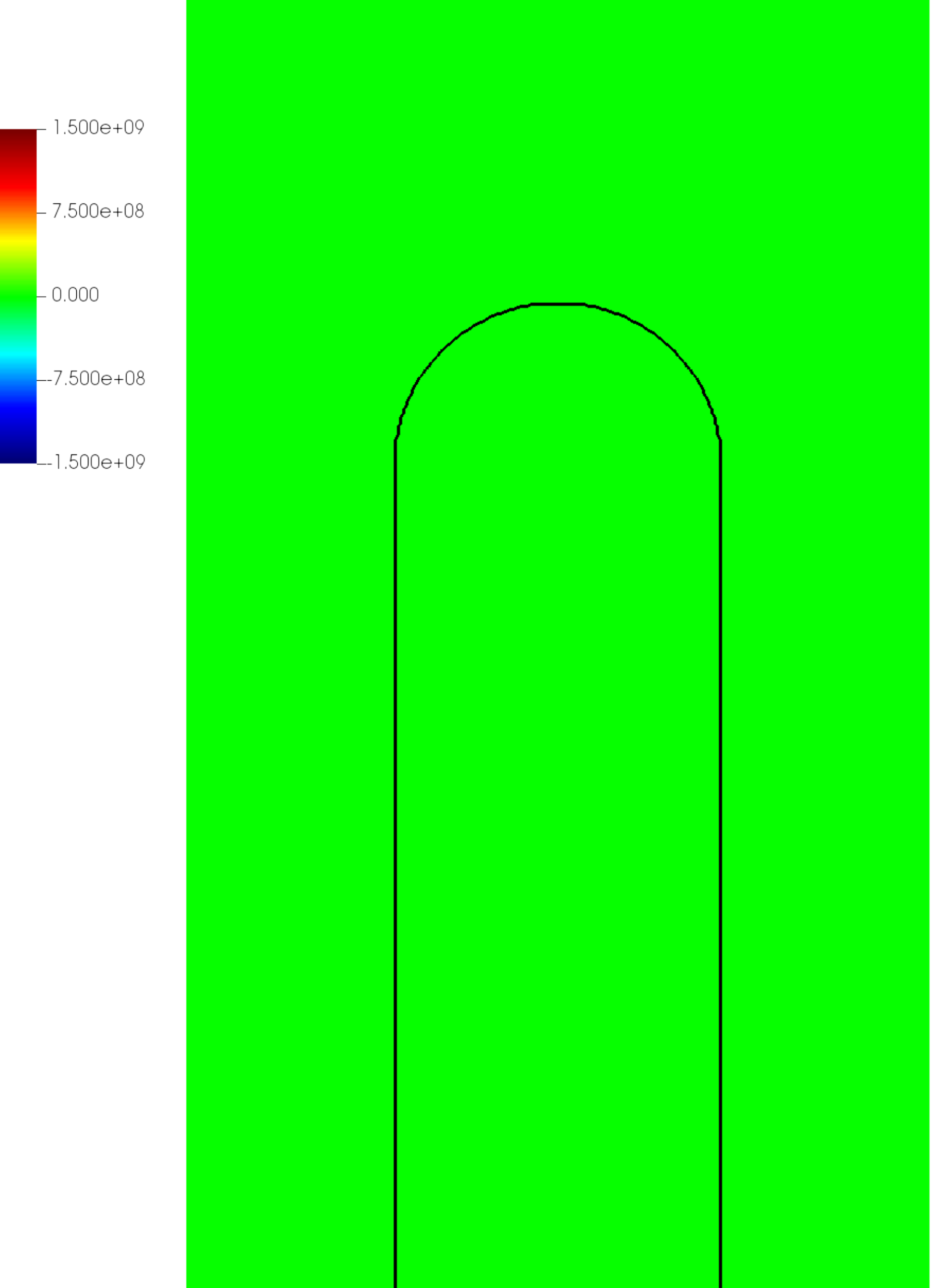}%
    \includegraphics[height=3cm,clip,trim=6cm 0cm 0cm 0cm]
{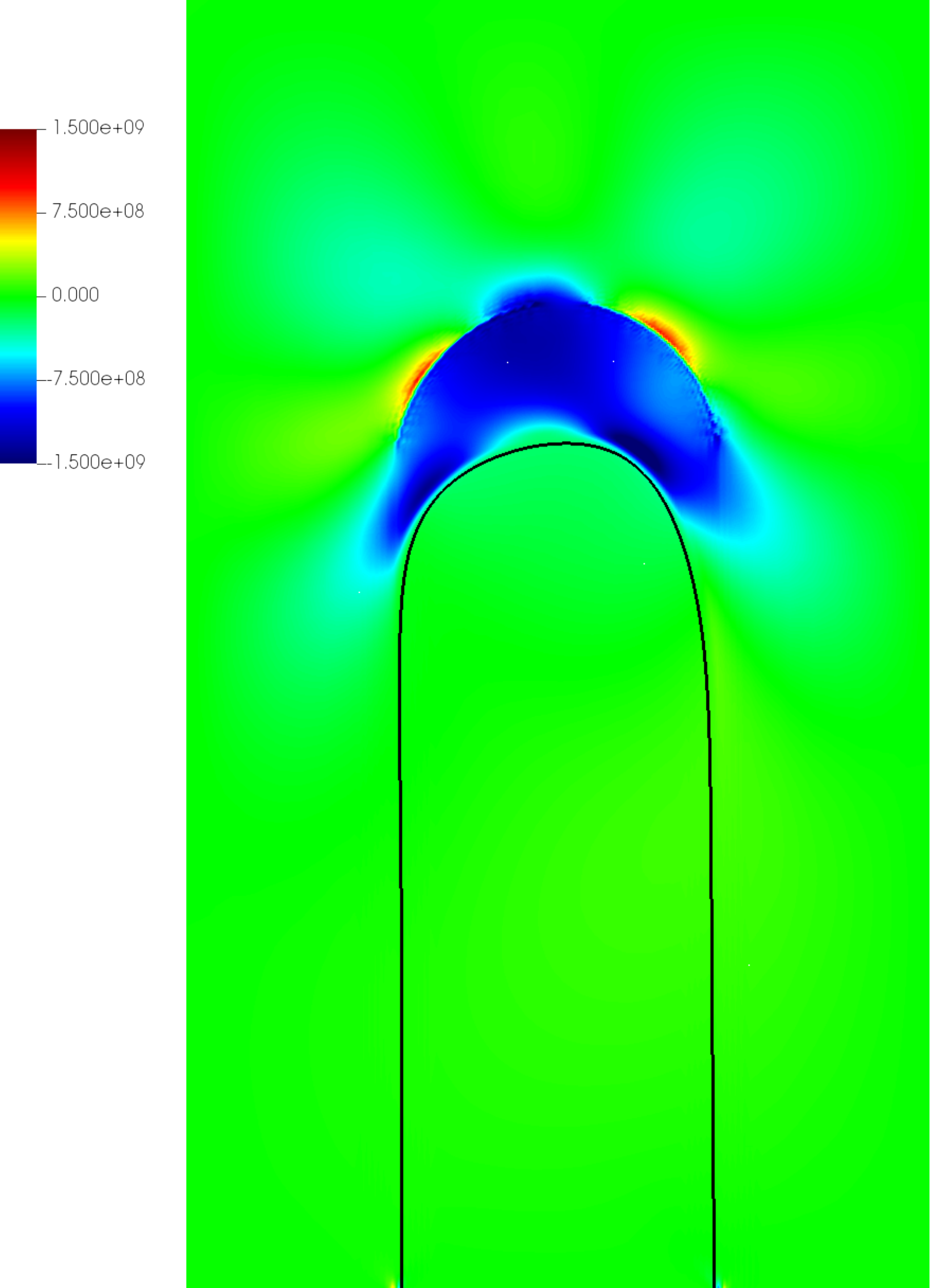}%
    \includegraphics[height=3cm,clip,trim=6cm 0cm 0cm 0cm]
{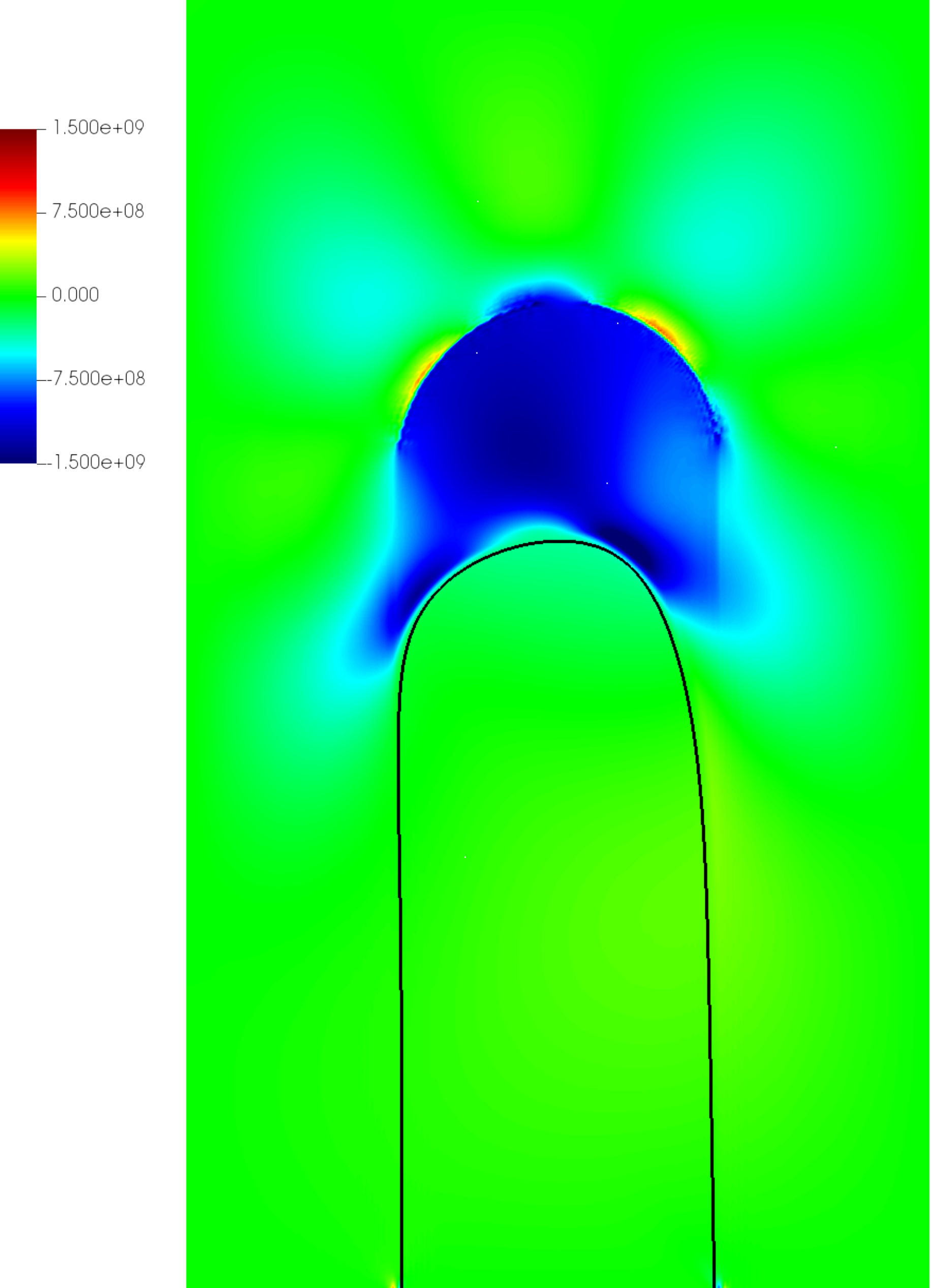}%
    \includegraphics[height=3cm,clip,trim=6cm 0cm 0cm 0cm]
{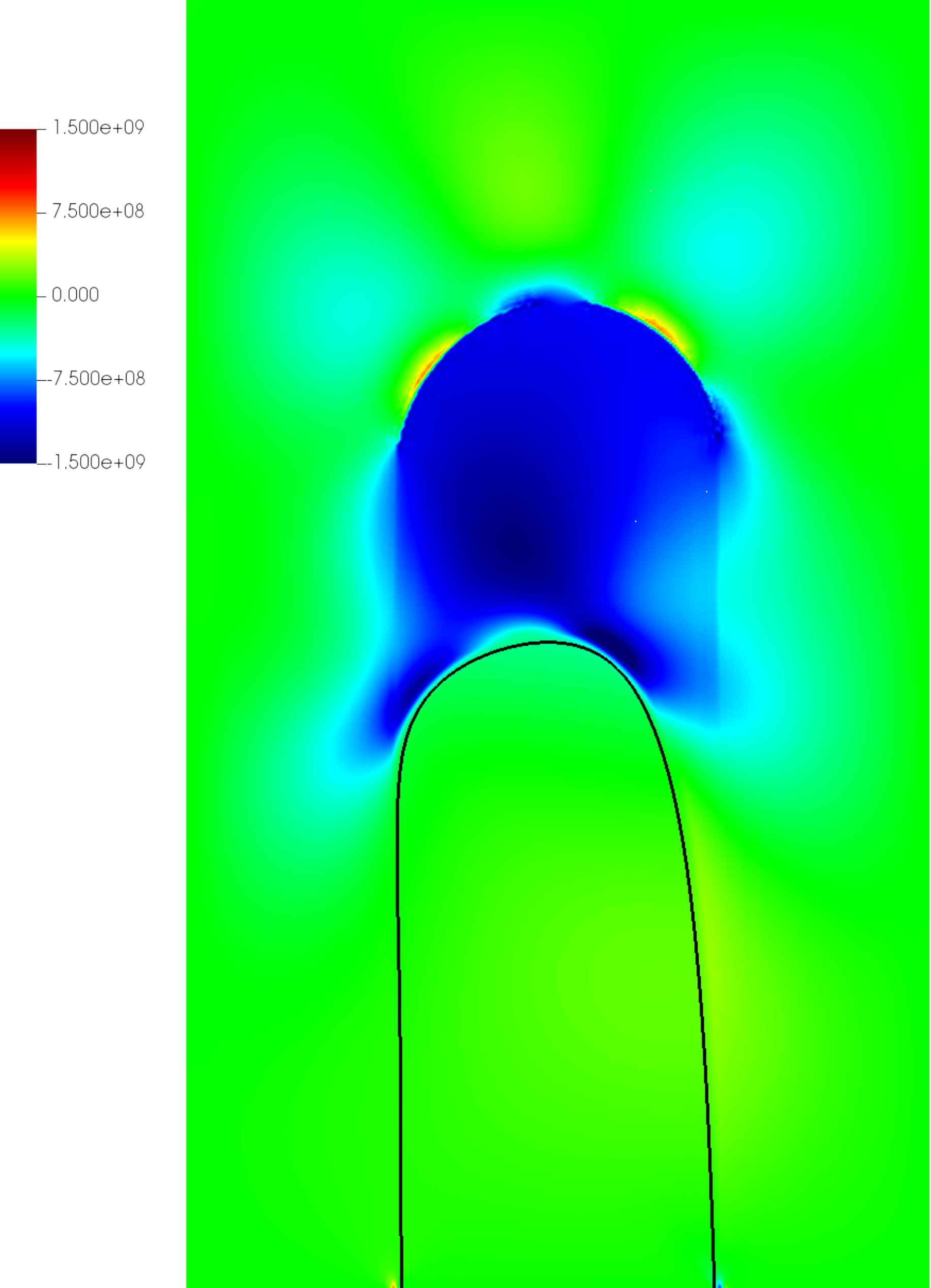}%
    \caption{Shear stress $\sigma_{xy}$ [Pa]}
    \label{fig:results/HalfLoop/output_reverse/sigma_xy}
  \end{subfigure}%
  \begin{subfigure}{0.5\linewidth}
    \includegraphics[width=1cm,clip,trim=0cm 27cm 25cm 3cm]
{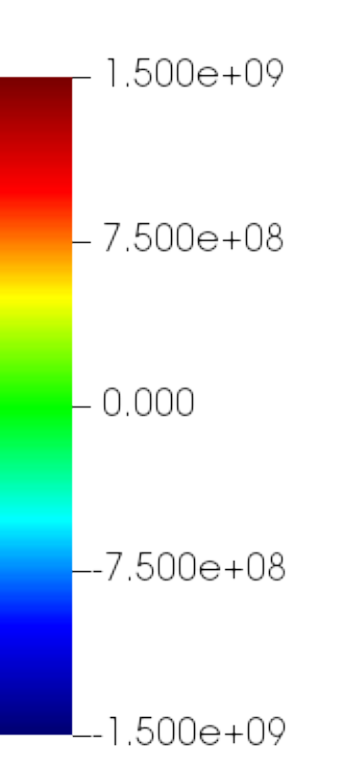}%
    \includegraphics[height=3cm,clip,trim=6cm 0cm 0cm 0cm]
{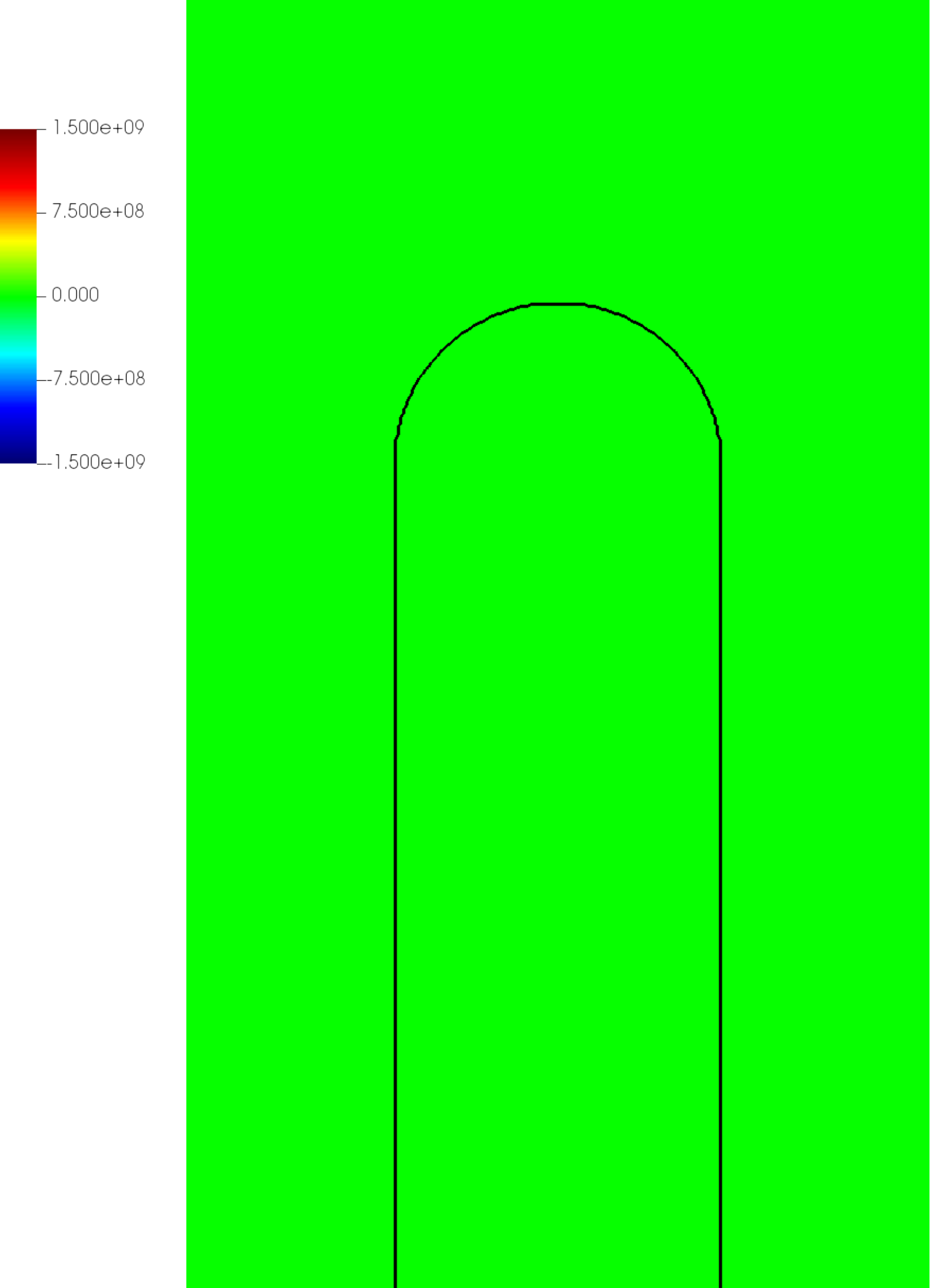}%
    \includegraphics[height=3cm,clip,trim=6cm 0cm 0cm 0cm]
{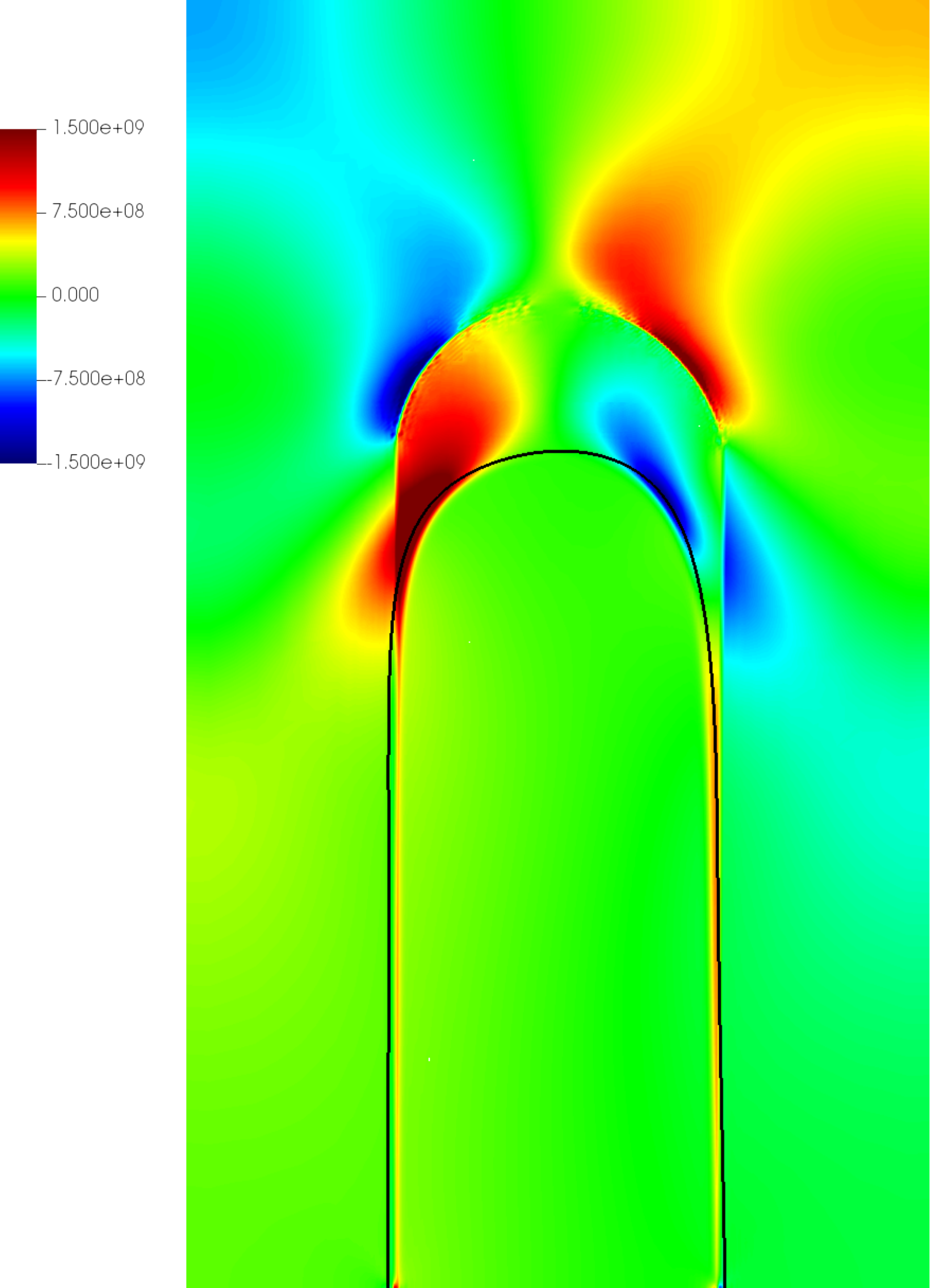}%
    \includegraphics[height=3cm,clip,trim=6cm 0cm 0cm 0cm]
{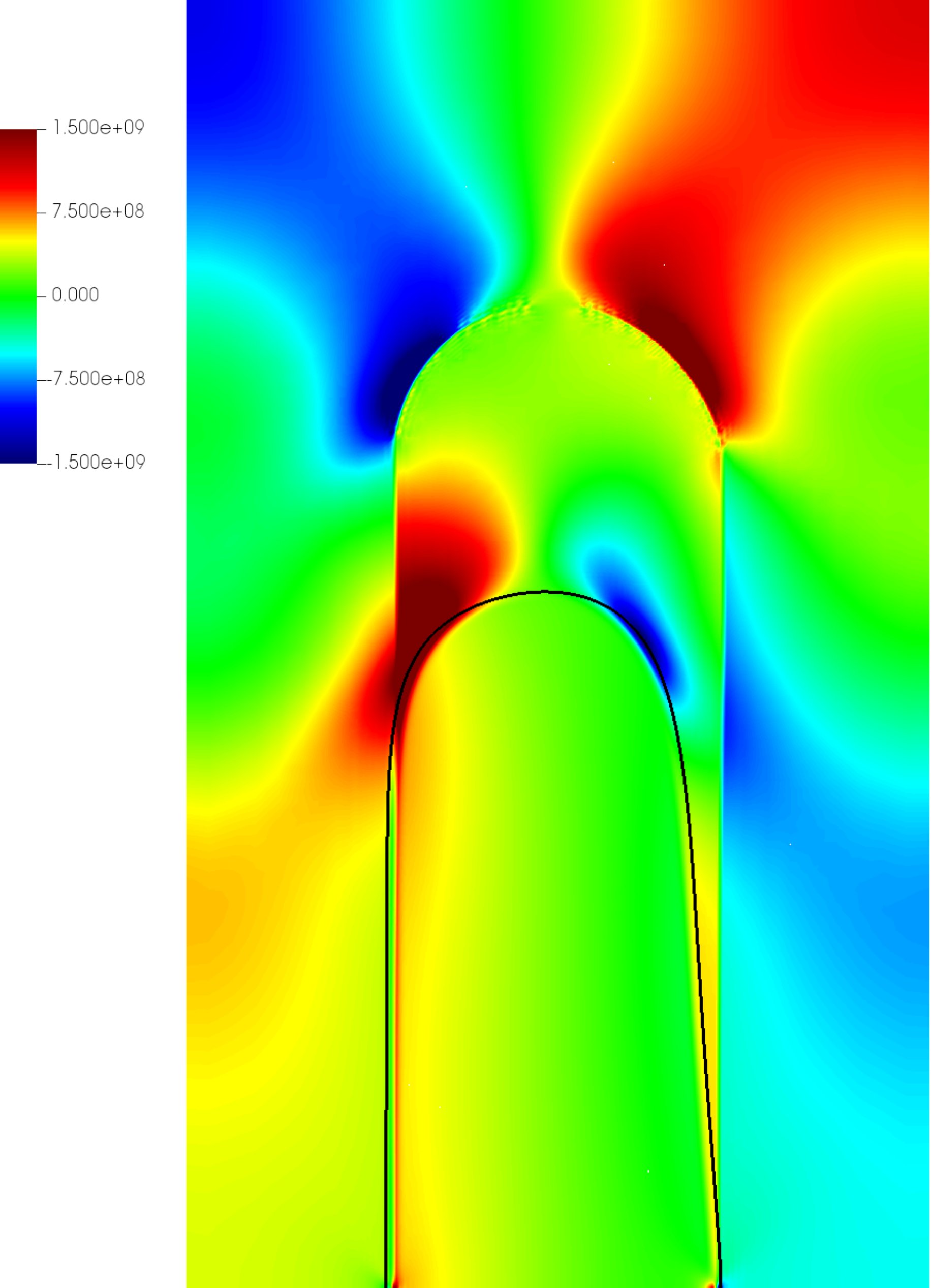}%
    \includegraphics[height=3cm,clip,trim=6cm 0cm 0cm 0cm]
{a6eba35c24d97e4e.pdf}\par
    \hspace{1cm}%
    \includegraphics[height=3cm,clip,trim=6cm 0cm 0cm 0cm]
{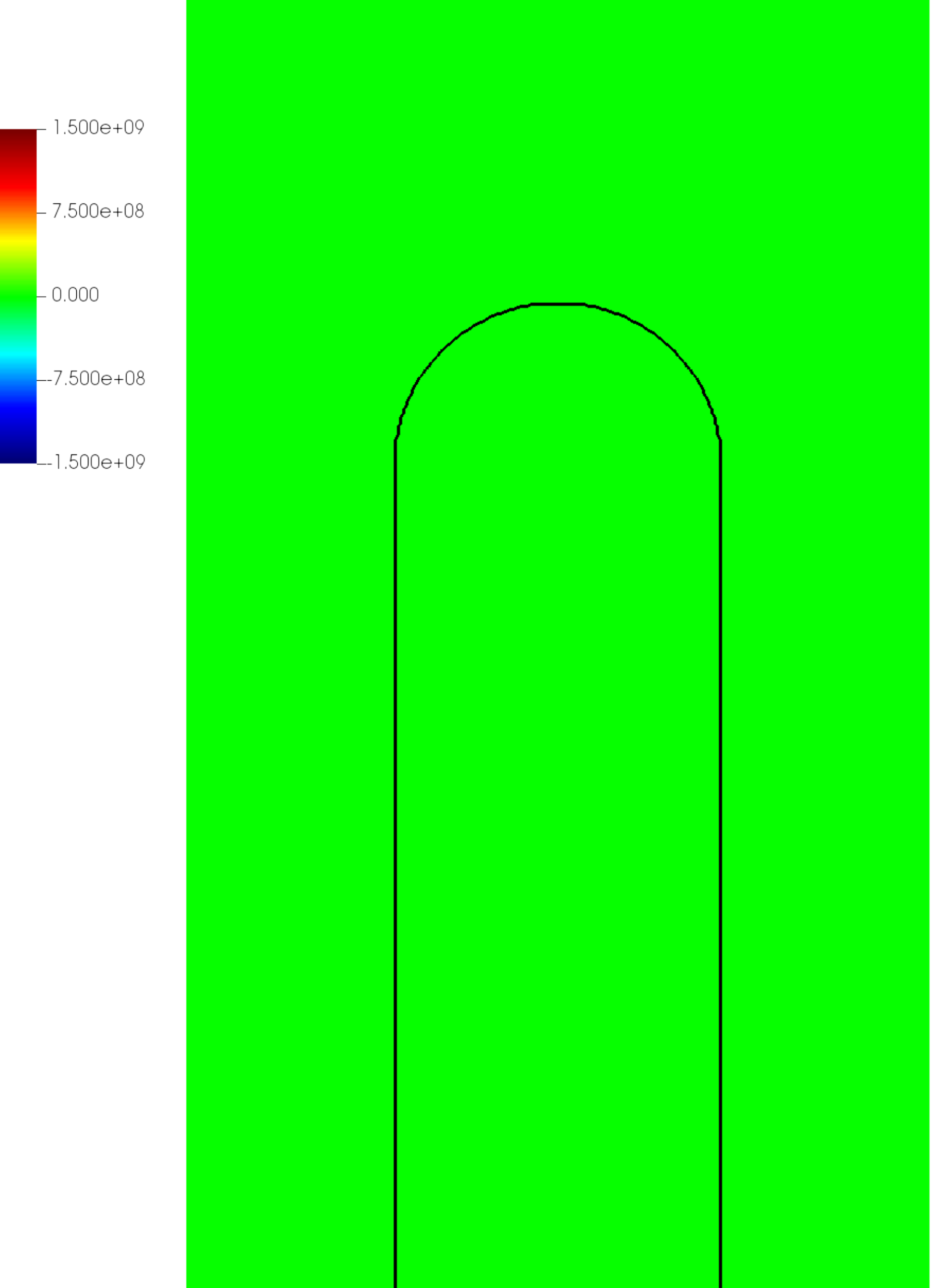}%
    \includegraphics[height=3cm,clip,trim=6cm 0cm 0cm 0cm]
{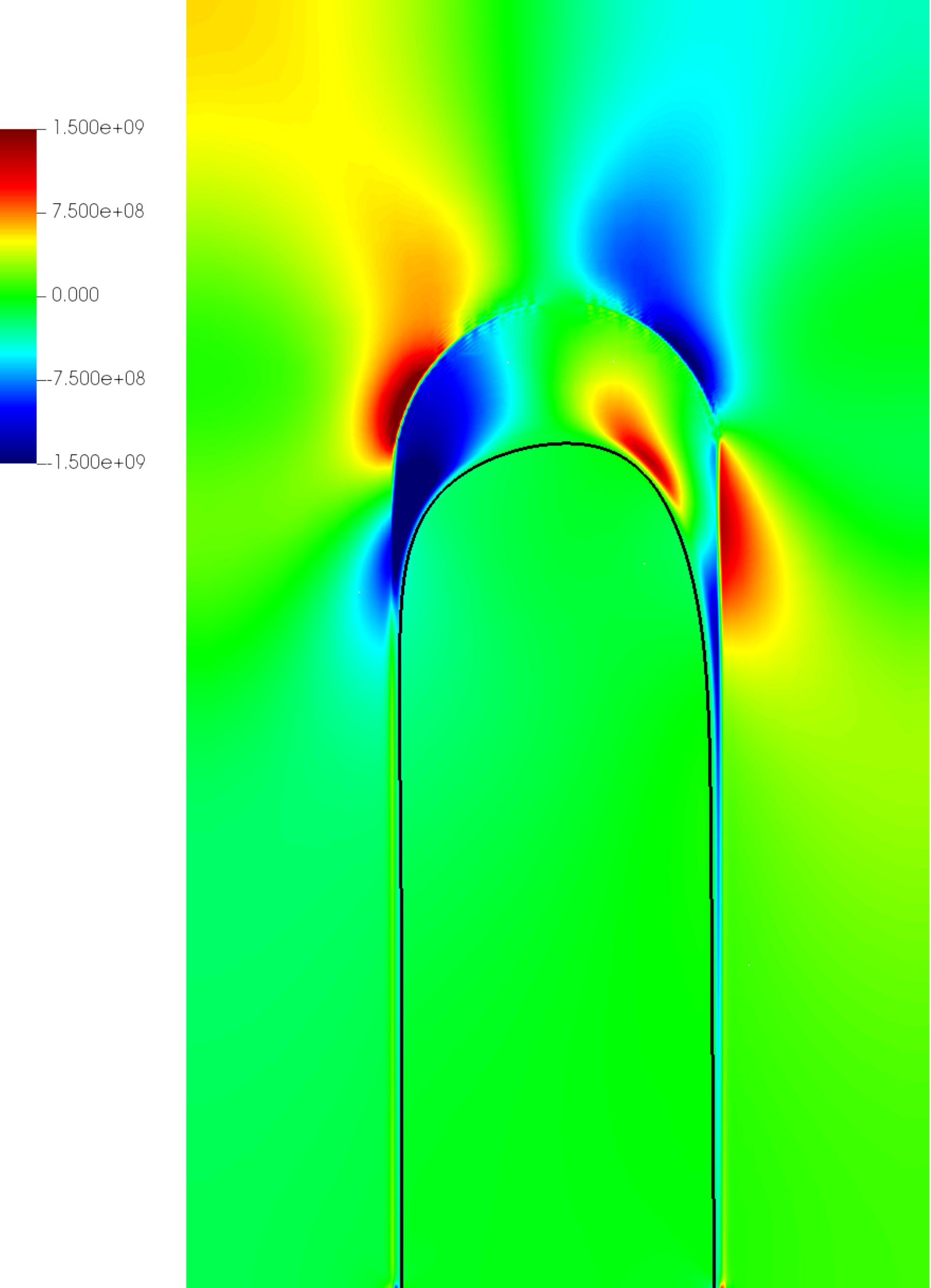}%
    \includegraphics[height=3cm,clip,trim=6cm 0cm 0cm 0cm]
{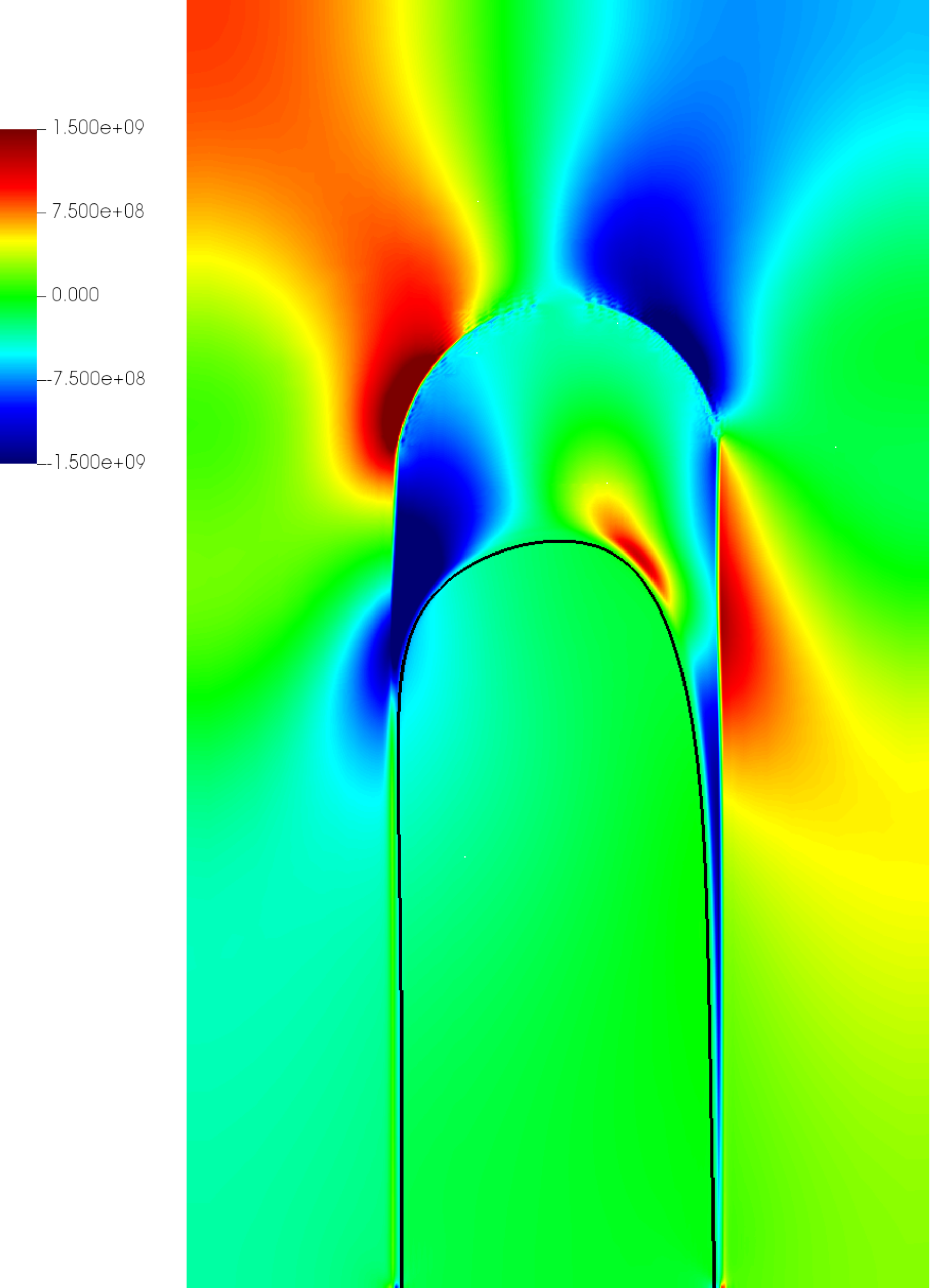}%
    \includegraphics[height=3cm,clip,trim=6cm 0cm 0cm 0cm]
{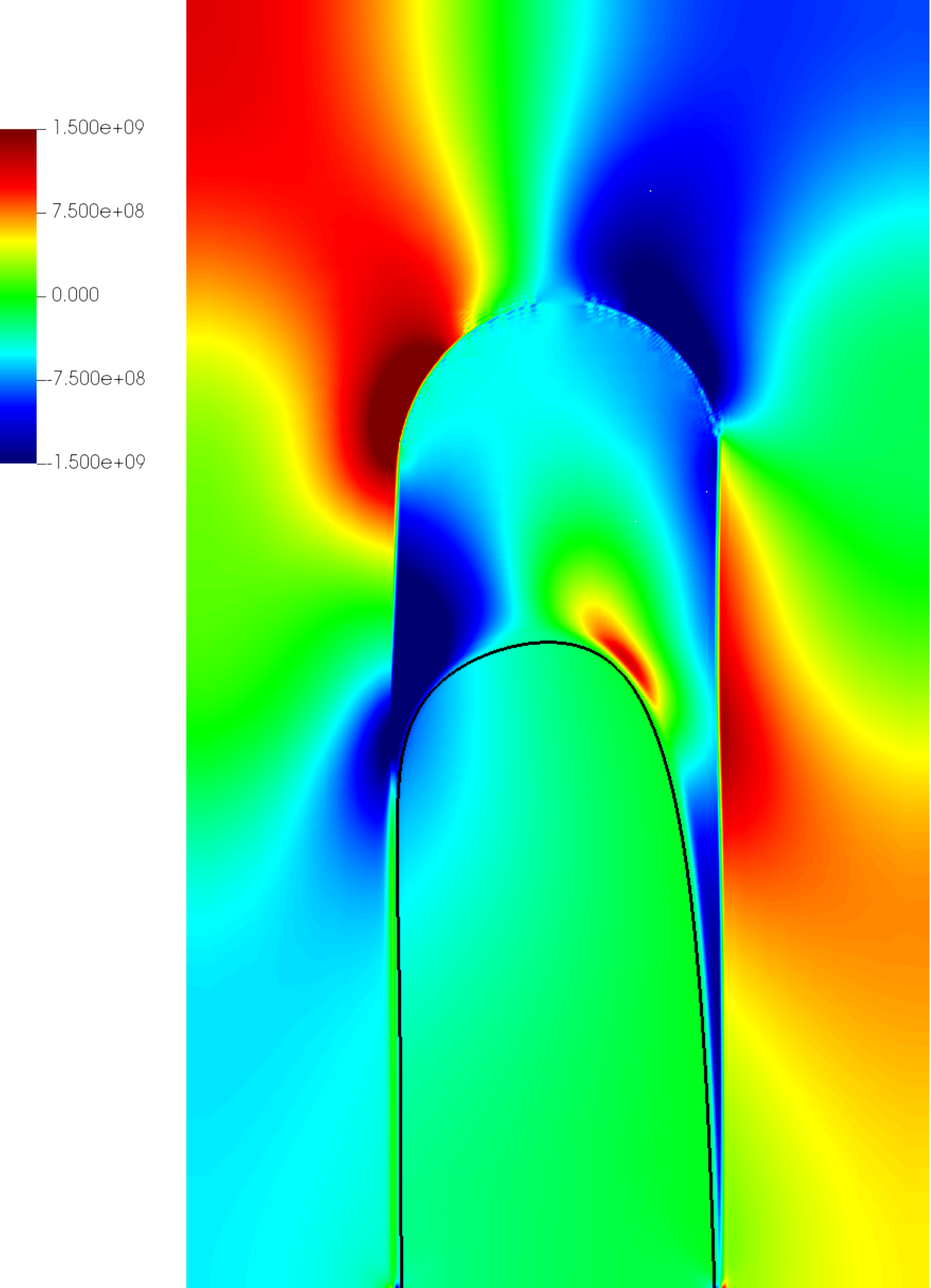}%
    \caption{Normal stress $\sigma_{yy}$ [Pa]}
    \label{fig:results/HalfLoop/output_reverse/sigma_yy}
  \end{subfigure}
  \caption{Back-stress distribution observed in curvature-driven ATGB migration.}
\end{figure}

The final example considered here is the evolution of a boundary with a ``half-loop'' geometry.
The use of a loop instead of a circle allows for the deliberate imposition of asymmetry into the system, as neither side of the half-loop are symmetric.
On the other hand, the only driving force present in the system is the curvature force from the curved part of the loop.
Here, we match as closely as possible the configuration presented in \cite[Fig. 2e]{qiu2024grain}.
The mechanical properties are the same as that in the previous results.
The STGB is a $\hkl<111>$ symmetric tilt boundary, which is inclined from the x axis by $\phi=13.9^\circ$ (\cref{fig:results/HalfLoop/drawing}).
The $\eta$ boundary condition is Neumann on all sides.
The mechanical boundary condition for displacement is Dirichlet in the normal direction (i.e. fixed displacement) and Neumann in the tangential direction.
As in \cite{qiu2024grain}, two cases are considered that are identical except for the placement of grain 1 vs grain 2, with the goal of determining whether the response changes.

In both cases (named ``grain 1 outside'' and ``grain 1 inside''), the simulation is allowed to relax and the volume of the first grain, as well as the location of the tip  is determined with respect to time (\cref{fig:results/HalfLoop/velocities}).
The average velocities are 0.47nm/$\mu$s and 0.34nm/$\mu$s, respectively, a difference of 27\%.
If the average curvature for both is assumed to be roughly similar (as is indeed apparent from visual inspection of the boundary migration), this corresponds to a similar percentage difference in the {\it effective mobility} as well.
As before, a constant, uniform value for mobility is used; the only change between the two simulations is the actual location of the grains (inside vs outside).  
Thus, the asymmetric response cannot be linked to an asymmetric mobility: rather, the explanation lies in the mechanics.

\subsubsection{Summary}

\replaced[id=R1,comment={1.3}]{This selection of results shows that, when directional dependency appears, it is not the result of an asymmetric mobility rate coefficient.
Rather, the SDF- and curvature-driven directional response can be explained by differences in back-stress generated from the incompatibility in shear coupling matrices, itself resulting from the asymmetry of the boundary.}{This selection of results shows that directional dependency is not the result of an asymmetric mobility rate coefficient.
Rather, this behavior can be entirely explained by the difference in back-stress generated from the incompatibility in shear coupling matrices, itself resulting from the asymmetry of the boundary.}
As the boundary moves, a jump in elastic deformation is required to maintain compatibility (\cref{eq:backstressjump}).
\replaced[id=R1,comment={1.3}]{Because the shear coupling matrix is different (and because of elastic anisotropy), this can produce differing responses based on the grain through which the boundary migrates.
This implicit change in driving force may manifest as direction-dependency in boundary response, although the mechanically driven examples show that the effect can be suppressed by the driving mode and boundary character.}{Because the shear coupling matrix is different (and because of elastic anisotropy), this produces differing responses based on the grain through which the boundary migrates.
This implicit change in driving force, which is direction dependent, manifests as direction-dependency in boundary response.}
Thus, a direction dependence of the \emph{apparent} driving force may be observed; however, the rate coefficient itself is actually perfectly symmetric.

This model considers boundary migration as the only deformation mechanism, meaning that the observed material exhibits higher stresses and less deformation than its realistic (or atomistic simulation) counterpart.
\added[id=R1]{Accordingly, the examples should be interpreted as isolating the mechanical consequences of incompatible shear-coupled migration rather than as complete predictions for situations in which lattice dislocation activity, climb, diffusion, or other relaxation mechanisms are active on comparable time scales.}
\replaced[id=R1]{In systems where those mechanisms are present, they may relax part of the back-stress, alter the threshold for motion, or change the apparent asymmetry; a coupled dislocation/disconnection or crystal-plasticity model would be required to make that quantitative.}{This, however, underlies the importance of the result, as there is no reason to believe that dislocation-mediated deformation would counteract the asymmetric response.}
\replaced[id=R1,comment={1.4}]{The intended domain of applicability is therefore mesoscale boundary-mediated plasticity in which boundary migration and shear coupling are the mechanisms of interest, and competing plastic mechanisms are either suppressed, secondary, or represented by an additional constitutive model.}{On the contrary, it is more likely that a combined dislocation/disconnection plasticity model will exacerbate the asymmetry, due to the asymmetric slip systems present across the boundary.}

\section{Conclusion}\label{sec:conclusion}

This work introduces a phase field formulation for boundary-mediated plasticity in which a spatially varying grain boundary eigendeformation evolves by a flow rule coupled to phase field kinetics.
The formulation removes the need to prescribe grain-wise eigenstrains, and allows eigendeformation to develop through boundary motion, enabling general initial conditions without Eshelby-like initialization artifacts.
This model equips phase field with a spatially varying eigendeformation whose evolution is governed by boundary motion, resembling a crystal-plasticity model in which slip is regulated by microstructural kinetics.

Across several test problems, the model reproduces behaviors that are difficult to capture with conventional phase field coupling.
Inclusions can be initialized stress-free and stabilized by incompatibility-induced back-stress.
Perturbed symmetric tilt boundaries develop threshold-like resistance, and leave defect-like residuals following migration.
For asymmetric tilt boundaries, forced incompatible motion produces large back-stress and can trigger a transition from quasi-planar migration to lamination at larger inclinations.

The framework also provides a mesoscale explanation for ``ratcheting''-like response of asymmetric boundaries.
\replaced[id=R1,comment={1.3}]{In synthetically driven and curvature-driven cases, asymmetric boundaries migrate at different apparent mobilities despite a strictly symmetric mobility coefficient; the mechanically driven cases serve as a complementary check showing that the magnitude of the effect depends on the driving mode and boundary character.}{In synthetically driven, mechanically driven, and curvature driven cases, asymmetric boundaries migrate at different apparent mobilities despite a strictly symmetric mobility coefficient.}
The asymmetry arises from direction-dependent back-stress generated by incompatibility, not intrinsically asymmetric kinetics.
More broadly, these results identify elastic incompatibility as a key factor in interface-mediated plasticity.

\replaced[id=R1]{Though not prohibited by the theory, all of the examples in this work consider boundary-mediated plasticity to be the sole inelastic deformation mechanism; dislocation plasticity is not included, although it would likely be present in many realistic microstructures.}{Though not prohibited by the theory, all of the examples in this work considered boundary-mediated plasticity to be the sole damage mechanism; dislocation plasticity is not considered, though it is likely that it would be present in many of the cases of interest.}
\replaced[id=R1]{This restriction is deliberate: the purpose of the examples is to expose the elastic-compatibility contribution of the proposed boundary model in isolation.}{This is by design, as the present goal is to demonstrate the mechanical model alone, but the model can be extended to include crystal plasticity if intended to model realistic systems.}
\added[id=R1,comment={1.4}]{For quantitative prediction in regimes where dislocation plasticity, diffusion-assisted relaxation, or grain-boundary structural transformations compete with shear-coupled migration, the present model should be coupled to the appropriate additional kinetics rather than used as a stand-alone description.}

In conclusion, this work presents a new model, and uses this new method to show that elastic compatibility plays a central role in governing boundary-mediated plasticity, acting not merely as a constraint but as an active source of resistance and hysteresis.
With this perspective, such phenomena as thresholding, defect emission, and asymmetry are emergent consequences of mechanics rather than intrinsic material properties.
This provides a natural bridge between atomistic observations and mesoscale modeling of boundary-driven deformation.

\subsection*{Acknowledgments}

This work was funded by the National Science Foundation through the CAREER program, grant \# 2341922.

\bibliography{main,zotero}

\end{document}